\documentclass[prx, twocolumn, nofootinbib, showpacs, amsmath, amssymb, floatfix, eqsecnum]{revtex4-1}
\usepackage{amsmath}
\usepackage{amssymb}
\usepackage{amsthm}
\usepackage{amsfonts}

\usepackage{graphicx,color,framed}
\usepackage{hyperref}
\usepackage{times}
\usepackage{enumerate}
\usepackage{lipsum}
\usepackage{slashed}
\usepackage{url}
\usepackage{bbm}
\hypersetup{
    colorlinks=true, 
    linktoc=all,     
    linkcolor=blue,  
}

\def \beq {\begin{equation}}
\def \eeq {\end{equation}}
\def \beqa {\begin{eqnarray}}
\def \eeqa {\end{eqnarray}}
\def \bseq {\begin{subequations}}
\def \eseq {\end{subequations}}

\newcommand \M {\mathcal{M}}

\newcommand{\<}{\langle}
\renewcommand{\>}{\rangle}

\begin{document}

\title{Classification of interacting Floquet phases with $U(1)$ symmetry in two dimensions
}

\author{Carolyn Zhang}
\author{Michael Levin}
\affiliation{Department of Physics, Kadanoff Center for Theoretical Physics, University of Chicago, Chicago, Illinois 60637,  USA}

\begin{abstract}
We derive a complete classification of Floquet phases of interacting bosons and fermions with $U(1)$ symmetry in two spatial dimensions. According to our classification, there is a one-to-one correspondence between these Floquet phases and rational functions $\pi(z) = a(z)/b(z)$ where $a(z)$ and $b(z)$ are polynomials obeying certain conditions and $z$ is a formal parameter. The physical meaning of $\pi(z)$ involves the stroboscopic \emph{edge} dynamics of the corresponding Floquet system: in the case of bosonic systems, $\pi(z) = \frac{p}{q} \cdot \tilde{\pi}(z)$ where $\frac{p}{q}$ is a rational number which characterizes the flow of quantum information at the edge during each driving period, and $\tilde{\pi}(z)$ is a rational function which characterizes the flow of $U(1)$ charge at the edge. A similar decomposition exists in the fermionic case. We also show that $\tilde{\pi}(z)$ is directly related to the time-averaged $U(1)$ current that flows in a particular geometry. This $U(1)$ current is a generalization of the quantized current and quantized magnetization density found in previous studies of non-interacting fermionic Floquet phases.
\end{abstract}

\maketitle

\section{Introduction}\label{sintroduction}

In recent years, it has become evident that periodically driven (``Floquet'') quantum many-body systems can exhibit a rich array of physical phenomena\cite{floquetreview, rudnerband}. These phenomena are particularly clear in Floquet systems that are many-body localized\cite{abanincolloquium, fate, ponte, abanintheory, bordia}.
Floquet systems of this kind are special in that they do not thermalize or thermalize very slowly. As a result, they can display interesting dynamical behavior over long time scales, unlike generic interacting Floquet systems which absorb energy from the drive and ultimately heat up to infinite temperature\cite{rigollongtime, lazaridesgeneric, ponteergodic}.

Once we specialize to many-body localized Floquet systems, it is possible to define a notion of a Floquet ``phase'' -- i.e. an equivalence class of Floquet systems with the same qualitative properties\cite{floquetreview, khemaniphase, keyserlingk1DI, keyserlingk1DII, cohomology, potter1D, roy1D}. 
In fact, there are several ways to define this concept (see Appendix~\ref{sphase}). The definition we will use in this paper is that two Floquet systems belong to the same phase if it is possible to construct a spatial boundary between the two systems that is many-body localized and preserves all relevant symmetries.\footnote{We explain this definition in more detail in Sec.~\ref{smbl} below.} 

An interesting example of a Floquet phase is the two dimensional (2D) ``SWAP circuit'' introduced in Refs.~\onlinecite{chiralbosons, harperorder}. SWAP circuits are Floquet systems that can be constructed out of either bosonic or fermionic degrees of freedom living on the sites of the square lattice. The most important property of these systems is their \emph{edge} dynamics: when a SWAP circuit is defined on a finite lattice with a boundary, one finds that the lattice sites near the edge undergo a unit translation during each driving period. Using this edge dynamics, Refs.~\onlinecite{chiralbosons, harperorder} argued that SWAP circuits are examples of non-trivial Floquet phases, independent of any symmetry. 

Going a step further, Ref.~\onlinecite{chiralbosons, fermionic, harperorder} derived a complete classification of 2D Floquet phases without symmetry. According to this classification, every Floquet phase is uniquely labeled by a single number that quantifies the flow of quantum information at the edge\cite{tracking}. In the case of bosonic phases, this number -- which is based on the ``GNVW index'' of Ref.~\onlinecite{GNVW} -- can take any positive rational value $p/q$. Likewise, fermionic Floquet phases are labeled by numbers of the form $\sqrt{2}^\zeta p/q$ with $\zeta = 0,1$.

The goal of this paper is to obtain a similarly systematic understanding of 2D Floquet phases with a $U(1)$ symmetry. Floquet phases of this kind were studied previously by several groups. In one line of research, Ref.~\onlinecite{kitagawa2010, anomalousedge, afai}
showed the existence of nontrivial $U(1)$ symmetric Floquet phases built out of non-interacting fermions. Refs.~\onlinecite{magnetization, kundubias} found simple physical signatures of these phases involving quantized currents and quantized magnetization densities, and Ref.~\onlinecite{nathanafi} argued that these phases are stable to weak interactions. In another line of work, Ref.~\onlinecite{eft} reproduced some of these results from the field theory perspective using the Keldysh formalism, and also derived invariants for certain strongly interacting systems. 

An important question raised by this body of work is whether new types of $U(1)$ symmetric Floquet phases can be realized in general interacting systems. Another question is whether there exists a more general invariant that unifies the quantized current and magnetization density of Ref.~\onlinecite{magnetization} with the GNVW index. 

In this paper we address these and other questions by deriving a complete classification of 2D $U(1)$ symmetric Floquet phases of interacting bosons and fermions. According to our classification, there is a one-to-one correspondence between Floquet phases of this kind and rational \emph{functions}  $\pi(z) = a(z)/b(z)$, where $a(z)$ and $b(z)$ are polynomials obeying certain conditions and $z$ is a formal parameter. Our invariant $\pi(z)$ contains two different pieces of information about the corresponding Floquet phase: in the case of bosonic systems, $\pi(z) = \frac{p}{q} \cdot \tilde{\pi}(z)$ where $\frac{p}{q}$ is the previously discussed GNVW index which characterizes the flow of quantum information at the edge, and $\tilde{\pi}(z)$ is a new invariant which characterizes the flow of $U(1)$ \emph{charge} at the edge. In the fermionic case the invariant has a similar structure but with the bosonic index $\frac{p}{q}$ replaced by its fermionic counterpart, $\sqrt{2}^\zeta \frac{p}{q}$.

In addition to our classification results, we also discuss the physical signatures of these $U(1)$ symmetric phases. In particular, we show that upon substituting $z=e^\mu$ where $\mu$ is a chemical potential, our invariant $\tilde{\pi}(z)$ is directly related to the time-averaged $U(1)$ current that flows in a particular geometry. The $U(1)$ current that flows in our setup is a generalization of the quantized current and magnetization density introduced in Ref.~\onlinecite{magnetization}.

We derive our classification using the same approach as Ref.~\onlinecite{chiralbosons, fermionic}. First, we use a bulk-boundary correspondence argument to map our classification problem onto a simpler problem of classifying 1D $U(1)$ symmetric locality preserving unitaries (LPUs). We then solve the latter problem with the help of the powerful mathematical machinery developed by Ref.~\onlinecite{GNVW}. 

We note that some of our results were anticipated by Ref.~\onlinecite{kirby}, which proposed invariants for classifying 1D LPUs with continuous and discrete symmetries. Our work is also connected to Ref.~\onlinecite{mpu}, which discussed the classification of 1D matrix product unitaries with discrete symmetries.

The rest of the paper is structured as follows: in Sec.~\ref{sex0} we preview our results with a simple example of a $U(1)$ symmetric Floquet phase. In Sec.~\ref{smbl}, we explain our definition of Floquet phases and we review the connection between the classification of 2D Floquet phases and 1D LPUs. In Sec~\ref{sclassification}, we present our classification result in the case of bosonic systems and we illustrate it with several examples in Sec.~\ref{sexamples}. In Sec.~\ref{sdefs}, we define our invariant $\pi(z)$ and in Sec.~\ref{sproofs}, we derive our (bosonic) classification result. In Sec.~\ref{smagnetization}, we show that the invariant $\tilde{\pi}(z)$ is directly related to the $U(1)$ current that flows in a particular geometry. In Sec.~\ref{sfermionic}, we extend our classification to fermionic systems. Finally in Sec.~\ref{scohomology}, we discuss the relationship between our classification and the cohomology classification of Floquet symmetry protected topological (SPT) phases\cite{keyserlingk1DI,keyserlingk1DII, cohomology, potter1D, roy1D, alldimensions}. We discuss some remaining questions and extensions in Sec.~\ref{sdiscussion}. Technical details and additional proofs can be found in the Appendices.

\section{Preview: Example of a $U(1)$ symmetric Floquet phase}\label{sex0}

\begin{figure}[tb]
   \centering
   \includegraphics[width=.9\columnwidth]{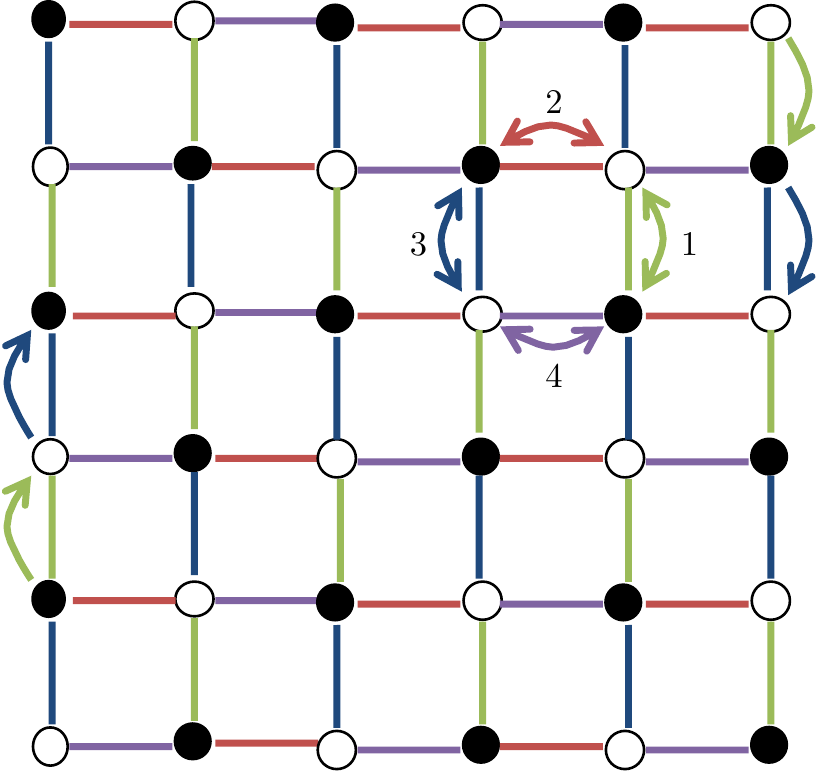} 
   \caption{The four steps of the SWAP circuit. Each step involves a set of disjoint SWAP gates, along the bond type $1\text{ (green)},2\text{ (red)},3\text{ (blue)},$ or $4\text{ (purple)}$ indicated above. After all four steps, the spin operators in the bulk return to their original positions, while the spins on the edge are translated by a unit cell.}
   \label{fig:swapcircuit}
\end{figure}

We begin with an example that illustrates some of our main results. This example is based on the ``SWAP circuit" introduced in Refs.~\onlinecite{chiralbosons, harperorder}, so we begin by reviewing this circuit. Consider a spin system with $d$-state spins located on the sites of the square lattice. The Hamiltonian $H(t)$ is periodic with period $T$ and with the following structure: for $0 \leq t \leq T/4$, we turn on a two-spin interaction on all the ``1" (green) bonds in Fig.~\ref{fig:swapcircuit}, where this interaction is chosen so that it generates a SWAP gate on each pair of spins. Then for $T/4 \leq t \leq T/2$, we turn on a two-spin interaction on all the ``2" (red) bonds, again implementing a SWAP gate on the corresponding pairs of spins. We then repeat this for the ``3" (blue) bonds for $T/2 \leq t \leq 3T/4$, and the ``4" (purple) bonds for $3T/4 \leq t \leq T$. 

To understand the dynamics of the SWAP circuit, consider the Heisenberg evolution of a (single site) spin operator during a period $T$. In the bulk, each single site spin operator undergoes four swaps, ultimately returning to its original position; it follows that the Floquet unitary acts like the identity operator in the bulk. On the other hand, at the edge, the spin operators undergo a \emph{translation} (Fig.~\ref{fig:swapcircuit}). (More precisely, the spins on one sublattice undergo a two site translation, while the spins on the other sublattice are left alone. All together this corresponds to a unit translation with a two site unit cell).

The above edge translation is significant because it is ``anomalous'': it cannot be generated by a strictly one dimensional local Hamiltonian. Starting from this observation, one can argue that the SWAP circuit belongs to a nontrivial Floquet phase, i.e. a different phase from the trivial Floquet system $H(t) = 0$. According to the classification of Refs.~\onlinecite{chiralbosons,harperorder}, 2D Floquet phases without symmetry are labeled by rational numbers $p/q$; in this classification scheme, the SWAP circuit is labeled by $p/q = d$ while the trivial phase is labeled by $p/q=1$.

With this background we can now present an example of a Floquet phase protected by $U(1)$ symmetry. Consider a 2D square lattice with a four state spin on each site. Suppose that the four states on each site carry $U(1)$ charges $0,1,2,3$ respectively. This structure can be summarized by a $4\times 4$ charge matrix $Q$:
\begin{equation}
Q=\begin{pmatrix}
0 & 0 & 0 & 0\\
0 & 1 & 0 & 0\\
0 & 0 & 2 & 0\\
0 & 0 & 0 & 3
\end{pmatrix}
\end{equation}
Notice that $Q$ can be decomposed as a sum of the form
\begin{equation}
Q=Q_1\otimes\mathbbm{1}+\mathbbm{1}\otimes Q_2
\end{equation}
where $\mathbbm{1}$ denotes the $2 \times 2$ identity matrix and where
\begin{equation}
Q_1=\begin{pmatrix}
0 & 0\\
0 & 1\end{pmatrix}
\qquad
Q_2=\begin{pmatrix}
0 & 0\\
0 & 2\end{pmatrix}
\end{equation}
This decomposition implies that we can factor each four state site into $2$ two-level systems describing hard core bosons carrying charge $1$ and charge $2$ respectively. We then consider a Floquet system which performs a SWAP circuit on the charge $2$ bosons, and another SWAP circuit on the charge $1$ bosons but with the opposite chirality (i.e. steps 1-4 reversed). During each period, the two types of bosons will undergo (edge) translations in opposite directions.

The above Floquet system has nontrivial charge flow on the edge, so one might suspect that it is non-trivial. Indeed, we will show that this system belongs to a Floquet phase which is non-trivial in the presence of $U(1)$ symmetry, but trivial if this symmetry is broken. More generally, we will show that $U(1)$ symmetric Floquet phases are classified by rational functions $\pi(z) = a(z)/b(z)$ where $z$ is a formal parameter. This particular example is labeled by $\pi(z)=\frac{1+z^2}{1+z}$, which is distinct from the trivial phase which is labeled by $\pi(z)=1$.

\section{2D Floquet phases and 1D locality preserving unitaries}\label{smbl}

In this section we define the concept of a phase in 2D Floquet systems and we show that the classification of these phases is closely related to the classification of 1D locality preserving unitaries. We begin by reviewing these ideas for 2D Floquet systems without any symmetry\cite{chiralbosons}; we then explain how the story generalizes to the case with $U(1)$ symmetry.

\subsection{Review: No symmetry case}

\subsubsection{Definitions}
We consider \emph{bosonic}\footnote{See Sec.~\ref{sfermionic} for a discussion of the fermionic case.} Floquet systems, built out of a two-dimensional lattice of $d$-state spins. The Hamiltonian $H(t)$ can be arbitrary with the only restrictions being (i) $H(t)$ is \emph{local} in the sense that it includes only finite-range spin interactions, and (ii) $H(t)$ is periodic in time:
\begin{equation}
H(t+T)=H(t)
\end{equation}
where $T$ is the period. The stroboscopic dynamics of these systems is determined by the Floquet unitary, which gives the time-evolution over one period:
\begin{equation}
U_F=\mathcal{T}e^{-i\int_0^T H(t)dt}
\end{equation}
Here $\mathcal{T}$ denotes time ordering. 

In a typical interacting system, one expects that the stroboscopic dynamics described by $U_F$ will lead to thermalization at infinite temperature since energy is not conserved\cite{rigollongtime, lazaridesgeneric,ponteergodic}. To avoid this fate, we restrict our attention to Floquet systems in which $U_F$ is ``many-body localized.'' In the present context, this amounts to the requirement that $U_F$ can be written as a product of mutually commuting quasi-local unitaries\cite{chiralbosons}. That is:
\begin{align}
U_F = \prod_r U_r, \quad \quad [U_r , U_{r'}] = 0
\label{mblcond}
\end{align}
where each $U_r$ denotes a unitary that is supported within a distance of $\xi$ of site $r$ (possibly with exponentially decaying tails) where $\xi$ is a fixed localization length that does not vary with the system size. We will refer to (\ref{mblcond}) as the ``MBL condition''. 

We are now ready to define the concept of a Floquet phase: we say that two (MBL) Floquet systems, $H_A(t)$ and $H_B(t)$ belong to the same phase if the 1D boundary between $H_A(t)$ and $H_B(t)$ can be many-body localized. More precisely, let $H_A^-(t)$ be the restriction of the Hamiltonian $H_A$ to the lower half-plane $\mathbf{R}^- = \{r: r_y \leq 0\}$ -- that is, $H_A^-(t)$ contains all the terms in $H_A(t)$ that are supported entirely within $\mathbf{R}^-$. Similarly, let $H_B^+(t)$ be the restriction of $H_B(t)$ to the upper half-plane $\mathbf{R}^+ = \{r: r_y > 0\}$ -- that is, $H_B^+(t)$ contains all the terms in $H_B(t)$ that are supported entirely within $\mathbf{R}^+$. We will say that $H_A(t)$ and $H_B(t)$ belong to the same phase if there exists at least one choice of ``boundary Hamiltonian'' $H_{\mathrm{bd}}(t)$, supported within a finite distance from the $x$-axis, such that the composite system with Hamiltonian 
\begin{align*}
H_{\mathrm{tot}} (t)= H_A^- (t)+ H_B^+ (t)+H_{\mathrm{bd}}(t)
\end{align*}
has a Floquet unitary $U_{F,\mathrm{tot}}$ that obeys the MBL condition (\ref{mblcond}). We emphasize that we allow $H_{\mathrm{bd}}(t)$ to be arbitrary, as long as it is local, periodic in time and supported within a finite distance of the $x$-axis; the key question is whether there exists \emph{any} boundary Hamiltonian $H_{\mathrm{bd}}(t)$ such that $U_{F,\mathrm{tot}}$ is many-body localized.

A few comments about the above definition of a Floquet phase: first, we should mention that while we have chosen a particular way to define a Floquet phase, there are other possible definitions. In general, different definitions may lead to different classifications. We compare our definition with two other definitions common in the literature in Appendix~\ref{sphase}. 

Another important comment is that one can make a rough analogy between Floquet phases and gapped (zero temperature) phases in stationary systems, by thinking of the MBL condition (\ref{mblcond}) as the analog of the spectral gap condition. Following this analogy, our definition of Floquet phases is similar to defining two gapped systems to belong to the same phase if the boundary between them can be gapped.

\subsubsection{Mapping to 1D locality preserving unitaries}
\label{1dunitrev}
One of the most powerful tools for analyzing 2D Floquet phases is the bulk-boundary mapping introduced in Ref.~\onlinecite{chiralbosons}. This mapping associates a 1D unitary $U_{\mathrm{edge}}$ to every 2D Floquet system $H(t)$. Roughly speaking, $U_{\mathrm{edge}}$ describes the stroboscopic \emph{edge} dynamics of $H(t)$. 

To define $U_{\mathrm{edge}}$ precisely, let $H(t)$ be a 2D Floquet system defined in an infinite plane geometry and let $H^-(t)$ be the restriction of $H(t)$ to the lower half-plane $\mathbf{R}^-$: $H^-(t)$ contains all the terms in $H(t)$ that are supported entirely within $\mathbf{R}^-$. We then define a unitary $U_{\mathrm{edge}}$ by\cite{chiralbosons}
\begin{equation}\label{edgeu}
U_{\mathrm{edge}}=\mathcal{T}e^{-i\int_0^T H^{-}(t)dt} \cdot \prod_{r\in \mathbf{R}^-}U_r^\dagger
\end{equation}
where the $U_r$ operators are those that appear in the decomposition\footnote{In principle there could be multiple choices of $U_r$ that obey (\ref{urdef}), and different choices could lead to different unitaries $U_{\mathrm{edge}}$. However, this ambiguity is not important for our purposes: it follows from the proofs in Appendix~\ref{sUr} that different choices of $\{U_r\}$ lead to the same $U_{\mathrm{edge}}$ up to composition with a 1D FDLU, and ultimately we will only be interested in the value of $U_{\mathrm{edge}}$ \emph{modulo} 1D FDLUs, as we will see below.}
\begin{align}
U_F = \mathcal{T}e^{-i\int_0^T H(t)dt} = \prod_r U_r
\label{urdef}
\end{align}
The unitary $U_{\mathrm{edge}}$ has several important properties. First, it is supported within a finite distance from the edge of $\mathbf{R}^-$, i.e. the $x$-axis. To see this, note that the first term, $\mathcal{T}e^{-i\int_0^T H^{-}(t)dt}$ describes the stroboscopic dynamics of the restricted Hamiltonian $H^-(t)$, while $\prod_{r\in \mathbf{R}^-} U_r$ describes the \emph{restriction} of stroboscopic dynamics of $H(t)$. These two unitaries must coincide except near the edge of $\mathbf{R}^-$, so $U_{\mathrm{edge}}$ is supported near the edge region. 

Another important property of $U_{\mathrm{edge}}$ is that it is ``locality preserving'': that is, for any operator $O_r$ acting on site $r$, the conjugated operator $U_{\mathrm{edge}}^\dagger O_r U_{\mathrm{edge}}$ is supported within a finite distance of $r$ up to an exponentially small error term. To see this, note that $\mathcal{T}e^{-i\int_0^T H^{-}(t)dt}$ is locality preserving (by Lieb-Robinson bounds) while $\prod_{r\in \mathbf{R}^-}U_r^\dagger$ is also clearly locality preserving. 

Putting this all together, we have constructed a mapping from 2D Floquet systems $H(t)$ to 1D locality preserving unitaries $U_{\mathrm{edge}}$. Why is this mapping useful for classifying Floquet phases? The reason is the following result: in Appendix~\ref{sfdlu} we show that $H_A(t)$ and $H_B(t)$ belong to the same Floquet phase if and only if the corresponding 1D locality preserving unitaries (LPUs) $U_{A,\mathrm{edge}}$ and $U_{B,\mathrm{edge}}$ differ by a 1D finite-depth local unitary (FDLU):
\begin{align}
U_{A,\mathrm{edge}} = U_{B,\mathrm{edge}} \cdot (\text{1D FDLU})
\end{align}
Here, a 1D FDLU is a unitary that is generated by the time evolution of a local 1D Hamiltonian over a finite time $T$, i.e. a unitary of the form $ \mathcal{T} \exp( \int_0^T H_{1D}(t) dt)$. 

We conclude that the classification of 2D Floquet phases is equivalent to classifying 1D LPUs modulo 1D FDLUs:\footnote{Here we are using the fact that the mapping between Floquet systems and equivalence classes of 1D locality preserving unitaries is \emph{surjective}. This surjectivity can be demonstrated explicitly using SWAP circuits, as in Sec.~\ref{sconstructunitary}.}
\begin{align}
\{\text{2D Floquet phases} \} \leftrightarrow \frac{\{\text{1D LPUs}\}}{\{\text{1D FDLUs}\}}
\label{classmapping}
\end{align}

\subsubsection{Classification of 1D locality preserving unitaries}
The classification of 1D LPUs modulo 1D FDLUs was solved by Gross, Nesme, Vogts, and Werner in Ref.~\onlinecite{GNVW}. The authors showed that for a $d$-state spin chain, there is a one-to-one correspondence between equivalence classes of 1D LPUs and rational numbers $p/q$ of the form
\begin{align}
\frac{p}{q} = p_1^{n_1} p_2^{n_2} \cdots p_k^{n_k}, \quad n_i \in \mathbb{Z}
\label{pqform}
\end{align}
where $p_1,...,p_k$ are the prime factors of $d$. (Note that the $n_i$'s can be positive or negative integers, so the product on the right hand side is generally a rational number, rather than an integer). The authors also showed how compute the rational number $p/q$ for any 1D LPU $U$: 
\begin{align}
\frac{p}{q} = \mathrm{ind}(U)
\end{align}
where $\mathrm{ind}(U)$ is defined by the explicit formula in Appendix~\ref{GNVWformapp}. In what follows we will refer to $\mathrm{ind}(U)$ as the ``GNVW index.''

An intuitive way to understand this classification result is that it says that the only non-trivial 1D LPUs are \emph{translations} or combinations of translations. Each translation or combination of translations can be associated with a rational number as follows: for any combination of translations, the corresponding rational number $p/q$ can be obtained by letting $p$ be the total dimension of the Hilbert space that is translated in the positive direction, and $q$ be the total dimension of the Hilbert space that is translated in the negative direction.

\subsubsection{Classification of 2D Floquet phases without symmetry}
We now have everything we need to derive the classification of 2D Floquet phases without symmetry\cite{chiralbosons}: combining the GNVW classification of 1D LPUs with the bulk-boundary correspondence (\ref{classmapping}), it follows that there is a one-to-one correpondence between Floquet phases without symmetry and rational numbers $p/q$ of the form (\ref{pqform}). In this paper, we will apply a similar logic to classify 2D Floquet phases with $U(1)$ symmetry.

\subsection{$U(1)$ symmetric case}
\subsubsection{Setup}
\label{u1symmsetup}
As in the no-symmetry case, we consider \emph{bosonic} Floquet systems built out of a two-dimensional lattice of $d$-state spins. We assume that each spin transforms in the same way under the $U(1)$ symmetry, and we describe this $U(1)$ symmetry transformation using a $d \times d$ ``charge matrix'' $Q$ of the form
\begin{align}
Q = \begin{pmatrix} q_1 & 0 & \cdots & 0\\ 0 & q_2 & \cdots & 0 \\ \vdots & \vdots & \cdots & 0 \\ 0 & 0 & \cdots & q_d \end{pmatrix}
\label{Qdef}
\end{align}
where the $q_i$'s are the $U(1)$ charges of the different spin states. We will assume without loss of generality that the $q_i$'s are all \emph{non-negative} integers and that $\min_i q_i = 0$.\footnote{We do not lose any generality with this assumption since we can replace $Q \rightarrow \lambda Q + c \mathbbm{1}$ without affecting the classification.} 

In addition to the charge matrix $Q$, we also find it useful to define a charge \emph{operator} $Q_r$ associated to each lattice site $r$: this operator $Q_r$ measures the charge on site $r$ and can be thought of as a tensor product $Q_r = Q \otimes \mathbbm{1}$ where $Q$ acts on site $r$ and $\mathbbm{1}$ acts on the other sites. The \emph{total} $U(1)$ charge is then given by $Q_{tot} = \sum_r Q_r$. 

With this notation, we are now ready to explain our setup. We consider Hamiltonians $H(t)$ that are local, time-periodic, and $U(1)$ symmetric, i.e. $[H(t),Q_{\mathrm{tot}}]=0$. Just as in the case with no symmetry, we restrict to Floquet systems obeying the MBL condition (\ref{mblcond}) but with the additional requirement that each $U_r$ in (\ref{mblcond}) is $U(1)$ symmetric, i.e. $[U_r, Q_{\mathrm{tot}}] = 0$. 

Our definition of Floquet phases is similar to the case with no symmetry: we say that two $U(1)$ symmetric Floquet systems $H_A(t)$ and $H_B(t)$ belong in the same phase if the 1D boundary between $H_A(t)$ and $H_B(t)$ can be many-body localized while preserving the $U(1)$ symmetry. More precisely, let $H_A^-(t)$ be the restriction of the Hamiltonian $H_A$ to the lower half plane $\mathbf{R}^-$ and let $H_B^+(t)$ be the restriction of $H_B(t)$ to the upper half plane $\mathbf{R}^+$. We will say that $H_A$ and $H_B$ belong to the same phase if there exists at least one choice of a $U(1)$ symmetric boundary Hamiltonian $H_{\mathrm{bd}}(t)$ such that the composite system $H_{\mathrm{tot}} = H_A^- (t)+ H_B^+ (t)+H_{\mathrm{bd}}(t)$ has a Floquet unitary $U_{F,\mathrm{tot}}$ that obeys the MBL condition (\ref{mblcond}) where each  $U_r$ in (\ref{mblcond}) is $U(1)$ symmetric.

Note that we can already see that the classification of Floquet phases will depend crucially on the structure of $Q$. For example, if $Q\sim\mathbbm{1}$, then the $U(1)$ symmetry does not lead to any additional constraints, so the classification of phases in such systems is the same as in the case without symmetry. In general, though, the $U(1)$ symmetry makes the classification finer.

\subsubsection{Mapping to 1D  locality preserving unitaries}
\label{edgeU1}
As in the no-symmetry case, it is helpful to map each 2D Floquet system to a 1D edge unitary $U_{\mathrm{edge}}$. We do this in exactly the same way as before: we define the 1D unitary $U_{\mathrm{edge}}$ using (\ref{edgeu}). By construction, the unitary operator $U_{\mathrm{edge}}$ is locality preserving and $U(1)$ symmetric.

By the same logic as in the no-symmetry case, two Floquet systems $H_A(t)$ and $H_B(t)$ belong to the same Floquet phase if and only if the corresponding 1D unitaries, $U_{A,\mathrm{edge}}$ and $U_{B,\mathrm{edge}}$, differ by a 1D $U(1)$ symmetric FDLU:
\begin{equation}
U_{A,\mathrm{edge}}=U_{B,\mathrm{edge}} \cdot (\text{1D $U(1)$ symmetric FDLU})
\end{equation}
Here, by 1D $U(1)$ symmetric FDLU, we mean a unitary of the form $ \mathcal{T} \exp( \int_0^T H_{1D}(t) dt)$ where $H_{1D}(t)$ is a 1D local, $U(1)$ symmetric Hermitian Hamiltonian.

Putting this all together, it follows that the classification of 2D $U(1)$ symmetric Floquet phases is equivalent to the classification of 1D $U(1)$ symmetric LPUs modulo 1D $U(1)$ symmetric FDLUs:
\begin{align}
&\{\text{2D $U(1)$ symm. Floquet phases} \} \leftrightarrow \nonumber \\
&\frac{\{\text{1D $U(1)$ symm. LPUs}\}}{\{\text{1D $U(1)$ symm. FDLUs}\}}
\label{equivclass}
\end{align}
In this paper we work out the classification on the right hand side of Eq.\ref{equivclass}, and in this way derive a complete classification of 2D $U(1)$ symmetric Floquet phases. We explain our classification result in the next section. 

\section{Main result: classification of 2D $U(1)$ symmetric Floquet phases}\label{sclassification}
In this section we present our main result: a complete classification of 2D $U(1)$ symmetric Floquet phases that can be realized using $d$-state spins with a fixed charge matrix $Q$ (\ref{Qdef}).

In order to explain our result, it is useful to define a generating function $f_Q(z)$ which encodes the eigenvalue spectrum of $Q$. Specifically, define
\begin{equation}
f_Q(z) = \text{Tr}(z^Q) = \sum_{i=1}^d z^{q_i}
\end{equation}
where $z$ is a formal parameter and $q_1,...,q_d$ are the eigenvalues of $Q$. Given our assumption that the $q_i$'s are non-negative integers and that $\min_i q_i = 0$, it follows that $f_Q(z)$ is always a \emph{non-negative integer} polynomial, i.e. a polynomial with non-negative integer coefficients. 

At this point, we should mention a property of $f_Q(z)$ which will play an important role in this paper: $f_Q(z)$ is \emph{multiplicative} under tensor product. Suppose $\mathcal{H}_1, \mathcal{H}_2$ are two finite dimensional Hilbert spaces and $Q_1, Q_2$ are the corresponding charge matrices. If we denote the tensor product by $\mathcal{H} = \mathcal{H}_1 \otimes \mathcal{H}_2$, and $Q = Q_1 \otimes \mathbbm{1} + \mathbbm{1} \otimes Q_2$, then it is easy to check that
\begin{align}
f_Q(z) = f_{Q_1}(z) f_{Q_2}(z)
\end{align}
This property is important because it gives a simple criterion for when a $U(1)$ symmetric Hilbert space can be factored into two pieces: such a factorization exists if and only if the polynomial $f_Q(z)$ can be factored into two smaller polynomials with non-negative integer coefficients.

We are now ready to state our classification result: the Floquet phases occurring in systems with charge matrix $Q$ are in one-to-one correspondence with rational functions $\pi(z)=\frac{a(z)}{b(z)}$ with the property that
\begin{align}\label{requ}
\begin{split}
[f_Q(z)]^{N_1}\pi(z)&= \text{non-negative integer polynomial} \\
\frac{[f_{Q}(z)]^{N_2}}{\pi(z)}&=\text{non-negative integer polynomial}
\end{split}
\end{align}
for some integers $N_1, N_2 \geq 0$. Furthermore, two systems corresponding to $\pi_1(z)$ and $\pi_2(z)$ belong to the same phase in the \emph{absence} of any symmetry if and only if $\pi_1(1) = \pi_2(1)$.  

This result deserves a few comments: \\

\textbf{1}. It is not hard to show that any $\pi(z)$ that obeys (\ref{requ}) must be of the form
\begin{align}
\pi(z) = p_1(z)^{n_1} p_2(z)^{n_2} \cdots p_M(z)^{n_M}
\label{pifact}
\end{align}
where $p_1(z),...,p_M(z)$ are the prime (irreducible) factors of $f_Q(z)$, and where $n_j$ is an integer (which may be positive or negative) for each $j$. One way to see this is to multiply together the two equations in (\ref{requ}). This calculation reveals that the product of the two polynomials appearing on the right hand side of (\ref{requ}) is $[f_Q(z)]^{N_1+ N_2}$. It follows that every prime factor of these polynomials must also be a prime factor of $f_Q(z)$. The claim then follows by substituting this prime factorization into either of the two equations in (\ref{requ}).  \\

\textbf{2}. For any $Q$, Eq.~\ref{requ} always has the solution $\pi(z) = f_Q(z)$. We will see below that this solution corresponds to the SWAP circuit. \\

\textbf{3}. It is easy to check that the set of $\pi(z)$'s obeying condition (\ref{requ}) is closed under multiplication and inverses, and therefore forms a group (under multiplication). Furthermore, by property \textbf{1} above, the group of allowed $\pi(z)$'s has at most $M$ generators where $M$ is the number of prime factors of $f_Q(z)$. Also, by property \textbf{2}, it has at least one generator. Hence, charge conserving Floquet phases with charge matrix $Q$ are classified by a group of the form $\mathbb{Z}^m$ where $1 \leq m \leq M$. \\

The above comments suggest a general procedure for determining the classification for a given charge matrix $Q$. The first step is to find the prime factors of $f_Q(z)$. The next step is to construct all $\pi(z)$ of the form (\ref{pifact}). Finally, for each $\pi(z)$ of this form, one needs to check whether it satisfies Eq.~\ref{requ}. Once one has found all possible $\pi(z)$'s, we can also read off the classification in the absence of charge conservation symmetry by evaluating the polynomials $\pi(z)$ at $z=1$. Below we illustrate this recipe with a few examples.

\section{Examples}\label{sexamples}
\subsection{$Q=\mathrm{diag}(0,1)$}\label{sex1}
We begin with the simplest nontrivial example: two-state spins with a charge matrix $Q = \mathrm{diag}(0,1)$. We wish to find the classification of Floquet phases built out of such two-state spins. To do this, we follow the recipe outlined above. First, we note that $f_Q(z) = 1+z$, which is prime (i.e. irreducible). Next, using Eq.~(\ref{pifact}) we deduce that the most general possibility for $\pi(z)$ is $\pi(z)=(1+z)^n$. To complete the analysis, we need to check whether $\pi(z) = (1+z)^n$ satisfies condition (\ref{requ}) for each $n$. Indeed, it is easy to see that $(1+z)^n$ obeys this condition for \emph{all} $n$, so we conclude that the Floquet phases are classifed by rational functions of the form $\pi(z) = (1+z)^n$, with $n \in \mathbb{Z}$. In other words, there is a $\mathbb{Z}$ classification in this case. 

Given the above $\mathbb{Z}$ classification, the next question is whether any of the above Floquet phases collapse to the same phase in the absence of $U(1)$ symmetry. To answer this question, we evaluate each $\pi(z)$ at $z=1$. In particular, since $\pi(z=1) = 2^n$ takes different values for each $n$, we conclude that these phases are all distinct even if the $U(1)$ symmetry is broken.

What is the physical picture for these Floquet phases? As we will see in Sec.~\ref{sconstructunitary}, the two cases $\pi(z) = (1+z)^{\pm 1}$ can be realized by SWAP circuits with opposite chiralities. More generally, $\pi(z) = (1+z)^n$ can be realized by a `(SWAP)$^n$ circuit' -- i.e. a SWAP circuit, of the appropriate chirality, that is executed $|n|$ times within each Floquet period. 
\subsection{$Q=\mathrm{diag}(0,1,2,3)$}\label{sex3}
Next we consider the example sketched in Sec.~\ref{sex0}, where the local Hilbert space is four dimensional and the charge matrix is $Q = \mathrm{diag}(0, 1,2,3)$. Following the same recipe as above, the first step is to compute $f_Q(z)$ which in this case is $f_Q(z) =1+z+z^2+z^3=(1+z)(1+z^2)$. Next we note that $f_Q(z)$ has two prime factors so the most general possibility for $\pi(z)$ is $\pi(z)=(1+z)^{n_1}(1+z^2)^{n_2}$. The last step is to check whether $\pi(z)=(1+z)^{n_1}(1+z^2)^{n_2}$ obeys Eq.~\ref{requ}. Indeed, it is easy to check that (\ref{requ}) is satisfied for \emph{every} $n_1, n_2$, so the classification is $\mathbbm{Z}\times\mathbbm{Z}$ in this case.

To determine whether any of these phases are equivalent in the absence of $U(1)$ symmetry, we need to evaluate each $\pi(z)$ at $z=1$. Observing that $\pi(1) = 2^{n_1+n_2}$, we conclude that the phases with the same $n_1+ n_2$ are equivalent in the absence of $U(1)$ symmetry. In particular, this means that the phase with $\pi(z) = (1+z)^{-1} (1+z^2)$ is an example of nontrivial $U(1)$ symmetric Floquet phase that is trivial in the absence of the symmetry. In other words, the nontrivial properties of this phase are completely due to $U(1)$ symmetry. This is true for any phase described by $\pi(z)$ with $n_1=-n_2$. 

Again, one may ask about the physical realizations of these phases. For the case $\pi(z) = (1+z)^{-1} (1+z^2)$, we presented a physical realization in Sec.~\ref{sex0}: the basic idea is to factor the four dimensional Hilbert into a tensor product of $2$ two-dimensional Hilbert spaces with charge matrices $Q = \mathrm{diag}(0,1)$ and $Q = \mathrm{diag}(0,2)$. After this factorization, the Floquet phase can be realized by two decoupled SWAP circuits, each acting on one of the two-dimensional Hilbert respectively, but with opposite chiralities. This construction can be straightforwardly generalized to any $\pi(z)=(1+z)^{n_1}(1+z^2)^{n_2}$.  

\subsection{$Q=\mathrm{diag}(0,1,2,\cdots,d-1)$}\label{sex5}
Generalizing the previous two examples, we now consider the case where the local Hilbert space is $d$ dimensional and the charge matrix is $Q = \mathrm{diag}(0, 1,...,d-1)$. To determine the phase classification, the first step is to find the prime factors of the polynomial
\begin{equation}
f_Q(z)=1+z+z^2+\cdots z^{d-1}
\end{equation}
This prime factorization is known and is given by
\begin{equation}
f_Q(z)=\prod_{k|d,k\neq 1}\phi_k(z)
\end{equation}
where the product runs over all $k \neq 1$ that divide into $d$, and where $\phi_k(z)$ denotes the $k$th cyclotomic polynomial. Given this factorization, the most general possibility for $\pi(z)$ is
\begin{align}
\pi(z) = \prod_{k|d,k\neq 1}\phi_k(z)^{n_k}
\label{picycl}
\end{align}
The last step is to determine which of the above $\pi(z)$'s satisfy Eq.~\ref{requ}. This is not obvious, but we show in Appendix~\ref{sex5app} that in fact \emph{all} the $\pi(z)$'s of the form (\ref{picycl}) satisfy Eq. \ref{requ}. We conclude that the classification of $U(1)$ symmetric Floquet phases with $Q = \mathrm{diag}(0, 1,...,d-1)$ is $\mathbb{Z}^{n_d-1}$ where $n_d$ is the number of divisors of $d$.  

It is interesting to compare this classification to the case without symmetry: in the latter case, the classification of Floquet phases with a $d$ dimensional Hilbert space is $\mathbb{Z}^{n_p}$ where $n_p$ is the number of distinct \emph{prime} factors of $d$. It is easy to check that $n_d-1$ is strictly larger than $n_p$ unless $d$ is prime, so we conclude that the classification of $U(1)$ symmetric phases is richer than the no-symmetry case -- except when $d$ is prime, in which case the two classifications are the same.

\subsection{$Q=\mathrm{diag}(0,2,3,5)$}\label{sex4}
Finally we consider an example in which the constraints in Eq.~\ref{requ} reduce the classification of phases from $\mathbb{Z}^{M} \rightarrow \mathbb{Z}^{\tilde{M}}$, where $M$ is the number of prime factors of $f_Q(z)$ and $\tilde{M}$ is strictly smaller than $M$. 

Specifically, we consider $Q=\mathrm{diag}(0,2,3,5)$ which corresponds to 
\begin{align*}
f_Q(z)&=1+z^2+z^3+z^5 \nonumber \\
&=(1+z)(1-z+z^2)(1+z^2)
\end{align*}
Given this factorization, the most general possibility for $\pi(z)$ is 
\begin{align}
\pi(z)=(1+z)^{n_1}(1-z+z^2)^{n_2}(1+z^2)^{n_3}
\label{pigenex}
\end{align}
We will now show that the only $\pi(z)$ that satisfy Eq.~\ref{requ} are those with $n_1=n_2$, so the classification of phases is $\mathbb{Z}^2$ rather than $\mathbb{Z}^3$.

To see this, we substitute (\ref{pigenex}) into the first equation in (\ref{requ}), and rewrite the result as
\begin{align}
[f_Q(z)]^{N_1} \pi(z) &= \nonumber \\
(1+z^3)^{N_1+ n_1}&(1-z+z^2)^{n_2 -n_1}(1+z^2)^{N_1+n_3}
\end{align}
If $n_2 > n_1$, then it is easy to see that the right hand side is a polynomial with \emph{negative} coefficients (for example, consider the coefficient of $z$) so we conclude that $n_1 \geq n_2$. Similarly, substituting (\ref{pigenex}) into the second equation in (\ref{requ}) gives
\begin{align}
[f_Q(z)]^{N_2}/\pi(z) &= \nonumber \\
(1+z^3)^{N_2 - n_1}&(1-z+z^2)^{n_1 -n_2}(1+z^2)^{N_2-n_3}
\end{align}
If $n_2 < n_1$, then the right hand side is again a polynomial with negative coefficients (for large enough $N_2$), so we conclude that $n_2 \geq n_1$. Combining these two inequalities, we deduce that $n_1 = n_2$. Thus, the allowed $\pi(z)$'s are labeled by two integers, $n_1=n_2$ and $n_3$, and are of the form $\pi(z)=(1+z^3)^{n_1}(1+z^2)^{n_3}$, as claimed above. 

\section{Definition of $\pi(z)$}\label{sdefs}
We now explain how to compute the rational function $\pi(z)$, given a 2D $U(1)$ symmetric Floquet system $H(t)$. This procedure can be thought of as the \emph{definition} of $\pi(z)$.

The first step is the construction described in Sec.~\ref{edgeU1}: given any 2D Floquet system $H(t)$ we can construct a corresponding 1D unitary $U_{\text{edge}}$ (\ref{edgeu}) that describes the stroboscopic edge dynamics of $H(t)$. In what follows, we will denote $U_{\text{edge}}$ by $U$, for brevity. 

As we explained in Sec.~\ref{edgeU1}, the unitary $U$ is guaranteed to be locality preserving in the sense that it transforms local operators into operators that are locally supported with tails that decay exponentially (or faster). In fact, to state our definition, we will assume that $U$ is \emph{strictly} locality preserving: for any operator $O_r$ supported on site $r$, the operator $U^\dagger O_r U$ is supported within the interval $[r-\ell, r+\ell]$ for some finite $\ell$ that depends only on $U$. We will refer to $\ell$ as the ``operator spreading length'' of $U$. We do not expect to lose any generality with this assumption since we can approximate any locality preserving unitary with a strictly locality preserving unitary with arbitrarily small error.

We now explain how to define the rational function $\pi(z)$ in terms of the 1D unitary $U$. To do this, we first define two other quantities associated with $U$: (i) a rational number $\frac{p}{q}$ and (ii) a rational function $\tilde{\pi}(z)$ with the property that $\tilde{\pi}(1) = 1$. We then define $\pi(z)$ as the product
\begin{equation}
\pi(z)=\frac{p}{q} \cdot \tilde{\pi}(z)
\label{pidef}
\end{equation}
To complete the story, we need to explain how $\frac{p}{q}$ and $\tilde{\pi}(z)$ are defined in terms of $U$. We define $\frac{p}{q}$ by
\begin{align}
\frac{p}{q} = \mathrm{ind}(U)
\end{align}
where $\mathrm{ind}(U)$ is the GNVW index that classifies 1D locality preserving unitaries in the \emph{absence} of symmetry\cite{GNVW}. We review the definition of $\mathrm{ind}(U)$ in Appendix~\ref{GNVWformapp}, but roughly speaking, $\mathrm{ind}(U)$ can be thought of as measuring the quantum information flow associated with the unitary $U$\cite{tracking}. In contrast, $\tilde{\pi}(z)$ characterizes the \emph{charge} flow associated with $U$. 

\begin{figure}[tb]
   \centering
   \includegraphics[width=0.9\columnwidth]{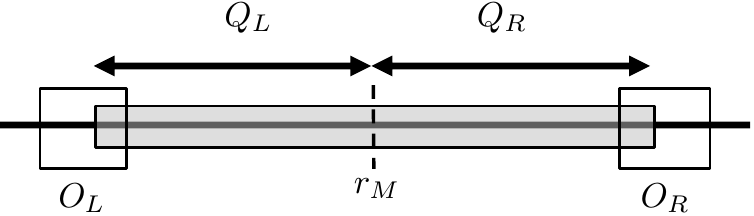} 
   \caption{The interval $A$ is indicated by the light grey shaded region. The unitary $U$ transforms $Q_A=Q_L+Q_R$ into $U^\dagger Q_AU = Q_L + Q_R + O_L+O_R$ where $O_L$ and $O_R$ are supported near the left and right endpoints of $A$, respectively. }
   \label{fig:chargeinterval}
\end{figure}

To define $\tilde{\pi}(z)$, let $A = [r_L, r_R]$ be any interval with length $r_R - r_L \geq 2 \ell - 1$ and let $Q_A$ denote the total charge in $A$: 
\begin{align}
Q_A = \sum_{r \in A} Q_r.
\end{align}
Since $U$ is locality preserving and $U(1)$ symmetric, it is easy to show that $U^\dagger Q_A U - Q_A$ is supported near the two ends of $A$. More precisely, one can show that (Fig.~\ref{fig:chargeinterval})
\begin{equation}
U^\dagger Q_AU=Q_A+O_L+O_R
\label{UQU}
\end{equation}
where $O_L$ and $O_R$ are operators that are supported within the intervals $[r_L - \ell, r_L +\ell-1]$ and $[r_R -\ell+1, r_R + \ell]$ (see Appendix~\ref{sUQU} for a proof). A similar decomposition to Eq.~\ref{UQU} was also used in Ref.~\cite{bachmanncharge} in a related context. 

Next, choose a point $r_M$ in the middle of $A$, or more specifically, choose $r_M$ so that $r_L + \ell -1 \leq r_M < r_R - \ell +1$. Using $r_M$, we can split $A$ into two disjoint intervals: 
\begin{align}
A = A_L \cup A_R
\end{align}
where $A_L = [r_L, r_M]$ and $A_R = [r_M+1, r_R]$. 

Likewise, we can split $Q_A$ into two pieces: $Q_A = Q_L + Q_R$ where $Q_L$ and $Q_R$ denote the total charge in $A_L$ and $A_R$ respectively. Substituting $Q_A = Q_L + Q_R$ into Eq.~\ref{UQU}, we obtain:
\begin{equation}
U^\dagger(Q_{L}+Q_{R})U=Q_{L}+Q_{R}+O_L+O_R
\label{QIR}
\end{equation}

Next, we observe that Eq.~(\ref{QIR}) implies that 
\begin{align}
\mathrm{Spec}(Q_L + Q_R) = \mathrm{Spec}(Q_L + Q_R + O_L + O_R)
\label{specQid}
\end{align}
where $\mathrm{Spec}(O)$ is the eigenvalue spectrum of the operator $O$. This identity can be written more conveniently using generating functions: 
\begin{align}
\mathrm{Tr}\left(z^{Q_{L}+Q_{R}}\right) = \mathrm{Tr}\left(z^{Q_{L}+O_L+Q_{R}+O_R}\right)
\label{specid0}
\end{align}
where $z$ is a formal variable and where $\mathrm{Tr}(O)$ denotes the trace over the Hilbert space of the whole spin chain.\footnote{We can make the trace finite by taking the chain to be finite and periodic.}

Now, since $Q_L + O_L$ and $Q_R + O_R$ are supported on non-overlapping regions, as are $Q_L, Q_R$, Eq.~\ref{specid0} implies that
\begin{align}
\mathrm{Tr}\left(z^{Q_{L}}\right) \cdot \mathrm{Tr}\left(z^{Q_{R}}\right) = \mathrm{Tr}\left(z^{Q_{L}+O_L}\right) \cdot \mathrm{Tr}\left(z^{Q_{R}+O_R}\right)
\label{specid}
\end{align}
We then rewrite this identity as
\begin{equation}\label{RL}
\tilde{\pi}_L(z) \cdot \tilde{\pi}_R(z)=1
\end{equation}
where
\begin{equation}\label{pilrformula}
\tilde{\pi}_L(z)=\frac{\mathrm{Tr}(z^{Q_{L}+O_L})}{\mathrm{Tr}(z^{Q_{L}})}\qquad \tilde{\pi}_R(z)=\frac{\mathrm{Tr}(z^{Q_{R}+O_R})}{\mathrm{Tr}(z^{Q_{R}})} 
\end{equation}
The two functions $\tilde{\pi}_L(z), \tilde{\pi}_R(z)$ can be thought of as characterizing the charge flow associated with $U$ through the left and right end of $A$. Likewise, Eq.~(\ref{RL}) can be thought of as a kind of conservation law that relates the charge flow through the two ends of $A$. Notice that since we trace over the same number of sites in the numerator and denominator of $\tilde{\pi}_L(z)$ and $\tilde{\pi}_R(z)$, the two quantities satisfy $\tilde{\pi}_L(1)=\tilde{\pi}_R(1)=1$.

At this point we need to address an important subtlety: we have defined $\tilde{\pi}_L(z)$, $\tilde{\pi}_R(z)$ in terms of $O_L, O_R$ which are in turn defined by Eq.~(\ref{UQU}); however, Eq.~(\ref{UQU}) only fixes $O_L, O_R$ up to a constant shift of the form
\begin{align}
O_L \rightarrow O_L +c \mathbbm{1}, \qquad O_R \rightarrow O_R - c \mathbbm{1}. 
\label{shiftamb}
\end{align}
To fix this ambiguity, we choose $O_R, O_L$ so that the smallest eigenvalue of both $Q_R + O_R$ and $Q_L + O_L$ is $0$. (It is always possible to do this since the smallest eigenvalue of $Q_R + Q_L$ is $0$ by assumption -- see Sec.~\ref{u1symmsetup}). With this convention, $O_L, O_R$ are completely determined, as are $\tilde{\pi}_L(z)$, $\tilde{\pi}_R(z)$.

We are now ready to define $\tilde{\pi}(z)$:
\begin{align}
\tilde{\pi}(z) = \tilde{\pi}_R(z) = [\tilde{\pi}_L(z)]^{-1} 
\label{deftilpi}
\end{align}

To complete the discussion, we now show that $\tilde{\pi}(z)$ does not depend on the choice of the interval $A = [r_L, r_R]$ or on the point $r_M \in A$ that we used to define our partition $Q_A = Q_L + Q_R$. This is important because it establishes that $\tilde{\pi}(z)$ is a well-defined invariant associated with the locality preserving unitary $U$. To prove the first result -- i.e. $\tilde{\pi}(z)$ is independent of $A$ -- it suffices to show that $\tilde{\pi}(z)$ does not change if we move the left or right endpoint, $r_L$ or $r_R$. This invariance follows from Eq.~\ref{RL}: indeed, it is clear that $\tilde{\pi}_L(z)$ is independent of the choice of $r_R$, while $\tilde{\pi}_R(z)$ is independent of the choice of $r_L$. Therefore by Eq.~\ref{RL}, both quantities must be independent of both endpoints. To prove the second result -- i.e $\tilde{\pi}(z)$ is independent of the choice of $r_M \in A$ -- suppose we replace $r_M \rightarrow r'_M < r_M$. Then, $Q_R \rightarrow Q_{M} + Q_R$ where $Q_M = \sum_{r'_M < r \leq r_M} Q_r$. Then since the region of support of $Q_M$ is disjoint from $Q_R$ and $Q_R + O_R$, it is easy to see that
\begin{align}
\frac{\mathrm{Tr}(z^{Q_{R}+Q_{M}+O_R})}{\mathrm{Tr}(z^{Q_{R}+Q_{M}})} = \frac{\mathrm{Tr}(z^{Q_{R}+O_R})}{\mathrm{Tr}(z^{Q_{R}})} 
\end{align}
so that $\tilde{\pi}(z)$ does not change under this operation.

Before concluding this section, we now list two important properties of $\pi(z)$ which we will prove in Appendix~\ref{scomp}. These properties encode the fact that $\pi(z)$ is \emph{multiplicative} under both stacking (i.e. tensoring) and composition of unitaries:
\begin{enumerate}
\item{\textbf {Stacking}: $\pi_{U_1 \otimes U_2}(z) = \pi_{U_1}(z) \pi_{U_2}(z)$
}
\item{\textbf {Composition}: $\pi_{U_1 \cdot U_2}(z) = \pi_{U_1}(z) \pi_{U_2}(z)$ 
}
\end{enumerate}
Here, we use the notation $\pi_U(z)$ to denote the value of $\pi(z)$ corresponding to a LPU $U$. In the first property, $U_1 \otimes U_2$ is the LPU obtained by stacking (i.e. tensoring) two other LPUs $U_1$ and $U_2$. Likewise, in the second property, $U_1 \cdot U_2$ is the LPU obtained by composing two LPUs $U_1$ and $U_2$ acting on the same $Q$.

A final comment: our definition of $\tilde{\pi}(z)$ is similar to one of the ``symmetry protected indices'' of Ref.~\onlinecite{mpu}, which were used to classify 1D locality preserving unitaries with \emph{discrete} symmetries. In particular, we should compare $\tilde{\pi}(z)$ to the invariant defined in Eq.~(6) of Ref.~\onlinecite{mpu}. If we naively set $G = U(1)$, then the latter invariant is equivalent to $\log |\tilde{\pi}(z)|$ evaluated along the unit circle $|z| = 1$. This quantity carries some, but not all, the information in $\tilde{\pi}(z)$ due to the absolute value sign. 


\section{Proving the classification}\label{sproofs}

In the previous section, we showed that every one dimensional $U(1)$ symmetric LPU $U$ can be associated with a corresponding $\pi(z)$. We now establish three properties of this labeling scheme which, together, imply our classification result:
\begin{enumerate}
\item{For any $U$, the corresponding $\pi(z)$ is a rational function satisfying Eq.~\ref{requ}.}
\item{All rational functions $\pi(z)$ satisfying Eq.~\ref{requ} can be realized by some $U$.}
\item{$\pi_{U'}(z)=\pi_{U}(z)$ if and only if $U' U^{-1}$ is a $U(1)$ symmetric FDLU.}
\end{enumerate} 
To see why these properties imply our classification result, note that properties (1) and (2) prove that the set of possible $\pi(z)$'s is exactly the set of rational functions $\pi(z) = a(z)/b(z)$ obeying the conditions in Eq.~(\ref{requ}). Likewise property (3) proves that there is a one-to-one correspondence between equivalence classes of $U(1)$ symmetric LPUs and $\pi(z)$'s. 

We now prove properties (1), (2), and (3) in the next three sections.
\subsection{$\pi(z)$ satisfies Eq.~\ref{requ}}\label{ssatisfiesrequ}

In this section we prove property (1) above: we show that for any 1D $U(1)$ symmetric LPU $U$, the corresponding $\pi(z)$ is a rational function satisfying Eq.~\ref{requ}. 

Let $\ell$ be the operator spreading length for the (strictly) locality preserving unitary $U$. Next, choose $A = [-\ell+1, \ell]$ and $r_M = 0$ so that
\begin{align}
Q_L = \sum_{r= -\ell+1}^{0} Q_r, \quad \quad Q_R = \sum_{r=1}^{\ell} Q_r, 
\end{align}
We then rewrite the expression (\ref{pilrformula}) for $\tilde{\pi}_R(z)$ in a slightly different form: instead of evaluating the various traces over the entire spin chain, we evaluate them over a smaller interval $R = [1,2\ell]$ which is chosen so that it contains the region of support of both $Q_R$ and $Q_R + O_R$. The new expression for $\tilde{\pi}_R(z)$ is then
\begin{align}
\tilde{\pi}_R(z) &=\frac{\mathrm{Tr}_R(z^{Q_R+O_R})}{\mathrm{Tr}_R(z^{Q_R})}
\label{pirexp}
\end{align}
where the symbol $\mathrm{Tr}_R(O)$ denotes the trace of the \emph{restriction} of the operator $O$ to the interval $R$ (which is well-defined assuming $O$ is supported in $R$).

Next, consider the numerator and denominator of (\ref{pirexp}). We claim that the operator $Q_R + O_R$ in the numerator has only non-negative integer eigenvalues. To see this, observe that $Q_R + Q_L$ has integer eigenvalues, and hence by (\ref{specQid}), $Q_{L}+O_L + Q_{R}+O_R$ also has integer eigenvalues, that is:
\begin{align}
\mathrm{Spec}(Q_L + O_L + Q_R + O_R ) \subset \mathbb{Z}
\end{align}
We then deduce that
\begin{align}
\mathrm{Spec}(Q_L + O_L) + \mathrm{Spec}(Q_R + O_R) \subset \mathbb{Z}
\end{align}
since $Q_{R}+O_R$ and $Q_{L}+O_L$ act on disjoint regions. It follows as a corollary that the difference $\lambda_i -\lambda_j$ between any two eigenvalues $\lambda_i, \lambda_j$ of $Q_R + O_R$ must be an integer. Therefore, since the minimum eigenvalue of $Q_R + O_R$ is $0$ by our convention in Sec.~\ref{sdefs}, we conclude that all the eigenvalues of $Q_R + O_R$ are non-negative integers, as we wished to show.

Now, since all the eigenvalues of $Q_R + O_R$ are non-negative integers, it follows that
\begin{align}
\mathrm{Tr}_R(z^{Q_R+O_R}) = \text{non-negative integer polynomial} 
\end{align}
In fact, we can say more: in Appendix~\ref{sgnvw} we use results from Ref.~\onlinecite{GNVW} to show that the restriction of $Q_R + O_R$ to the interval $R$ can be written, in an appropriate basis, in the form 
\begin{align}
Q_R + O_R = O \otimes \mathbbm{1}
\label{qrorid}
\end{align}
where $O$ is a matrix of dimension $(p/q)d^{\ell}$ and $\mathbbm{1}$ is an identity matrix of dimension $(q/p)d^{\ell}$. Here, $p/q = \mathrm{ind}(U)$. In particular, this means that every eigenvalue of $Q_R + O_R$ has a degeneracy which is a multiple of $(q/p)d^{\ell}$ so that
\begin{align}
\mathrm{Tr}_R(z^{Q_R+O_R}) = \frac{q d^\ell}{p} \alpha(z)
\end{align}
for some non-negative integer polynomial $\alpha(z)$.

Moving on to the denominator of (\ref{pirexp}), we observe that
\begin{align}
\mathrm{Tr}_R(z^{Q_R}) = [f_Q(z)]^{\ell} d^{\ell}
\label{tracezqr}
\end{align}
where the factor of $f_Q(z)^\ell$ comes from the interval $[1, \ell]$ where $Q_R$ is supported, while the factor $d^\ell$ comes from the interval $[\ell+1, 2\ell]$ where $Q_R$ acts like the identity.

Putting this all together, we derive
\begin{align}
\tilde{\pi}_R(z) =  \frac{q}{p}\frac{\alpha(z)}{ [f_Q(z)]^{\ell}}
\end{align}
so that
\begin{align}
\pi(z) = \frac{p}{q} \tilde{\pi}_R(z) = \frac{\alpha(z)}{ [f_Q(z)]^{\ell}}
\end{align}

Hence
\begin{align}
[f_Q(z)]^{\ell} \pi(z) = \alpha(z)
\end{align}
This proves that $\pi(z)$ obeys the first condition in (\ref{requ}) with $N_1 = \ell$. 

The second condition follows from almost the same argument: first, we rewrite the expression for $\tilde{\pi}_L(z)$ as
\begin{align}
\tilde{\pi}_L(z) &=\frac{\mathrm{Tr}_L(z^{Q_L+O_L})}{\mathrm{Tr}_L(z^{Q_L})}
\label{pilexp}
\end{align}
where $L$ is the interval $[-2\ell+1, 0]$.

Next, we note that the eigenvalues of $Q_L + O_L$ are non-negative integers (by the same reasoning as above) so
\begin{align}
\mathrm{Tr}_L(z^{Q_L+O_L}) = \text{non-negative integer polynomial} 
\end{align}
In fact, using the same arguments as in Appendix~\ref{sgnvw}, one can show that the restriction of $Q_L + O_L$ to the interval $L$ can be written, in an appropriate basis, in the form $Q_L + O_L = O \otimes \mathbbm{1}$ where $\mathbbm{1}$ is an identity matrix of dimension $(p/q)d^{\ell}$. In particular, this means that every eigenvalue of $Q_L + O_L$ has a degeneracy which is a multiple of $(p/q) d^{\ell}$ so that
\begin{align}
\mathrm{Tr}_L(z^{Q_L+O_L}) = \frac{p d^\ell}{q} \beta(z)
\end{align}
for some non-negative integer polynomial $\beta(z)$.

Moving on to the denominator of (\ref{pilexp}), the same arguments as in (\ref{tracezqr}) imply
\begin{align}
\mathrm{Tr}_L(z^{Q_L}) = [f_Q(z)]^{\ell} d^{\ell}
\end{align}

Putting this all together, we derive
\begin{align}
\tilde{\pi}_L(z) = \frac{p}{q}\frac{\beta(z)}{ [f_Q(z)]^{\ell}}
\end{align}
It follows that
\begin{align}
\pi(z) = \frac{p}{q} (\tilde{\pi}_L(z))^{-1} = \frac{[f_Q(z)]^{\ell}}{\beta(z) }
\end{align}

Hence
\begin{align}
\frac{[f_Q(z)]^{\ell}}{\pi(z)} = \beta(z)
\end{align}
This proves that $\pi(z)$ obeys the second condition in (\ref{requ}) with $N_2 = \ell$.

\subsection{Constructing a unitary that realizes each $\pi(z)$}\label{sconstructunitary}
In this section, we prove property (2) above: we show that every $\pi(z)$ obeying Eq.~\ref{requ} can be realized by some $U(1)$ symmetric  LPU. Our proof is constructive: we explicitly construct a LPU for each $\pi(z)$.

To begin, we note that by Eq.~\ref{requ}, there exists non-negative integers $N_1, N_2$ such that
\begin{equation}
[f_Q(z)]^{N_1} \pi(z) = \alpha(z), \quad \frac{[f_Q(z)]^{N_2}}{\pi(z)} = \beta(z)
\label{alphabeta}
\end{equation}
for some non-negative integer polynomials $\alpha(z), \beta(z)$. 

Multiplying these equations together gives
\begin{equation}
[f_{Q}(z)]^{N_1+N_2} = \alpha(z)\beta(z)
\label{alphabetafact}
\end{equation}
This equation tells us that the polynomial $[f_{Q}(z)]^{N_1+N_2}$ can be factored into a product of two non-negative integer polynomials, $\alpha(z), \beta(z)$. Such a factorization of polynomials implies a corresponding factorization of Hilbert spaces, as we explained in Sec.~\ref{sclassification}. More specifically, let $\mathcal{H}^{N_1 + N_2}$ denote the Hilbert space for a cluster of $N_1 + N_2$ sites. Then
\begin{align}
\mathcal{H}^{N_1 + N_2} = \mathcal{H}_\alpha \otimes \mathcal{H}_\beta
\label{Hfact}
\end{align}
where $\mathcal{H}_\alpha$ and $\mathcal{H}_\beta$ are Hilbert spaces of dimension $d_\alpha = \alpha(1)$ and $d_\beta = \beta(1)$, with corresponding charge matrices $Q_\alpha, Q_\beta$ defined by
\begin{align}
\mathrm{Tr}(z^{Q_\alpha}) = \alpha(z),  \quad  \mathrm{Tr}(z^{Q_\beta}) = \beta(z)
\end{align}

We now use the above factorization to construct the desired LPU. We proceed in two steps. First, we cluster together groups of $2N_1+N_2$ sites into supersites with a Hilbert space $\mathcal{H}^{2N_1+ N_2}$. We then factor each supersite Hilbert space into a tensor product
\begin{align}
\mathcal{H}^{2N_1+ N_2} = \mathcal{H}^{N_1} \otimes \mathcal{H}_\alpha \otimes \mathcal{H}_\beta
\label{superfact}
\end{align}
using (\ref{Hfact}).

\begin{figure}[tb]
   \centering
   \includegraphics[width=0.9\columnwidth]{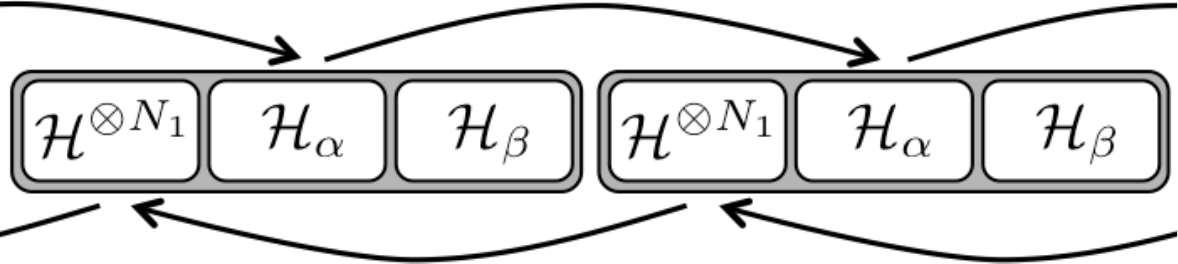} 
   \caption{The protocol for constructing an edge unitary, and therefore also a 2D Floquet circuit, that realizes a given $\pi(z)$: first cluster together $2N_1+N_2$ sites and factor them into $\mathcal{H}^{N_1}\otimes\mathcal{H}_{\alpha}\otimes\mathcal{H}_{\beta}$. Then the unitary that realizes $\pi(z)$ performs a unit translation of $\mathcal{H}_{\alpha}$ and $\mathcal{H}^{N_1}$ in opposite directions and does nothing to $\mathcal{H}_{\beta}$.}
   \label{fig:supersites}
\end{figure}

Now, consider the LPU that performs a unit translation on the $\mathcal{H}_{\alpha}$ sites, performs a unit translation on the $\mathcal{H}^{N_1}$ sites in the opposite direction, and that does nothing to the $\mathcal{H}_{\beta}$ sites (Fig.~\ref{fig:supersites}). It is easy to see that the corresponding $\pi_U(z)$ is
\begin{align}
\pi_U(z) = \frac{\alpha(z)}{[f_{Q}(z)]^{N_1}} 
\end{align}
Hence $\pi_U(z) = \pi(z)$ by (\ref{alphabeta}). This completes the proof: we have explicitly constructed a 1D $U(1)$ symmetric locality preserving unitary $U$ that realizes $\pi(z)$. 

In fact, it is not hard to go a step further and construct a 2D Floquet system that realizes $\pi(z)$. To do that, we again cluster sites together into supersites $\mathcal{H}^{2N_1+ N_2}$, which we then factor as in
Eq.~\ref{superfact}. We then consider a Floquet system that implements a SWAP circuit on the $\mathcal{H}_{\alpha}$ sites and a SWAP circuit with the opposite chirality on the $\mathcal{H}^{N_1}$ sites, and does nothing to the $\mathcal{H}_{\beta}$ sites. By construction, the edge dynamics of this Floquet system is described by the unitary $U$, and therefore the invariant corresponding to this circuit is $\pi_U(z) = \pi(z)$, as desired.

For an example of this construction, consider the charge matrix $Q=\mathrm{diag}(0,1,1,3,4,4)$. In this case, $f_Q(z)=1+2z+z^3+2z^4$ which has a prime factorization $f_Q(z)=(1+2z)(1+z)(1-z+z^2)$. Consider the Floquet phase corresponding to $\pi(z) = (1-z+z^2)$. This phase is an interesting example because it cannot be realized by factoring the \emph{single} site Hilbert space into smaller pieces and performing SWAP circuits on each piece. Instead, one needs to cluster multiple sites into supersites, and then factor these supersites, as in the general construction discussed above. To see how this works, note that the minimal values for $N_1$ and $N_2$ in this case are $N_1=2$ and $N_2=1$:
\begin{align}
[f_Q(z)]^2(1-z+z^2) &= (1+3z+z^2+4z^4)(1+z^3)^2 \nonumber \\ 
&= \alpha(z) \nonumber \\
\frac{f_Q(z)}{1-z+z^2} &= (1+2z)(1+z) \nonumber \\
&= \beta(z) 
\end{align}
Multiplying the two equations together, we conclude that the Hilbert space describing a cluster of $3$ sites, with total dimension $6^3$, can be factored into a tensor product of two Hilbert spaces -- one of dimension $\alpha(1) = 36$ and one of dimension $\beta(1) = 6$ -- with charge matrices corresponding to $\alpha(z), \beta(z)$. 

To construct the appropriate locality preserving unitary, we cluster groups of $2N_1 + N_2 = 5$ sites into supersites of dimension $6^5$. We then factor the supersite Hilbert space $\mathcal{H}^5$ into a Hilbert space $\mathcal{H}_\alpha$ of dimension $36$, a Hilbert space $\mathcal{H}_\beta$ of dimension $6$, and a $2$-site Hilbert space $\mathcal{H}^2$ of dimension $6^2$. The desired locality preserving unitary performs a unit translation on the $36$-dimensional Hilbert space $\mathcal{H}_\alpha$, and a unit translation with the opposite chirality on the $36$ dimensional Hilbert space $\mathcal{H}^2$, and does nothing to the $\mathcal{H}_\beta$ sites. This locality preserving unitary can, in turn, be realized by a 2D Floquet system that performs a SWAP circuit on the $\mathcal{H}_\alpha$ sites, a SWAP circuit with the opposite chirality on the $\mathcal{H}^2$ sites, and does nothing to the $\mathcal{H}_\beta$ sites.

\subsection{One-to-one correspondence}\label{sequivunitary}

In this section, we prove property (3) above: we show that there is a one-to-one correspondence between equivalence classes of one dimensional $U(1)$ symmetric LPUs and rational functions $\pi(z)$. More precisely, we prove the following: let $U, U'$ be $U(1)$ symmetric LPUs. We will show that $\pi_{U'}(z)=\pi_{U}(z)$ if and only if $U'U^{-1}$ is a $U(1)$ symmetric FDLU. 

\subsubsection{Modified definitions of locality preserving unitaries and FDLUs}
For technical reasons, our proof requires us to use slightly more restricted definitions of LPUs and FDLUs then the definitions presented in Sec.~\ref{1dunitrev}. Specifically, for the purposes of this proof, we will define a LPU to be a unitary $U$ that is \emph{strictly} locality preserving -- i.e. a unitary with the property that, for any single site operator $O_r$, the conjugated operator $U^\dagger O_r U$ is completely supported within the interval $[r-\ell, r+\ell]$ for some finite operator spreading length $\ell$. 

Also, for the purposes of this proof, we define a FDLU to be any unitary $U$ that can be written as a ``finite-depth quantum circuit.'' That is,
\begin{align}
U = U^{(1)}U^{(2)} \cdots U^{(D)}
\end{align}
where each $U^{(j)}$ of the form
\begin{align}
U_{j} = \begin{cases}
\prod_r U^{(j)}_{2r-1,2r} & \ \text{even $j$}  \\
\prod_r U^{(j)}_{2r,2r+1}  & \ \text{odd $j$}
\end{cases}
\end{align}
where each $U^{(j)}_{r,r+1}$ acts on two neighboring spins $r, r+1$. Here $D$ is the ``depth'' of the quantum circuit. Likewise, we will say that $U$ is a $U(1)$ \emph{symmetric} FDLU if $U$ can be written as a finite-depth quantum circuit in which every unitary gate $U^{(j)}_{r,r+1}$ is $U(1)$ symmetric. The above definition can be thought of as a discrete time analog of the continuous time evolution definition of an FDLU given in Sec.~\ref{1dunitrev}. The advantage of the new definition is that it guarantees that an FDLU is \emph{strictly} locality preserving with an operator spreading length $\ell \leq D$.

\subsubsection{Proof}
\label{proof11corr}
To start the proof, we make an observation which simplifies our problem considerably: we observe that, according to the composition property of $\pi(z)$ (see Sec.~\ref{sdefs}), the condition $\pi_{U'}(z)=\pi_{U}(z)$ is equivalent to $\pi_{U'U^{-1}}(z) = 1$. Therefore, the statement we wish to prove can be rephrased as follows: $\pi_{U'U^{-1}}(z) = 1$ if and only if $U'U^{-1}$ is a $U(1)$ symmetric FDLU. Equivalently, replacing $U' U^{-1} \rightarrow W$, it suffices to show that for any $U(1)$ symmetric, locality preserving $W$, the corresponding $\pi_W(z) = 1$ if and only if $W$ is a $U(1)$ symmetric FDLU. We now prove the latter statement. 

We start with the `if' direction. Suppose that $W$ is a $U(1)$ symmetric FDLU. We wish to show that $\pi_W(z) = 1$. The first step is to note that $\mathrm{ind}(W) = 1$ since $W$ is a FDLU\cite{GNVW}. Therefore, we only need to show that $\tilde{\pi}_W(z) = 1$. We will do this using the definition of $\tilde{\pi}(z)$ in Sec.~\ref{sdefs}. 

To apply the definition of $\tilde{\pi}(z)$, we need to choose an interval $A$ and a point $r_M \in A$. We choose $A$ to be $[-D+1, D]$ where $D$ denotes the depth of the quantum circuit corresponding to $W$, and we choose $r_M=0$ so
\begin{align}
Q_L = \sum_{r= -D+1}^{0} Q_r, \quad \quad Q_R = \sum_{r=1}^{D} Q_r, 
\end{align}
To find $\tilde{\pi}(z)$, we need to compute $W^\dagger(Q_L + Q_R)W$. 

\begin{figure}[tb]
   \centering
   \includegraphics[width=0.9\columnwidth]{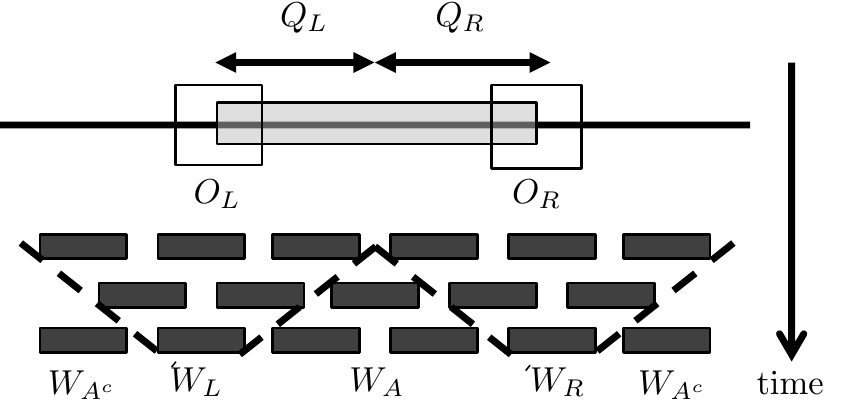} 
   \caption{Decomposition of a FDLU $W$ into $W=W_AW_{A^c}W_LW_R$. Each unitary $W_A, W_{A^c}, W_L, W_R$ is a time ordered product of two site unitaries, represented by the black rectangles. By construction, $[W_L,Q_R]=[Q_R,Q_L]=0$ because they act on disjoint regions. In addition, $[W_A,Q_L+Q_R]=0$ because $W_A$ acts entirely within $A$. }
   \label{fig:WLWAWR}
\end{figure}

To this end, we observe that since $W$ is a $U(1)$ symmetric finite-depth quantum circuit with depth $D$, we can write it as a product of four $U(1)$ symmetric unitaries
\begin{align}
W=  W_A W_{A^c} W_L W_R
\end{align}
where $W_A$ is supported in the interval $A$ and $W_{A^c}$ is supported in $A^c$ and $W_L, W_R$ are supported in the intervals $[-2D+1,0]$ and $[1,2D]$ respectively (see Fig.~\ref{fig:WLWAWR}).

Comparing the regions of support of these operators, we see immediately that
\begin{align}
[W_L,Q_R]=[W_R,Q_L]=0, \quad [W_{A^c}, Q_L + Q_R] = 0
\end{align}
Also,
\begin{align}
[W_A, Q_L + Q_R] =0
\end{align}
since $W_A$ is $U(1)$ symmetric and is supported within $A$.

Using the above commutation relations, we can simplify the action of $W$ on $Q_L+Q_R$:
\begin{align}
W^\dagger(Q_L+Q_R)W&=W_R^\dagger W_L^\dagger (Q_L+Q_R) W_LW_R \nonumber \\
&=W_L^\dagger Q_L W_L+W_R^\dagger Q_R W_R
\end{align}
Comparing this expression with Eq.~(\ref{QIR}), we derive
\begin{align}
W_L^\dagger Q_L W_L+W_R^\dagger Q_R W_R = Q_L + O_L + Q_R + O_R 
\end{align}
Next, observe that the `$L$' terms on both sides are supported within the interval $[-2D+1, 0]$ while the `$R$' terms are supported within the interval $[1, 2D]$. In particular, the `$L$' terms and `$R$' terms are supported on \emph{disjoint} intervals which implies that the $L$ terms and $R$ terms must be equal individually\footnote{Here, we also use the fact that the $L$ terms and $R$ terms on both sides have smallest eigenvalue $0$.}: 
\begin{align}
W_L^\dagger Q_LW_L &= Q_L + O_L  \nonumber \\
W_R^\dagger Q_RW_R &= Q_R + O_R 
\end{align}
It follows that
\begin{align}
\mathrm{Tr}(z^{Q_L+O_L}) = \mathrm{Tr}(z^{Q_L}) \nonumber \\
\mathrm{Tr}(z^{Q_R+O_R}) = \mathrm{Tr}(z^{Q_R})
\end{align}
Hence, $\tilde{\pi}_W(z) = 1$ as we wished to show. This completes our proof of the `if' direction.

Next, we prove the converse statement: if $W$ is a $U(1)$ symmetric locality preserving unitary with $\pi_W(z)=1$, then $W$ is a $U(1)$ symmetric FDLU. To begin, notice that $\pi_W(z)=1$ implies that $\mathrm{ind}(W)= \pi_W(1)=1$. Thus, by Ref.~\onlinecite{GNVW}, we know that $W$ is a FDLU; all we need to show is that this FDLU can be realized in a way that each unitary gate in $W$ is $U(1)$ symmetric. To do this, it is convenient to assume that $W$ has an operating spreading length of $\ell = 1$, i.e. $W^\dagger O_r W$ is supported on sites $\{r-1, r, r+1\}$ for any single site operator $r$. It is also convenient to assume that $W$ can be written as a depth-$2$ quantum circuit. We do not lose any generality with either of these assumptions since \emph{every} finite-depth quantum circuit can be written as a depth-$2$ circuit with an operator spreading length of $\ell=1$ by clustering sufficiently large groups of neighboring sites into supersites\cite{GNVW}. 

\begin{figure}[tb]
   \centering
   \includegraphics[width=0.9\columnwidth]{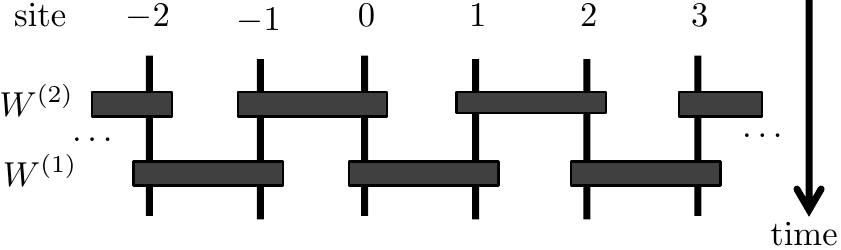} 
   \caption{$W$ written as a depth-2 quantum circuit. Each layer, $W^{(1)}$ and $W^{(2)}$, consists of a set of disjoint two-site gates, represented by black rectangles. }
   \label{fig:depth2}
\end{figure}

With these assumptions, we can write $W$ as
\begin{align}
W = W^{(1)} W^{(2)} = \prod_r W^{(1)}_{2r,2r+1} \prod_r W^{(2)}_{2r-1,2r}
\label{w1w2}
\end{align}
This decomposition of $W$ is illustrated in Fig.~\ref{fig:depth2}. 

Next, consider the action of $W$ on $(Q_{2r}+Q_{2r+1})$. By the definition of $O_L, O_R$, 
\begin{align}
W^\dagger(Q_{2r}+Q_{2r+1})W = Q_{2r}+Q_{2r+1} +O_L + O_R 
\end{align}
where $O_L$ is supported on sites $\{2r-1, 2r\}$ and $O_R$ is supported on sites $\{2r+1, 2r+2\}$. Using $W=W^{(1)}W^{(2)}$, we deduce
\begin{align}
W^{(1)\dagger}(Q_{2r}&+Q_{2r+1})W^{(1)} \nonumber \\
&= W^{(2)}(Q_{2r}+Q_{2r+1} + O_L + O_R)W^{(2)\dagger}
\label{OR}
\end{align}
Substituting the explicit form of $W^{(1)}$ and $W^{(2)}$ as products of two-site gates (\ref{w1w2}) gives
\begin{align}
W^{(1)\dagger}_{2r,2r+1}(Q_{2r}+Q_{2r+1})W^{(1)}_{2r,2r+1} = \tilde{Q}_{2r}+\tilde{Q}_{2r+1}
\label{tildeQ}
\end{align}
where
\begin{align}
\tilde{Q}_{2r} &=W^{(2)}_{2r-1,2r}(Q_{2r} + O_L) W^{(2)\dagger}_{2r-1,2r} \nonumber \\
\tilde{Q}_{2r+1} &= W^{(2)}_{2r+1,2r+2}(Q_{2r+1} + O_R)W^{(2)\dagger}_{2r+1,2r+2}, 
\label{tildeQdef}
\end{align}
Next, we observe that the left side of Eq \ref{tildeQ} is supported only on sites $\{2r, 2r+1\}$, while $\tilde{Q}_{2r}$ is supported on sites $\{2r-1, 2r\}$ and $\tilde{Q}_{2r+1}$ is supported on sites $\{2r+1, 2r+2\}$. It follows that $\tilde{Q}_{2r}$ must be supported on site $2r$ alone and $\tilde{Q}_{2r+1}$ must be supported on site $2r+1$ alone. Furthermore, by (\ref{tildeQdef}), $\tilde{Q}_{2r}$ has the same spectrum as $Q_{2r} + O_L$. The latter operator has the same spectrum as $Q_{2r}$ since $\pi_W(z) = 1$, so we conclude that $\tilde{Q}_{2r}$ must have the same spectrum as $Q_{2r}$. The same argument shows that $\tilde{Q}_{2r+1}$ must have the same spectrum as $Q_{2r+1}$. Putting this together, we conclude that there exists single site operators $V^{(2r)}, V^{(2r+1)}$ such that 
\begin{align}
V_{2r}^{\dagger}\tilde{Q}_{2r} V_{2r}=Q_{2r}, \quad V_{2r+1}^{\dagger}\tilde{Q}_{2r+1} V_{2r+1}=Q_{2r+1}. 
\label{Vexist}
\end{align}
Using these single site operators, we can define new 2-site gates:
\begin{align}
\overline{W}^{(1)}_{2r,2r+1} \equiv  W^{(1)}_{2r,2r+1} V_{2r}V_{2r+1} \nonumber \\
\overline{W}^{(2)}_{2r-1,2r} \equiv V_{2r-1}^{\dagger} V_{2r}^{\dagger}W^{(2)}_{2r-1,2r}
\label{Wbardef}
\end{align}
By construction
\begin{align}
\prod_r \overline{W}^{(1)}_{2r,2r+1} \prod_r \overline{W}^{(2)}_{(2r-1,2r)} &=W
\label{w1w2bar}
\end{align}
so the 2-site gates $\overline{W}^{(1)}_{2r,2r+1}$ and $\overline{W}^{(2)}_{2r-1,2r}$ provide another way to write the unitary $U$ as a depth-$2$ quantum circuit. Furthermore, we will now
show that the $2$-site gates $\overline{W}^{(1)}_{2r,2r+1}$ and $\overline{W}^{(2)}_{2r-1,2r}$ are $U(1)$ symmetric.

To see that $\overline{W}^{(1)}_{2r,2r+1}$ is $U(1)$ symmetric, notice that (\ref{tildeQ}) implies that
\begin{align}
\overline{W}^{(1)\dagger}_{(2r,2r+1)} &(Q_{2r}+Q_{2r+1}) \overline{W}^{(1)}_{2r,2r+1} \nonumber \\
&= V_{2r+1}^{\dagger} V_{2r}^{\dagger} (\tilde{Q}_{2r}+\tilde{Q}_{2r+1}) V_{2r} V_{2r+1} \nonumber \\
&=Q_{2r}+Q_{2r+1}
\end{align}
Thus, $\overline{W}^{(1)}_{2r,2r+1}$ commutes with $Q_{2r} + Q_{2r+1}$ which means it also commutes with the total charge $\sum_r Q_r$.

Likewise, to see that $\overline{W}^{(2)}_{2r-1,2r}$ is $U(1)$ symmetric, note that $W = W^{(1)} W^{(2)}$ is $U(1)$ symmetric so
\begin{align}
W^{(2)\dagger} W^{(1)\dagger}\left(\sum_r Q_r\right) W^{(1)} W^{(2)}= \sum_r Q_r
\end{align}
and therefore
\begin{align}
W^{(2)\dagger}  \left(\sum_r \tilde{Q}_r \right) W^{(2)} = \sum_r Q_r
\end{align}
Next, using the decomposition of $W^{(2)}$ into a product of 2-site gates (\ref{w1w2}), we deduce that
\begin{align}
\sum_r W^{(2)\dagger}_{2r-1,2r} (\tilde{Q}_{2r-1} + \tilde{Q}_{2r}) W^{(2)}_{2r-1,2r} = \sum_r (Q_{2r-1} + Q_{2r})
\end{align}
Given that each side of the equation is a sum of terms supported on non-overlapping pairs of sites $\{2r-1,2r\}$ the terms
must be individually equal:
\begin{align}
W^{(2)\dagger}_{2r-1,2r} (\tilde{Q}_{2r-1} + \tilde{Q}_{2r}) W^{(2)}_{2r-1,2r} =  Q_{2r-1} + Q_{2r}
\end{align}
It then follows from (\ref{Vexist}) and (\ref{Wbardef}) that
\begin{align}
\overline{W}^{(2) \dagger}_{2r-1,2r} (Q_{2r-1}+Q_{2r}) \overline{W}^{(2)}_{2r-1,2r}
&=Q_{2r-1}+Q_{2r}
\end{align}
We conclude that $\overline{W}^{(2)}_{2r-1,2r}$ is $U(1)$ symmetric, as we wished to show.

This completes our proof that $W$ is a $U(1)$ symmetric FDLU: we have explicitly constructed a depth-$2$ unitary circuit (\ref{w1w2bar}) that realizes $W$ and has the property that each unitary gate is individually $U(1)$ symmetric.

\section{Connection between $\tilde{\pi}(z)$ and $U(1)$ current}\label{smagnetization}

In this section we derive a relationship between the invariant $\tilde{\pi}(z)$ and the time-averaged $U(1)$ current that flows in a particular geometry. This relationship provides a physical interpretation for $\tilde{\pi}(z)$ as well as a scheme for measuring it.

\begin{figure}[tb]
   \centering
   \includegraphics[width=0.9\columnwidth]{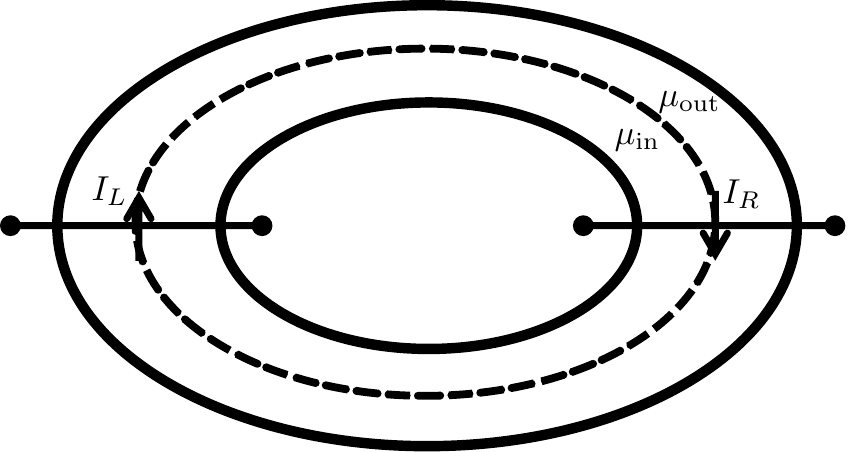} 
   \caption{Our setup for measuring $\tilde{\pi}(z)$. We consider an annulus $A$ which is initially in a mixed state $\rho$ (\ref{rho}) where the inner and outer edges are held at chemical potentials $\mu_{\mathrm{in}}$ and $\mu_{\mathrm{out}}$, with $\mu$ varying between these two values in the bulk of $A$. We define $I_L$ and $I_R$ to be the $U(1)$ currents flowing through the left and right cuts. We show that the time-averaged current $\overline{\< I\>} = \overline{\< I_L\>} = \overline{\< I_R\>}$ is given by (\ref{avgcurrent}).
}
   \label{fig:annulus}
\end{figure}

\subsection{Statement of result}
Our setup is as follows. Consider a 2D $U(1)$ symmetric Floquet system in an annulus geometry as shown in Fig.~\ref{fig:annulus}.  Suppose that at time $t=0$ the system is in a mixed state $\rho$ of the form
\begin{equation}
\rho=\frac{1}{Z} e^{\sum_{r}\mu_rQ_r}, \qquad Z=\mathrm{Tr}\left[e^{\sum_{r}\mu_rQ_r}\right]
\label{rho}
\end{equation}
where $\mu_r$ is some (real-valued) function of $r$ that can be thought of as a site-dependent chemical potential. 

More specifically, consider the case where the site-dependent chemical potential $\mu_r$ takes on constant values, $\mu_r = \mu_{\mathrm{in}}$ and $\mu_r = \mu_{\mathrm{out}}$, near the inner and outer edges of the annulus, and that $\mu_r$ interpolates between $\mu_{\mathrm{in}}$ and $\mu_{\mathrm{out}}$ somewhere deep in the middle of the annulus. We will show that, for an initial state of this kind, the \emph{time-averaged} $U(1)$ current $\overline{\langle I\rangle}$ that flows around the annulus takes a universal value that depends only on $\mu_{\mathrm{in}},  \mu_{\mathrm{out}}$ and the invariant $\tilde{\pi}(z)$ describing the Floquet system. In particular,
\begin{align}
\overline{\< I\>}&=
\frac{1}{T}\left(\frac{d}{d\mu}\log \tilde{\pi}(e^{\mu})|_{\mu_{\mathrm{in}}}-\frac{d}{d\mu}\mathrm{log}\tilde{\pi}(e^{\mu})|_{\mu_{\mathrm{out}}}\right)
\label{avgcurrent}
\end{align}
Here the time-averaged current is defined by
\begin{align}
\overline{\< I\>}&=\mathrm{lim}_{n\to\infty}\left[ \frac{1}{nT}\int_0^{nT}\langle I\rangle dt\right]
\end{align}
where $T$ is the Floquet period and $I$ is the $U(1)$ current operator defined in Eqs.~\ref{Idef1} and \ref{Idef2} below.

A few comments about this result: \\

\textbf{1}. Eq.~\ref{avgcurrent} suggests a scheme for \emph{measuring} the invariant $\tilde{\pi}(z)$: we can hold $\mu_{\mathrm{out}}$ at $-\infty$ and sweep $\mu_{\mathrm{in}}$ from $-\infty$ to $\infty$. By measuring the time-averaged current, we get a function $\overline{\<I\>}(\mu_{\mathrm{in}})$, which can be integrated to find $\tilde{\pi}(z)$ up to multiplication by a constant. This constant can be determined using the fact that $\tilde{\pi}(1)=1$. (Note that this scheme only allows one to measure $\tilde{\pi}(z)$, not $\pi(z)$: to get the latter quantity, one could also need a way to measure the GNVW index -- a challenging problem\cite{tracking}). \\

\textbf{2}. Eq.~\ref{avgcurrent} also gives a physical interpretation to the invariant $\tilde{\pi}(z)$: evidently this quantity is directly related to the time-averaged current that flows at the boundary between two regions with different chemical potentials. The argument $z$, which we originally introduced as a formal variable, corresponds to the \emph{fugacity} of the $U(1)$ charge, $z = e^\mu$. \\

\textbf{3}. It is interesting to consider the special case when $\mu_{\mathrm{in}} = \infty$, $\mu_{\mathrm{out}} = -\infty$. This case corresponds to a maximally filled region near the inner edge and empty region near the outer edge. Substituting into Eq.~\ref{avgcurrent}, we see that the resulting current is \emph{quantized}:
\begin{align}
 \overline{\< I\>} = \frac{n}{T}
\label{quantcurr}
\end{align}
where the integer $n$ is given by taking the difference between the degree of the numerator and degree of the denominator of $\tilde{\pi}(z)$. This quantized current is a direct generalization of the quantized current discussed in Ref.~\onlinecite{magnetization}. In Ref.~\onlinecite{magnetization} it was shown that, for Floquet systems built out of free fermions, there is a quantized time-averaged current that flows along the boundary between a completely filled region and a completely empty region. Eq.~\ref{avgcurrent} generalizes this result to interacting boson and fermion\footnote{Although our derivation of Eq.~\ref{avgcurrent} in written within the framework of bosonic systems, it is clear that the same derivation goes through in the fermonic case, as well.} systems, and to boundaries between two regions at different chemical potentials, $\mu_{\mathrm{in}}$ and $\mu_{\mathrm{out}}$. We can see that in the more general case, the current is not necessarily quantized, but instead is a universal function of the two chemical potentials, $\mu_{\mathrm{in}}$ and $\mu_{\mathrm{out}}$.  \\

\textbf{4}. Ref.~\onlinecite{magnetization} interpreted the quantized current that flows along the boundary between a fully filled region and a completely empty region as coming from a quantized \emph{magnetization density} in the filled region. In a similar fashion, we can interpret the current in Eq.~\ref{avgcurrent} as coming from a $\mu$-dependent magnetization density which generalizes the quantized magnetization density to interacting systems. We will discuss this magnetization density in more detail in a separate work.\\

\textbf{5}. Note that in our language, the free fermion case studied in Ref.~\onlinecite{magnetization} corresponds to $Q = \mathrm{diag}(0,1)$. In this case, the most general possibility for $\pi(z)$ is $\pi(z) = (1+z)^n$ with $\tilde{\pi}(z) = \pi(z)/\pi(1)$. In particular, there is a $\mathbb{Z}$ classification in this case and the quantized current in Eq.~\ref{quantcurr} contains all the information about the nature about the Floquet phase. In more general interacting systems, the quantized current only contains \emph{partial} information about the Floquet phase: to uniquely identify the Floquet phase one needs to know the current for more general chemical potentials, $\mu_{\mathrm{in}}$ and $\mu_{\mathrm{out}}$ (in addition to the GNVW index).\\

\textbf{6}. In addition to $\frac{d}{d\mu}\log\tilde{\pi}(e^\mu)$, higher order derivatives $\frac{d^n}{d\mu^n}\log\tilde{\pi}(e^\mu)$ must also correspond to physical quantities which are topological invariants. Therefore, $\log\tilde{\pi}(e^\mu)$ gives rise to an infinite family of topological invariants. In general, $\frac{d^n}{d\mu^n}\log\tilde{\pi}(e^\mu)$ would have units of charge to the $n$th power, so one could have guessed that the conserved current would take the form of Eq.~\ref{avgcurrent}. \\

\textbf{7}. As we mentioned above, the time averaged current (\ref{avgcurrent}) is not quantized in general. This may be surprising to some readers, since our setup is similar to that of a Thouless pump\cite{Thouless} and it has been proven quite generally that a Thouless pump always transports an integer amount of charge in each cycle\cite{bachmanncharge}. However, there is no contradiction here since our initial state $\rho$ is a \emph{mixed} state and the previous quantization results apply only to pure states.

\subsection{Calculation of time-averaged current}\label{sedge}

Before calculating anything, we first need to give a precise definition of the current operator $I$. To do this, we need to introduce some notation. Let $A$ be an annulus centered at $r_x=r_y=0$. We define $A^+$ to be the top half of the annulus and we define $Q^+$ to be the total charge in $A^+$, that is
\begin{align}
A^+ = \{r\in A : r_y > 0\}, \quad \quad Q^+ = \sum_{r \in A^+} Q_r
\end{align}

Next, we write the Hamiltonian for the Floquet system as a sum
\begin{align}
H(t) = H^+(t) + H^-(t) + H_L(t) + H_R(t), 
\end{align}
where $H^+$ consists of all terms in $H$ that are supported entirely within $A^+$, and $H^-$ consists of all terms that are supported entirely within $A^- = \{r\in A : r_y \leq 0\}$, and where $H_L + H_R$ consists of all terms supported in \emph{both} $A^+$ and $A^-$, with $H_L$ containing those terms that straddle the left boundary between $A^+$ and $A^-$, and $H_R$ containing the terms that straddle the right boundary between $A^+$ and $A^-$. 

We then define two operators, $I_L(t), I_R(t)$ by
\begin{align}
I_L(t) &=  \frac{1}{i} U(t)^\dagger \cdot  [Q^+, H_L(t)] \cdot U(t), \nonumber \\
I_R(t) &= -  \frac{1}{i} U(t)^\dagger \cdot [Q^+, H_R(t)] \cdot U(t)
\label{Idef1}
\end{align}
 where $U(t) = \mathcal{T} e^{-i \int_0^t dt' H(t')}$ is the time evolution operator to time $t$.
The operators $I_L(t), I_R(t)$ can be thought of as Heisenberg-evolved operators that measure the current through the left/right boundaries between $A^+$ and $A^-$ in the clockwise direction (Fig.~\ref{fig:annulus}). To see this, note that the time dependence of $Q^+$ in the Heisenberg picture is given by
\begin{align}
\frac{d}{dt} \left( U(t)^\dagger Q^+ U(t) \right)= I_L(t) - I_R(t)
\label{curcons}
\end{align}
since $[Q^+, H^+(t)] = 0$ due to the fact that $H^+$ is charge conserving, and $[Q^+, H^-(t)] =  0$ due to the fact that the two operators are supported on nonoverlapping regions. Motivated by this fact, we define the (Heisenberg-evolved) current operator $I(t)$ to be 
\begin{align}
I(t) \equiv I_R(t).
\label{Idef2}
\end{align} 
(Note that we could equally well have defined $I(t) \equiv I_L(t)$).

Having defined the current operator $I \equiv I_R$, the next step is to compute the time-average, $\overline{\<I_R\>}$. To this end, we integrate Eq.~(\ref{curcons}) between times $t=0$ and $t = nT$ which gives
\begin{align}
\int_0^{nT} [I_L(t) - I_R(t)] dt &= [U(T)]^{n\dagger} Q^+ [U(T)]^{n} - Q^+ 
\label{ILRn}
\end{align}
We then define time-averaged current operators
\begin{align}
I_L^{(n)} = \frac{1}{nT} \int_0^{nT} I_L(t) dt \nonumber \\
I_R^{(n)} = \frac{1}{nT} \int_0^{nT} I_R(t) dt
\end{align}
In this notation, (\ref{ILRn}) becomes
\begin{align}
I_L^{(n)} - I_R^{(n)} =
\frac{1}{nT} \left([U(T)]^{n\dagger} Q^+ [U(T)]^{n} - Q^+\right) 
\end{align}

To proceed further, we use the fact that $U(T)$ is MBL (in the bulk) to write
\begin{align}
U(T) = U_{\mathrm{edge}} \cdot \prod_r U_r
\end{align}
where $U_{\mathrm{edge}}$ is supported near the two edges of the annulus and where $U_r$ are mutually commuting local unitaries. We claim that only the $U_{\mathrm{edge}}$ term contributes to the time-averaged current flow. More precisely:
\begin{align}
I_L^{(n)} - I_R^{(n)} = \frac{1}{nT} \left(U_{\mathrm{edge}}^{n\dagger} Q^+ U_{\mathrm{edge}}^{n} - Q^+ \right) + \mathcal{O}\left(\frac{1}{n}\right)
\label{ilrid}
\end{align}
where the $\mathcal{O}(1/n)$ term on the right hand side denotes an \emph{operator} whose norm is bounded by $C/n$ for some  constant $C$ that does not depend on $n$. We defer the proof of this result to Appendix~\ref{ilridproof}, but the intuition behind this claim is easy to understand: the $\prod_r U_r$ term cannot generate any charge transport since it is built out of mutually commuting local unitaries. 

Next we write
\begin{align}
U_{\mathrm{edge}} = U_{\mathrm{in}} \cdot U_{\mathrm{out}}, 
\label{Uedgeinout}
\end{align}
where $U_{\mathrm{in}}, U_{\mathrm{out}}$ are supported near the inner and outer edges. We then decompose the annulus $A$ into three disjoint regions: 
\begin{align}
A = A_{\mathrm{in}} \cup A_{\mathrm{out}} \cup A_{\mathrm{bulk}}, 
\end{align}
Here $A_{\mathrm{in}}$ and $A_{\mathrm{out}}$ are finite-width strips near the inner and outer edges of annulus, chosen so that they are wide enough to contain the regions of support of $U_{\mathrm{in}}$ and $U_{\mathrm{out}}$, but narrow enough that the site-dependent chemical potential takes the constant values $\mu_r = \mu_{\mathrm{in}}$ and $\mu_r = \mu_{\mathrm{out}}$, within $A_{\mathrm{in}}$ and $A_{\mathrm{out}}$ respectively. The region $A_{\mathrm{bulk}}$ denotes the remainder of the annulus. 

Similarly, we decompose the upper half of the annulus into three regions
\begin{align}
A^+ = A_{\mathrm{in}}^+ \cup A_{\mathrm{out}}^+ \cup A_{\mathrm{bulk}}^+,
\label{Ainout}
\end{align}
and we define corresponding charge operators
\begin{align}
Q_{\mathrm{in}}^+ = \sum_{r \in A_{\mathrm{in}}^+} Q_r , \quad Q_{\mathrm{out}}^+ = \sum_{r \in A_{\mathrm{out}}^+} Q_r, \quad Q_{\mathrm{bulk}}^+ = \sum_{r \in A_{\mathrm{bulk}}^+} Q_r
\end{align}
We note that
\begin{align}
Q^+ = Q_{\mathrm{in}}^+ + Q_{\mathrm{out}}^+ + Q_{\mathrm{bulk}}^+
\label{Qplusinout}
\end{align}
by construction.

Substituting the above decompositions of $U_{\mathrm{edge}}$ and $Q^+$ (\ref{Uedgeinout}), (\ref{Qplusinout}) into Eq.~\ref{ilrid} and using the fact that $U_{\mathrm{in}}$ commutes with $Q_{\mathrm{out}}^+$ and $Q_{\mathrm{bulk}}^+$, and similarly for $U_{\mathrm{out}}$, we derive: 
\begin{align}
I_L^{(n)} - I_R^{(n)} = \frac{1}{nT} & \big(U_{\mathrm{in}}^{n\dagger} Q^+_{\mathrm{in}} U_{\mathrm{in}}^{n} - Q^+_\mathrm{in} \nonumber \\
&+ U_{\mathrm{out}}^{n\dagger} Q^+_{\mathrm{out}} U_{\mathrm{out}}^{n} - Q^+_\mathrm{out} \big) + \mathcal{O}\left(\frac{1}{n}\right)
\label{ilrid2}
\end{align}

At the same time, since $U_{\mathrm{in}}, U_{\mathrm{out}}$ are $U(1)$ symmetric LPUs, we know that
\begin{align}
U_{\mathrm{in}}^{n\dagger} Q^+_{\mathrm{in}} U_{\mathrm{in}}^n - Q^+_{\mathrm{in}} &= O_{L, \mathrm{in}}^{(n)} + O_{R, \mathrm{in}}^{(n)} \nonumber \\
U_{\mathrm{out}}^{n\dagger} Q^+_{\mathrm{out}} U_{\mathrm{out}}^n -Q^+_{\mathrm{out}} &= O_{L, \mathrm{out}}^{(n)} + O_{R, \mathrm{out}}^{(n)}
\label{olrn}
\end{align}
where $O_{L, \mathrm{in}}^{(n)}, O_{R , \mathrm{in}}^{(n)}$ are operators supported within $A_{\mathrm{in}}$ and near the left, right boundaries of $A^+$ respectively, and similarly for $O_{L, \mathrm{out}}^{(n)}, O_{R , \mathrm{out}}^{(n)}$. Each of these operators is well-defined up to shifting by a scalar $c \mathbbm{1}$; to fix this ambiguity, we choose $O_{L, \mathrm{in}}^{(n)}, O_{R , \mathrm{in}}^{(n)}$ so that the smallest eigenvalue of $Q^+_{\mathrm{in}} + O_{L, \mathrm{in}}^{(n)} $ and $Q^+_{\mathrm{in}} + O_{R, \mathrm{in}}^{(n)}$ is $0$, and similarly for $O_{L, \mathrm{out}}^{(n)}, O_{R , \mathrm{out}}^{(n)}$. Note that this the same convention as in Sec.~\ref{sdefs}.

Combining (\ref{olrn}) with (\ref{ilrid2}), we derive
\begin{align}
I_L^{(n)} - I_R^{(n)} = \frac{1}{nT}(O_{L, \mathrm{in}}^{(n)} &+ O_{L, \mathrm{out}}^{(n)} + O_{R, \mathrm{in}}^{(n)} + O_{R, \mathrm{out}}^{(n)}) \nonumber \\
&+ \mathcal{O}\left(\frac{1}{n}\right)
\end{align}
The `$L$' and `$R$' terms on both sides must agree up to addition by a scalar, so we deduce in particular that
\begin{align}
I_R^{(n)} = -\frac{1}{nT}(O_{R, \mathrm{in}}^{(n)} + O_{R, \mathrm{out}}^{(n)} ) + c_n \mathbbm{1} + \mathcal{O}\left(\frac{1}{n}\right)
\label{Irn0}
\end{align}
for some constant $c_n$ that may depend on $n$. In fact, one can show that the constant $c_n$ is at most of order $\mathcal{O}(1/n)$. \footnote{To see this, take the expectation value of Eq.~\ref{Irn0} in the density matrix $\rho$ (\ref{rho}) in the special case $\mu_{\mathrm{in}} = \mu_{\mathrm{out}} = 0$. Notice that $\<I_R^{(n)}\> = 0$ in this case since $\rho \propto \mathbbm{1}$ and $I_R^{(n)}$ is a traceless operator. At the same time, we have $\<O_{R, \mathrm{in}}^{(n)}\> + \<O_{R, \mathrm{out}}^{(n)}\> = 0$ using Eqs.~\ref{orout}-\ref{orin}.}  Therefore, we can absorb $c_n$ into the $\mathcal{O}(1/n)$ term, giving
\begin{align}
I_R^{(n)} = -\frac{1}{nT}(O_{R, \mathrm{in}}^{(n)} + O_{R, \mathrm{out}}^{(n)} ) + \mathcal{O}\left(\frac{1}{n}\right)
\end{align}
Taking the expectation value with respect to the density matrix $\rho$ (\ref{rho}) gives
\begin{align}
\<I_R^{(n)}\> = -\frac{1}{nT}(\<O_{R, \mathrm{in}}^{(n)}\> + \<O_{R, \mathrm{out}}^{(n)}\>) + \mathcal{O}\left(\frac{1}{n}\right)
\label{irexp}
\end{align}
where now the $\mathcal{O}(1/n)$ term denotes a \emph{scalar} whose absolute value is bounded by $C/n$ for some constant $C$.

\begin{figure}[tb]
   \centering
   \includegraphics[width=.9\columnwidth]{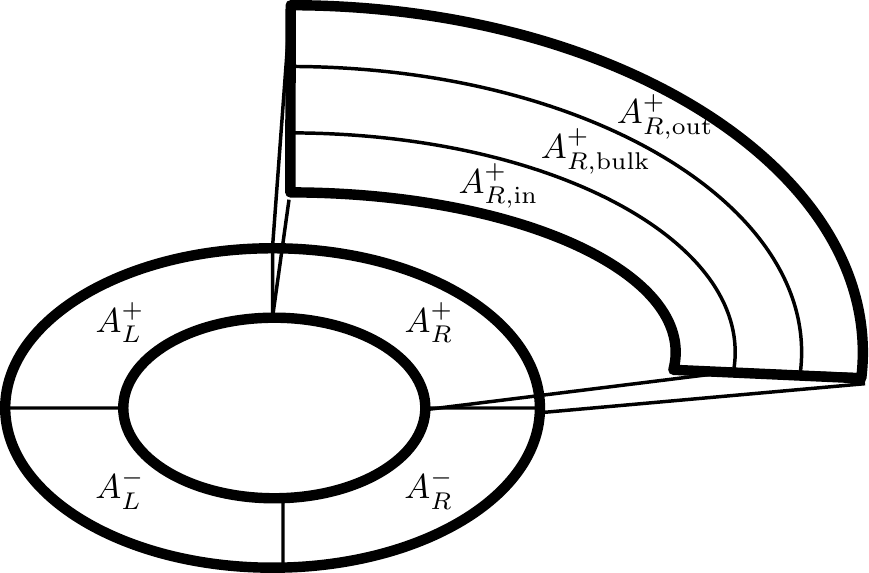} 
   \caption{The annulus $A$ is divided into 4 quadrants $A_R^+,A_L^+,A_R^-,$ and $A_L^-$ (\ref{quadrants}). Each quadrant is divided three smaller regions as written in (\ref{inoutbulk}).}
   \label{fig:Aquad}
\end{figure}

The next step is to evaluate the two terms, $\<O_{R, \mathrm{in}}^{(n)}\>$ and $\<O_{R, \mathrm{out}}^{(n)}\>$. Before we do this, we need to introduce some notation for denoting the $4$ quadrants of the annulus. First, we write
\begin{align}\label{quadrants}
A = A_R^+ \cup A_L^+ \cup A_R^- \cup A_L^-
\end{align}
where $A_R^+$ denotes the upper-right quadrant, and $A_L^+$ denotes the upper-left quadrant and so on. We then decompose each of the $4$ quadrants into three smaller regions, similarly to (\ref{Ainout}). For example, we write (Fig.~\ref{fig:Aquad})
\begin{align}\label{inoutbulk}
A_R^+ = A_{R,\mathrm{in}}^+ \cup A_{R,\mathrm{out}}^+ \cup A_{R,\mathrm{bulk}}^+ 
\end{align}
and we define corresponding charge operators, $Q_{R,\mathrm{in}}^{+}$, $Q_{R,\mathrm{out}}^{+}$ and $Q_{R,\mathrm{bulk}}^{+}$. 

With this notation, we are now ready to evaluate $\<O_{R, \mathrm{in}}^{(n)}\>$ and $\<O_{R, \mathrm{out}}^{(n)}\>$. We start with $\<O_{R, \mathrm{out}}^{(n)}\>$.  First, we write $\<O_{R, \mathrm{out}}^{(n)}\>$ as a difference of two terms:
\begin{align}
\<O_{R, \mathrm{out}}^{(n)}\> & = \mathrm{Tr}[(Q^+_{R, \mathrm{out}} + O_{R, \mathrm{out}}^{(n)})\rho] - \mathrm{Tr}[Q^+_{R, \mathrm{out}} \rho] 
\label{qoq}
\end{align}
Next, we evaluate the two terms on the right hand side of (\ref{qoq}). We begin with the second term. To evaluate this term, we note that
\begin{align}
\mathrm{Tr}[Q^+_{R, \mathrm{out}} \ \rho] 
&= \frac{\mathrm{Tr}[Q^+_{R, \mathrm{out}} \ e^{\mu_{\mathrm{out}} Q^+_{R, \mathrm{out}}}]}{\mathrm{Tr}[e^{\mu_{\mathrm{out}} Q^+_{R, \mathrm{out}}}]} \nonumber \\
&= \frac{d}{d\mu} \log \mathrm{Tr}[e^{\mu Q^+_{R, \mathrm{out}}}] |_{\mu_{\mathrm{out}}}
\label{qoq2}
\end{align}
where the first equality follows from tracing out all spins that are outside the region $A^+_{R, \mathrm{out}}$.

Likewise to evaluate the first term in (\ref{qoq}), we note that 
\begin{align}
\mathrm{Tr}&[(Q^+_{R, \mathrm{out}}+O_{R, \mathrm{out}}^{(n)})\rho] \nonumber \\
&= \frac{\mathrm{Tr}[(Q^+_{R, \mathrm{out}}+O_{R, \mathrm{out}}^{(n)}) e^{\mu_{\mathrm{out}} Q_{R, \mathrm{out}}}]}{\mathrm{Tr}[e^{\mu_{\mathrm{out}} Q_{R, \mathrm{out}}}]} \nonumber \\
&= \frac{\mathrm{Tr}[(Q^+_{R, \mathrm{out}}+O_{R, \mathrm{out}}^{(n)}) e^{\mu_{\mathrm{out}}(Q^+_{R, \mathrm{out}}+O_{R, \mathrm{out}}^{(n)})}]}{\mathrm{Tr}[e^{\mu_{\mathrm{out}}(Q^+_{R, \mathrm{out}}+ O_{R, \mathrm{out}}^{(n)})}]} \nonumber \\
&= \frac{d}{d\mu} \log \mathrm{Tr}[ e^{\mu (Q^+_{R, \mathrm{out}}+O_{R, \mathrm{out}}^{(n)})}] |_{\mu_{\mathrm{out}}}
\label{qoq1}
\end{align}
Here the first equality follows from tracing out all spins that are outside $A_{R, \mathrm{out}} = A_{R, \mathrm{out}}^+ \cup A_{R, \mathrm{out}}^-$. The second equality also follows from tracing out certain degrees of freedom, but its justification is more subtle. To explain this step, let $\mathcal{H}_{R, \mathrm{out}}$ denote the Hilbert space describing the spins in region $A_{R, \mathrm{out}}$. In Appendix~\ref{sgnvw} we show that $\mathcal{H}_{R, \mathrm{out}}$ can be written as a tensor product of two smaller Hilbert spaces $\mathcal{H}_1 \otimes \mathcal{H}_2$ and that, in this representation, the two operators $(Q^+_{R, \mathrm{out}}+O_{R, \mathrm{out}}^{(n)})$ and  $(Q^-_{R, \mathrm{out}}-O_{R, \mathrm{out}}^{(n)})$ take the form
\begin{align}
Q^+_{R, \mathrm{out}}+O_{R, \mathrm{out}}^{(n)} = \mathcal{O}_1 \otimes \mathbbm{1}, \nonumber \\
Q^-_{R, \mathrm{out}}-O_{R, \mathrm{out}}^{(n)} = \mathbbm{1} \otimes \mathcal{O}_2
\label{gnvwsplit}
\end{align}
for some operators $\mathcal{O}_1, \mathcal{O}_2$.
Therefore, since $Q_{R, \mathrm{out}}$ can be written as a sum
\begin{equation}
Q_{R, \mathrm{out}}=(Q^+_{R, \mathrm{out}}+O_{R, \mathrm{out}}^{(n)})+(Q^-_{R, \mathrm{out}}-O_{R, \mathrm{out}}^{(n)})
\end{equation}
we can derive the second equality in (\ref{qoq1}) by tracing out the degrees of freedom in $\mathcal{H}_2$.

Substituting (\ref{qoq2}) and (\ref{qoq1}) into (\ref{qoq}) gives
\begin{align}
\<O_{R, \mathrm{out}}^{(n)}\>&= 
\frac{d}{d\mu} \log \mathrm{Tr}[e^{\mu (Q^+_{R, \mathrm{out}}+O_{R, \mathrm{out}}^{(n)})}] |_{\mu_{\mathrm{out}}} \nonumber \\
&- \frac{d}{d\mu} \log \mathrm{Tr}[ e^{\mu Q^+_{R, \mathrm{out}}}] |_{\mu_{\mathrm{out}}}  \nonumber \\
&=\frac{d}{d\mu}\log \tilde{\pi}^{(n)}(e^{\mu})|_{\mu_{\mathrm{out}}}
\label{orout}
\end{align}
where $\tilde{\pi}^{(n)}(z)$ is the invariant corresponding to the edge unitary $U_{\mathrm{out}}^n$ (see Eq.~\ref{olrn}). Repeating the exercise for the inner edge gives
\begin{align}
\<O_{R, \mathrm{in}}^{(n)}\>&=-\frac{d}{d\mu}\log \tilde{\pi}^{(n)}(e^{\mu})|_{\mu_{\mathrm{in}}}
\label{orin}
\end{align}
where the minus sign comes from the fact that the inner edge has the opposite orientation of the outer edge.

Substituting (\ref{orout}) and (\ref{orin}) into (\ref{irexp}), we derive
\begin{align}
\<I_R^{(n)}\>=  \frac{1}{nT} &\left(\frac{d}{d\mu}\log \tilde{\pi}^{(n)}(e^{\mu})|_{\mu_{\mathrm{in}}}  -\frac{d}{d\mu}\log \tilde{\pi}^{(n)}(e^{\mu})|_{\mu_{\mathrm{out}}} \right) \nonumber \\
& + \mathcal{O}\left(\frac{1}{n}\right)
\end{align}
The next step is to note that $\tilde{\pi}^{(n)}(z) =  [\tilde{\pi}(z)]^n$ by the composition law for $\pi$ discussed in Sec.~\ref{sdefs}. Using this identity we derive
\begin{align}
\<I_R^{(n)}\>=\frac{1}{T}\left(\frac{d}{d\mu}\log \tilde{\pi}(e^{\mu})|_{\mu_{\mathrm{in}}}-\frac{d}{d\mu}\mathrm{log}\tilde{\pi}(e^{\mu})|_{\mu_{\mathrm{out}}}\right) \nonumber \\
+ \mathcal{O}\left(\frac{1}{n}\right)
\end{align}
If we now take the limit $n \rightarrow \infty$, we derive Eq.~\ref{avgcurrent}, the desired formula for $\overline{\< I\>}$.

\subsection{Where does the current flow?}\label{swhere}
In the previous section we computed the \emph{total} time-averaged current $\overline{\<I\>}$ that flows around the annulus. We now study the spatial distribution of this current -- that is, we study the current $I_{rr'}$ that flows between each pair 
of sites $r,r'$. We ask: where is $\overline{\<I_{rr'}\>} \neq 0$ -- that is, where does the current flow? One might guess that the current flows along the two edges of the annulus, but we will show below that the current actually flows along the \emph{boundary} between the two regions with different chemical potentials $\mu_{\mathrm{in}}$, $\mu_{\mathrm{out}}$.\footnote{More precisely, we prove that the current has this spatial distribution for \emph{our} definition of the current operator $I_{rr'}$; it may not be true for other definitions.} We note that a similar result was derived in Ref.~\cite{magnetization} in the case of free fermion systems with chemical potentials $\mu_{\mathrm{in}} = + \infty$ and $\mu_{\mathrm{out}} = -\infty$: this section can be viewed as a generalization of this result to interacting systems and general chemical potentials.


First we need to define the current operator $I_{rr'}$. Unlike the total current $I$, there is no canonical definition of $I_{rr'}$ -- there are many equally good definitions. Our definition starts by writing the Hamiltonian as a sum of local terms,
\begin{align}
H(t) = \sum_r H_r(t)
\end{align}
where the index $r$ runs over the different sites of the lattice, and where $H_r$ is a charge-conserving operator whose region of support contains site $r$. Such a decomposition of $H$ always exists for any charge-conserving Hamiltonian with local interactions, though it is not unique (this non-uniqueness is directly related to the fact that there is no canonical definition of $I_{rr'}$). Once we fix a decomposition of $H$, we define the (Heisenberg-evolved) current operator $I_{rr'}$ as
\begin{align}
I_{rr'}(t) = \frac{1}{i} U(t)^\dagger  \left([Q_{r'}, H_r(t)] - [Q_r, H_{r'}(t)] \right) U(t)
\end{align}
This definition is reasonable because (i) $I_{rr'}$ is a local operator supported near sites $r, r'$; (ii) $I_{rr'}$ is anti-symmetric in the sense that $I_{rr'} = -I_{r'r}$; and (iii) $I_{rr'}$ obeys the current conservation law
\begin{align}
\frac{d}{dt} \left( U(t)^\dagger Q_r U(t) \right) = -\sum_{r'} I_{rr'}
\end{align}

Having defined $I_{rr'}$, we need to explain the relationship between $I_{rr'}$ and the total current $I$. These quantities are related in a very simple and intuitive way: the total current $I \equiv I_R$ can be written as a sum over all $I_{rr'}$ where $r,r'$ lie on different sides of the cut that we use to define $I$. More specifically,
\begin{align}
I(t) = \sum_{r \in A^+_R} \sum_{r' \in A^-_R} I_{rr'}(t)
\end{align}
where $A^+_R$ denotes the upper right quadrant of the annulus, i.e. $A^+_R = \{r \in A : r_x, r_y > 0\}$ and $A^-_R$ denote the lower right quadrant, $A^-_R = \{r \in A : r_x > 0, r_y \leq 0\}$.
It is straightforward to show that this expression agrees with our original definition from Eq.~\ref{Idef2}.

We are now ready to compute the expectation value $\<I_{rr'}(t)\>$. First, consider the special case where the entire annulus is at constant chemical potential $\mu$: i.e.  $\rho = e^{-\mu Q}/ \mathrm{Tr}(e^{-\mu Q})$ where $Q = \sum_r Q_r$. In this case, we can see that the current vanishes exactly at every time $t$:
\begin{align}
&\<I_{rr'}(t)\> = \mathrm{Tr}\left( \frac{1}{i} U^\dagger(t)  ([Q_{r'}, H_r(t)] - [Q_r, H_{r'}(t)]) U(t) \rho \right) \nonumber \\
 &= \frac{\mathrm{Tr}\left(\frac{1}{i} U^\dagger(t)  ([Q_{r'}, H_r(t)] - [Q_r, H_{r'}(t)] ) U(t) e^{-\mu Q}\right)}{\mathrm{Tr}\left(e^{-\mu Q}\right)} \nonumber \\
&=  \frac{\mathrm{Tr}\left( \frac{1}{i} ([Q_{r'}, H_r(t)] - [Q_r, H_{r'}(t)] ) e^{-\mu Q}\right)}{\mathrm{Tr}\left(e^{-\mu Q}\right)} \nonumber \\
&= 0
\label{Iabzero}
\end{align}
Here the third equality follows from the fact that $U(t)$ commutes with the total charge $Q$ and the last equality follows from the fact that $Q_r$ and $H_{r'}(t)$ commute with $e^{-\mu Q}$.

Next consider the general  case where $\mu_{\mathrm{in}} \neq \mu_{\mathrm{out}}$. Given Eq.~\ref{Iabzero} it is clear that $\<I_{rr'}(t)\> = 0$ deep within any region with constant chemical potential. In particular, $\<I_{rr'}(t)\> = 0$ near the inner and outer edges of the annulus where $\mu$ takes the constant values $\mu_{\mathrm{in}}$, $\mu_{\mathrm{out}}$. We conclude that the current must be localized at the \emph{boundary} between the two regions at chemical potentials $\mu_{\mathrm{in}}, \mu_{\mathrm{out}}$. This proves the claim.


\section{Fermionic systems}\label{sfermionic}
We now extend our results to systems with fermionic degrees of freedom. Our main result is that fermionic systems can be classified using almost the same framework as bosonic systems: there is a one-to-one correspondence between 2D $U(1)$ symmetric fermionic Floquet phases and rational functions $\pi^f(z)$, satisfying a modified version of Eq.~\ref{requ} given by Eq.~\ref{requu1}.

\subsection{Review: Fermionic no symmetry case}\label{sfermionicnosymm}
We begin by reviewing the classification of fermionic Floquet systems \emph{without} symmetry, a problem that was studied in Ref.~\onlinecite{fermionic}. 

We consider fermionic Floquet systems that are built out of a two-dimensional lattice. We assume that each lattice site $r$ is described by a $d$-dimensional Hilbert space, $\mathcal{H}_r$, with a $\mathbb{Z}_2$-graded structure associated with fermion parity: that is, 
\begin{equation}
\mathcal{H}_r=\mathcal{H}_{0r}\oplus \mathcal{H}_{1r}
\end{equation}
where $\mathcal{H}_{0r}$ is the subspace spanned by states with \emph{even} fermion parity, and $\mathcal{H}_{1r}$ is the subspace spanned by states with \emph{odd} fermion parity. The $\mathbb{Z}_2$ graded structure can be characterized by two non-negative integers, $d_0 = \text{dim}(\mathcal{H}_0)$ and $d_1 = \text{dim}(\mathcal{H}_1)$, and we will denote it by  $\mathbb{C}^{d_0|d_1}$. 
We assume that all lattice sites are identical and have the same structure $\mathbb{C}^{d_0|d_1}$. In this language, a conventional spinless fermion is described by a $2$ dimensional Hilbert space, $\mathbb{C}^{1|1}$, while the bosonic systems we discussed earlier have $d$ dimensional Hilbert spaces of the form $\mathbb{C}^{d|0}$. 

Similarly to the bosonic case, we require the Hamiltonian $H(t)$ to be local, periodic in time, and fermion parity even, and we require the Floquet unitary $U_F$ to obey the MBL condition (\ref{mblcond}) where each $U_r$ is fermion parity even. The definition of a fermionic Floqet \emph{phase} is similar to the bosonic case, but with one technical difference: we define two fermionic Floquet systems, $H_A(t)$ and $H_B(t)$, to belong to the same phase if their boundary can be many-body localized in the presence of \emph{ancillas}. That is, when determining whether the boundary between $H_A(t)$ and $H_B(t)$, can be many-body localized, one is allowed to attach a one-dimensional chain of \emph{ancilla} lattice sites at the boundary between $H_A$ and $H_B$ which one can then couple to the other nearby sites with an arbitrary (local) fermion-parity even Hamiltonian $H_{\mathrm{bd}}$. Crucially, we allow these ancilla sites to have any Hilbert space structure $\mathbb{C}^{m|n}$, which need not be the same as the Hilbert space structure structure of the other lattice sites. The motivation for including these ancillas is that they can help many-body localize certain boundaries which are otherwise not localizable; as a result, including ancillas in the definition leads to a coarser (and simpler) classification of fermionic Floquet phases. This (coarser) notion of equivalence is sometimes called ``stable equivalence.''\footnote{In the bosonic case, it turns out that adding ancillas has no effect on whether a boundary can be many-body localized, and for that reason we omit them from the definition.}

Using similar arguments to the bosonic case, one can show that the classification of 2D fermionic Floquet phases is equivalent to the classification of 1D fermionic LPUs modulo fermionic FDLUs, in the presence of ancillas. The latter classification problem was studied in Ref.~\cite{fermionic}. In that work, the authors showed that there is a one-to-one correspondence between equivalence classes of 1D fermionic LPUs and real numbers of the form
\begin{align}
\sqrt{2}^{\zeta}\frac{p}{q}
\end{align}
where $p, q$ are products of prime factors of $d = d_0 + d_1$, and where $\zeta = 0$ unless $d_0 = d_1$, in which case $\zeta$ can be either $0$ or $1$. The authors also showed how to compute this number given a locality preserving unitary $U$:
\begin{equation}
\sqrt{2}^{\zeta}\frac{p}{q} = \mathrm{ind}^f(U)
\end{equation}
where $\mathrm{ind}^f(U)$ is defined by an explicit formula very similar to the one reviewed in Appendix~\ref{GNVWformapp}.

Similarly to the bosonic case, one way to interpret this classification result is that the only possible locality preserving unitaries in 1D fermionic systems are \emph{translations}. More specifically, fermionic systems can support two types of translations: conventional translations and ``Majorana'' translations. A general locality preserving unitary $U$ is a combination of a conventional translation and (possibly) a Majorana translation. The conventional translation can be labeled by a rational number $p/q$ just as in the bosonic case, while the presence or absence of a Majorana translation is encoded in the $\mathbb{Z}_2$ index, $\sqrt{2}^\zeta$.

\subsection{Fermionic $U(1)$ symmetric case}\label{sfermionicu1}

Moving on to  the $U(1)$ symmetric case, we now consider systems in which each lattice site is described by a $d$-dimensional $\mathbb{Z}_2$-graded Hilbert space with a $U(1)$ symmetry transformation. Such systems are naturally characterized by two $d \times d$ diagonal matrices: $P = \mathrm{diag}(p_1,...,p_d)$ and $Q = \mathrm{diag}(q_1,...,q_d)$. Here $p_i$ and $q_i$ describe the fermion parity and $U(1)$ charge of the $i$th state of a single lattice site. We will use a convention where the $p_i$'s take values in $\{\pm 1\}$ with $+1$ and $-1$ corresponding to even and odd fermion parity, respectively. Also, we will assume without loss of generality that all the $q_i$'s are non-negative integers and that $\mathrm{min}_i q_i = 0$. 

In addition to the two matrices $P, Q$, we also find it useful to define \emph{operators} $P_r, Q_r$ associated with lattice site $r$. These operators can be thought of as $P_r = P \otimes \mathbbm{1}$ and $Q_r = Q  \otimes\mathbbm{1}$ where $P, Q$ act on site $r$ and $\mathbbm{1}$ acts on the other sites. The total fermion parity is then given by $P_{\mathrm{tot}} = \prod_r P_r$, while the total $U(1)$ charge is $Q_{\mathrm{tot}} = \sum_r Q_r$.

With this notation, we are now ready to explain our setup. Similarly to the no-symmetry case, we require the Hamiltonian $H(t)$ to be local, periodic in time, fermion parity even, and $U(1)$ symmetric. Also we require the Floquet unitary $U_F$ to obey the MBL condition (\ref{mblcond}) where each $U_r$ is fermion parity even and $U(1)$ symmetric. The definition of a fermionic Floqet \emph{phase} is similar to the no-symmetry case discussed above: we define two Floquet systems, $H_A(t)$ and $H_B(t)$, to belong to the same phase if their boundary can be many-body localized in the presence of ancillas. These ancillas can have any (finite dimensional) Hilbert space structure and any $U(1)$ symmetry transformation (i.e. any $P, Q$) which need not be the same as the other lattice sites. 

Using similar arguments to the bosonic case, one can show that the classification of 2D $U(1)$ symmetric fermionic Floquet phases is equivalent to the classification of 1D locality preserving unitaries modulo FDLUs (in the presence of ancillas). In the remainder of this section, we solve the latter classification problem and in this way, we 
derive a complete classification of 2D $U(1)$ symmetric fermionic Floquet phases.

To describe our main result, it is convenient to introduce two generating functions:
\begin{align}
f_Q(z)=\mathrm{Tr}\left(z^{Q}\right), \quad \quad f_{\chi}(z)=\mathrm{Tr}\left(z^{Q}P\right)
\end{align}
Given our assumptions about the eigenvalues of $P,Q$, it follows that $f_Q(z)$ is a polynomial with non-negative integer coefficients and $f_\chi(z)$ is a polynomial with integer coefiicents.


Our main result is that the Floquet phases that can be realized in a system with a given $P$ and $Q$ have a one-to-one correspondence to rational functions $\pi^f(z)$ which satisfy
\begin{align}
\left[f_Q(z)\right]^{N_1}\pi^f(z)&=\sqrt{2}^\zeta[\alpha_0(z)+\alpha_1(z)] \nonumber \\
\left[f_Q(z)\right]^{N_2}\frac{1}{\pi^f(z)}&=\sqrt{2}^\zeta[\beta_0(z)+\beta_1(z)] \label{requu1} \\
\left[f_{\chi}(z)\right]^{N_1+N_2}=[\alpha_0&(z)-\alpha_1(z)][\beta_0(z)-\beta_1(z)]\delta_{\zeta,0} \nonumber
\end{align}
for some integers $N_1, N_2 \geq 0$, and some non-negative integer polynomials $\alpha_0(z),\alpha_1(z),\beta_0(z),\beta_1(z)$ and some $\zeta \in \{0,1\}$. Note that the Kronecker symbol $\delta_{\zeta,0}$ in the last equation implies that $\zeta = 0$ unless $f_{\chi}(z)=0$.

To better understand this result and its relation with our bosonic classification, it is helpful to divide $U(1)$ symmetric fermionic systems into three classes, according to the structures of $P, Q$: \\

\textbf{Case 1:} $P = (-1)^Q$, i.e $\mathbb{Z}_2$ fermion parity symmetry is incorporated as a subgroup of $U(1)$ symmetry. In this case, $f_{\chi}(z) = f_Q(-z)  \neq 0$, so we must have $\zeta = 0$ according to the third constraint in (\ref{requu1}).  We claim that any $\pi^f(z)$ that satisfies the first two equations in (\ref{requu1}) also satisfies the third equation. To see this, note that for any solution $\pi^f(z)$ to the first two equations, we can always choose the corresponding $\alpha_0(z), \alpha_1(z)$ to be even and odd polynomials, respectively, and similarly for $\beta_0(z), \beta_1(z)$. Then the third equation in (\ref{requu1}) is simply the product of the first two equations after the substitution $z\to -z$, so it is automatically satisfied. Hence, we can ignore the third equation, and the first two equations reduce to the constraints on $\pi(z)$ in the bosonic case (\ref{requ}). We conclude that, in this case, fermionic phases with a given $Q$ have the \emph{same} classification as their bosonic counterparts: $\{\pi^f(z)\}=\{\pi(z)\}$. \\

\textbf{Case 2:} $P  \neq (-1)^Q$ and $f_{\chi}(z)=0$. In this case, $\zeta$ can be either $0$ or $1$. If $\zeta = 1$ then the third equation in (\ref{requu1}) drops out, and the first two equations provide identical constraints to the bosonic case (\ref{requ}) except for an additional factor of $\sqrt{2}$. Therefore, the set of allowed $\pi^f(z)$ is simply $\{\pi^f(z)\}=\{\sqrt{2}\pi(z)\}$. On the other hand if $\zeta = 0$, then the third equation implies that $\alpha_0(z) = \alpha_1(z)$ or $\beta_0(z) = \beta_1(z)$. It is not hard to show that the latter equations do not imply any additional constraints on $\pi^f(z)$\footnote{The coefficients of $f_Q(z)$ are all even (since $f_{\chi}(z)=0$), so it is always possible to choose $\alpha_0(z) = \alpha_1(z)$ by taking $N_1$ sufficiently large.} while the first two equations are identical to the bosonic case (\ref{requ}). Therefore, the set of allowed $\pi^f(z)$ is $\{\pi^f(z)\}=\{\pi(z)\}$. Combining the $\zeta = 0$ and $\zeta = 1$ case, we conclude that fermionic phases with a given $Q$ have the same classification as their bosonic counterparts, except that there is an additional fermionic phase whose edge unitary is a neutral Majorana translation: $\{\pi^f(z)\}=\{\pi(z)\}\cup\{\sqrt{2}\pi(z)\}$. \\

\textbf{Case 3:} $P  \neq (-1)^Q$ and $f_{\chi}(z) \neq  0$. In this case, $\zeta = 0$, so the first two equations in (\ref{requu1}) reduce to the bosonic constraints (\ref{requ}). However, the third equation in  (\ref{requu1}) cannot be eliminated in general: this equation places additional constraints on the set of allowed $\pi^f(z)$. Therefore, all we can say is that the set of allowed $\{\pi^f(z)\}$ is in general a subgroup of the set of allowed $\{\pi(z)\}$ with the same $Q$. 

\subsubsection{Definition  of $\pi^f(z)$}
 
We define $\pi^f(z)$ in the same way that we defined $\pi(z)$ in the bosonic case: for any $U(1)$ symmetric fermionic Floquet system with an edge unitary $U = U_{\mathrm{edge}}$, we define $\pi^f(z)$ to be the product
\begin{equation}\label{pifdef}
\pi^f(z)=\mathrm{ind}^f(U)\cdot\tilde{\pi}^f(z)
\end{equation}
where $\mathrm{ind}^f(U)$ is the ``no-symmetry'' index discussed above, and $\tilde{\pi}^f(z)$ is defined in exactly the same way as $\tilde{\pi}(z)$ (\ref{deftilpi}). 
 
To prove the classification, we need to establish three claims: (1) $\pi^f(z)$ is always a rational function satisfying Eq.~\ref{requu1}; (2) All rational functions satisfyng Eq.~\ref{requu1} can be realized by some $U$; (3) $\pi^f_{U'}(z) = \pi^f_{U}(z)$ if and only if $U' U^{-1}$ can be written as a $U(1)$ symmetric FDLU in the presence of ancillas. In the following sections, we sketch proofs of these claims.

\subsubsection{$\pi^f(z)$ satisfies Eq.~\ref{requu1}}\label{ssatisfiesfermionic}

In this section we prove property (1): we show that $\pi^f(z)$ is always a rational function satisfying Eq.~\ref{requu1} for any 1D $U(1)$ symmetric locality preserving unitary $U$. 

Let $U$ be a 1D $U(1)$ symmetric locality preserving unitary. We first prove the claim in the case $\zeta = 0$ where $\zeta$ is defined by the no-symmetry index: $\mathrm{ind}^f(U) = \sqrt{2}^\zeta p/q$.

Our proof is very similar to the bosonic case. Let $\ell$ be the operator spreading length for $U$, and let $A_L, A_R$ denote the two adjacent intervals, $A_L = [-\ell+1,0]$ and $A_R = [1, \ell]$. Also, let $Q_L, Q_R$ denote the total charge in $A_L, A_R$, and let  $P_L, P_R$ denote the parity operator restricted to $A_L, A_R$. As in the bosonic case, we know that
\begin{align}
U^\dagger (Q_L + Q_R) U = Q_L + O_L + Q_R + O_R
\label{QLRid}
\end{align}
where $O_L, O_R$ are operators supported in $L = [-2\ell+1,0]$ and $R = [1, 2\ell]$, respectively. Similarly, one can show that
\begin{align}
U^\dagger (P_L P_R) U = (P_L Y_L)(P_R Y_R)
\label{PLRid}
\end{align}
where $Y_L, Y_R$ are operators supported in $L, R$. 

We now pause to explain our conventions for defining $O_L, O_R$ and $Y_L, Y_R$. Just as in the bosonic case, $O_L, O_R$ are ambiguous up to adding/subtracting a scalar (\ref{shiftamb}). We fix this ambiguity using the same prescription as in the bosonic case: we demand that the smallest eigenvalue of $Q_R+O_R$ is $0$. Similarly $Y_L, Y_R$ are ambiguous up to multiplication/division by a scalar; we fix the latter ambiguity by demanding that $P_R Y_R$ has eigenvalues $\pm 1$.\footnote{There is still a residual sign ambiguity, $Y_L \rightarrow - Y_L$, $Y_R \rightarrow -Y_R$ but this ambiguity will not be important below.}

With these conventions, we define
\begin{align}
\alpha_0(z) &= \frac{p}{d^\ell q} \mathrm{Tr}_R\left(z^{Q_R+O_R} \left[\frac{1+P_R Y_R}{2} \right]\right) \nonumber \\
\alpha_1(z) &= \frac{p}{d^\ell q}\mathrm{Tr}_R\left(z^{Q_R+O_R} \left[\frac{1-P_R Y_R}{2} \right]\right) \nonumber \\
\beta_0(z) &= \frac{q}{d^\ell p}\mathrm{Tr}_L\left(z^{Q_L+O_L} \left[\frac{1+P_L Y_L}{2} \right]\right) \nonumber \\
\beta_1(z) &= \frac{q}{d^\ell p} \mathrm{Tr}_L\left(z^{Q_L+O_L} \left[\frac{1-P_L Y_L}{2} \right]\right) 
\label{alphadef}
\end{align}
We claim that $\alpha_0(z), \alpha_1(z), \beta_0(z), \beta_1(z)$ are non-negative integer polynomials. To see this, let us first consider $\alpha_0(z)$. Notice that $(1 + P_R Y_R)/2$ is a \emph{projection} operator since $P_R Y_R$ has eigenvalues $\pm 1$. Next notice that $Q_R + O_R$ commutes with $(1+P_R Y_R)/2$: this follows from the fact that $Q$ and $P$ commute with each other. Now consider the eigenvalue spectrum of $Q_R +O_R$ within the projected subspace $P_R Y_R = 1$. To understand this eigenvalue spectrum, note that the eigenvalues of $Q_R + O_R$ are all non-negative integers by the same argument as in the bosonic case. Furthermore, using the same arguments as in the bosonic case, one can show that the restriction of $Q_R + O_R$ to the interval $R$ can be written, in an appropriate basis, as $Q_R + O_R = O \otimes \mathbbm{1}$ where $O$ is a matrix of dimension $(p/q)d^{\ell}$ and $\mathbbm{1}$ is an identity matrix of dimension $(q/p)d^{\ell}$. The same is true for the operator $(1+P_R Y_R)/2$. It follows that all the eigenvalues of $Q_R +O_R$ come with a degeneracy which is a multiple of $\frac{d^\ell q}{p}$ within the subspace $P_R Y_R = 1$. Putting this all together, it follows immediately that $\alpha_0(z)$ is a non-negative integer polynomial. In exactly the same way, we can show that $\alpha_1(z), \beta_0(z), \beta_1(z)$ are also non-negative integer polynomials.

We are now ready to show that $\pi^f(z)$ obeys Eqs.~\ref{requu1}. The first step is to note that, just as in the bosonic case,
\begin{align}
\tilde{\pi}^f(z) = \frac{\mathrm{Tr}_R(z^{Q_R + O_R})}{\mathrm{Tr}_R(z^{Q_R})} = \frac{\mathrm{Tr}_R(z^{Q_L})}{\mathrm{Tr}_R(z^{Q_L + O_L})}
\label{pifid}
\end{align}
Next, we use the first equality in (\ref{pifid}), together with $\mathrm{Tr}_R(z^{Q_R}) = d^\ell \left[f_Q(z)\right]^{\ell}$, to deduce :
\begin{align}
\left[f_Q(z)\right]^{\ell}\pi^f(z)&=\alpha_0(z)+\alpha_1(z) 
\end{align}
This is the first equation in (\ref{requu1}) with $N_1 = \ell$.  Likewise, using the second equality in (\ref{pifid}), together with $\mathrm{Tr}_L(z^{Q_L}) = d^\ell \left[f_Q(z)\right]^{\ell}$, we deduce
\begin{align}
\frac{\left[f_Q(z)\right]^{\ell}}{\pi^f(z)}&=\beta_0(z)+\beta_1(z)
\end{align}
This is the second equation in (\ref{requu1}) with $N_2 = \ell$.

To derive the last equation in (\ref{requu1}), we combine the four equations in (\ref{alphadef}) to derive
\begin{align}
d^{2\ell} &[\alpha_0(z) - \alpha_1(z)][[\beta_0(z) - \beta_1(z)]  \nonumber \\
&=\mathrm{Tr}_{L \cup R}\left(z^{Q_L + O_L+Q_R+O_R} P_L Y_L P_R Y_R \right) \nonumber \\
&=  \mathrm{Tr}_{L \cup R}\left(z^{Q_L+Q_R} P_L P_R \right) \nonumber \\
&= d^{2\ell} \left[f_{\chi}(z)\right]^{2\ell}
\end{align}
Here, the second equality follows from (\ref{QLRid}) and (\ref{PLRid}). Cancelling the factors of $d^{2\ell}$, we deduce 
\begin{align}
 \left[f_{\chi}(z)\right]^{2\ell}= [\alpha_0(z)-\alpha_1(z)][\beta_0(z)-\beta_1(z)]
\end{align}
This is the third equation in (\ref{requu1}) with $N_1 = N_2 = \ell$.

We now move on to the case where $\zeta = 1$. The first step is to show that $f_{\chi}(z) = 0$ in this case. We prove this result in Appendix~\ref{sfermionicGNVW}. (The basic idea of the proof is that $\zeta = 1$ implies the existence of a charge neutral Majorana operator, which in turn implies that $f_\chi(z) = 0$).

The next step is to reduce the $\zeta = 1$ case to the $\zeta = 0$ case. We do this by proving the following claim: given any locality preserving unitary $U$ with $\zeta=1$, defined for some choice of $P, Q$, we can construct two other locality preserving unitaries $U_{\pm}$ that are defined for the same $P, Q$ and that have $\zeta = 0$ and satisfy 
\begin{align}
\pi^f_{U_\pm}(z) = (\sqrt{2})^{\pm 1} \pi^f_{U}(z).
\label{UUprime}
\end{align}
Once we prove this claim, we can immediately derive Eqs.~\ref{requu1} since the $\zeta =1$ equations for $\pi^f_U$ follow immediately from the $\zeta = 0$ equations for $\pi^f_{U_\pm}$ together with (\ref{UUprime}). 

How do we construct unitaries $U_\pm$ obeying (\ref{UUprime})? The basic idea is to stack a neutral Majorana translation on top of $U$. In more detail: first, we factor the single site Hilbert space $\mathcal{H}$ into a tensor product of the form $\mathcal{H} = \mathbb{C}^{1|1} \otimes \mathcal{H}_{b}$ where $\mathbb{C}^{1|1}$ denotes a two-dimensional Hilbert space with $P = \mathrm{diag}(1,-1)$ and $Q = \mathrm{diag}(0,0)$, and where $\mathcal{H}_b$ is some other Hilbert space whose structure is not important. (Note that $\mathbb{C}^{1|1}$ is the conventional Hilbert space for a charge neutral fermion). Such a factorization is guaranteed to exist given that $f_\chi(z)=0$. Next, we cluster pairs of neighboring sites into supersites, and we factor the supersite Hilbert space as 
\begin{align}
\mathcal{H}^2 = \mathcal{H} \otimes \mathbb{C}^{1|1} \otimes \mathcal{H}_b
\end{align}
We then define $U_\pm$ to be the unitary operators that act like $U$ on the $\mathcal{H}$ part of the Hilbert space, act like a unit Majorana translation in the positive/negative direction on the $\mathbb{C}^{1|1}$ Hilbert space, and act like the identity on $\mathcal{H}_b$. Using the fact that $\pi^f$ is multiplicative under tensor products, it follows that 
$\pi^f_{U_\pm}(z) = (\sqrt{2})^{\pm 1} \pi^f_{U}(z)$, as required. It is also clear that $U_\pm$ have $\zeta = 0$. This proves the claim and establishes Eq.~(\ref{requu1}) for the case $\zeta = 1$.

\subsubsection{Constructing a unitary that realizes each $\pi^f(z)$}\label{sconstructunitaryfermionic}

In this section, we prove that every $\pi^f(z)$ that satisfies Eq.~\ref{requu1} can be realized by some $U(1)$ symmetric 1D locality preserving unitary. We construct this unitary in the same way as in Sec.~\ref{sconstructunitary}. To begin, we multiply the first two equations in Eq.~\ref{requu1} to obtain
\begin{align}
\left[f_{Q}(z)\right]^{N_1+N_2}= 2^\zeta (\alpha_0(z)+\alpha_1(z))(\beta_0(z)+\beta_1(z))
\label{prodrequ}
\end{align}
First consider the case where $\zeta = 0$. In that case, (\ref{prodrequ}) together with the third equation in Eq.~\ref{requu1} implies that we can factor the Hilbert space $\mathcal{H}^{N_1+N_2}$ for a cluster of $N_1 + N_2$ sites into a tensor product
\begin{equation}\label{zeta0}
\mathcal{H}^{N_1+N_2}=\mathcal{H}_\alpha\otimes\mathcal{H}_{\beta}
\end{equation}
Here $\mathcal{H}_\alpha, \mathcal{H}_\beta$ are Hilbert spaces of dimension $d_\alpha = \alpha_0(1) + \alpha_1(1)$ and $d_\beta = \beta_0(1) + \beta_1(1)$ with charge operators $Q_\alpha, Q_\beta$  and parity operators $P_\alpha, P_\beta$, defined by
\begin{align*}
\mathrm{Tr}(z^{Q_\alpha}) &= \alpha_0(z)+\alpha_1(z), \quad \mathrm{Tr}(z^{Q_\beta}) = \beta_0(z)+\beta_1(z)  \\
\mathrm{Tr}(z^{Q_\alpha} P_\alpha) &= \alpha_0(z)-\alpha_1(z), \quad \mathrm{Tr}(z^{Q_\beta}) = \beta_0(z)-\beta_1(z) 
\end{align*}
With the above factorization in mind, we cluster together $2N_1+N_2$ sites into supersites of dimension $d^{2N_1+N_2}$. We then factor each supersite Hilbert space into a tensor product
\begin{equation}
\mathcal{H}^{2N_1+N_2}=\mathcal{H}^{N_1}\otimes\mathcal{H}_\alpha\otimes\mathcal{H}_{\beta}
\end{equation}
To construct the desired locality preserving unitary, we consider the unitary that performs a unit translation on $\mathcal{H}_{\alpha}$ in the positive direction and a unit translation on $\mathcal{H}^{N_1}$ in the negative direction. This unitary realizes $\pi^f(z)$ since $\pi^f(z) = [\alpha_0(z) + \alpha_1(z)]/[f_Q(z)]^{N_1}$.

Now consider the case where $\zeta = 1$. In this case, the third equation in Eq.~\ref{requu1} implies that  $f_\chi(z) = 0$. This equation together with (\ref{prodrequ}) implies that 
\begin{equation}\label{zeta1}
\mathcal{H}^{N_1+N_2}= \mathbb{C}^{1|1} \otimes \mathcal{H}_\alpha\otimes\mathcal{H}_{\beta}
\end{equation}
where $\mathbb{C}^{1|1}$ is the two dimensional Hilbert space with $Q = \mathrm{diag}(0,0)$ and $P=\mathrm{diag}(1,-1)$. Again, we cluster $2N_1+N_2$ sites into supersites, and we factor each supersite into a tensor product
\begin{equation}
\mathcal{H}^{2N_1+N_2}=\mathcal{H}^{N_1}\otimes\mathbb{C}^{1|1}\otimes\mathcal{H}_\alpha\otimes\mathcal{H}_{\beta}
\end{equation}
To construct the desired locality preserving unitary, we consider the unitary that performs a unit translation on $\mathcal{H}_{\alpha}$ in the positive direction, a unit translation on $\mathcal{H}^{N_1}$ in the negative direction, and unit Majorana translation on $\mathbb{C}^{1|1}$ in the positive direction. This unitary realizes $\pi^f(z)$ since $\pi^f(z) = \sqrt{2} [\alpha_0(z) + \alpha_1(z)]/[f_Q(z)]^{N_1}$. 

So far we have shown that every $\pi_f(z)$ obeying Eq.~\ref{requu1} can be realized by some 1D locality preserving unitary. But we also need to show that every such $\pi_f(z)$ can be realized by a 2D Floquet system. The latter claim follows from the SWAP circuit construction, in the same way as in the bosonic case.

\subsubsection{One-to-one correspondence}\label{sonetoonefermionic}

In this section we prove that $\pi^f_{U'}(z) = \pi^f_{U}(z)$ if and only if $U' U^{-1}$ can be written as a $U(1)$ symmetric FDLU in the presence of ancillas. Just as in the bosonic case, the first step is to note that $\pi^f(z)$ is \emph{multiplicative} under composition of unitaries and therefore it suffices to prove a simpler claim: $\pi^f_{W}(z) = 1$ if and only if $W$ can be written as a $U(1)$ symmetric FDLU in the presence of ancillas.

The proof in the `if' direction is identical to the corresponding proof in the bosonic case (Sec.~\ref{proof11corr}) so we will not repeat it here. As for the `only if' direction, again the proof is essentially the same as the bosonic case -- except in two places. The first place occurs at the very beginning of the argument when we show that $W$ must be an FDLU. In the bosonic case, the argument goes as follows: since $\pi_{W}(z) = 1$, we know that $\mathrm{ind}(W) = 1$; it then follows from Ref.~\cite{GNVW} that $W$ can be written as a FDLU. In the fermionic case, we can follow the same logic, using the results of Ref.~\cite{fermionic}. However, instead of concluding that $W$ is an FDLU, we conclude that $W$ can be written as an FDLU in the presence of \emph{ancillas}, or more specifically, ancillas with the Hilbert space structure $\mathbb{C}^{1|1}$\cite{fermionic}. Note that the $U(1)$ symmetry does not play any role at this stage of the argument, so we are free to choose any charge matrix we like for these ancillas; here, we will choose $Q = \mathrm{diag}(0,0)$, which means that the ancillas can be thought of as neutral fermions.

The second place where the fermionic argument is different from the bosonic argument is when we derive the existence of a single site operator $V_{r}$ obeying Eq.~\ref{Vexist}, i.e.
\begin{align}
V_{r}^\dagger\tilde{Q}_{r} V_{r}=Q_{r}
\label{Vexist2}
\end{align}
In the bosonic case, we proved the existence of $V_{r}$ by establishing two properties of $Q_r, \tilde{Q}_r$: (i) $\tilde{Q}_{r}$ and $Q_{r}$ are both supported on site $r$; and (ii) $\tilde{Q}_{r}$ has the same spectrum as $Q_{r}$. We then argued that these two properties imply the existence of $V_{r}$. 

In the fermionic case, properties (i), (ii) hold just like in the bosonic case but we cannot use these properties to deduce the existence of $V_{r}$. The reason is that we need $V_{r}$ to be fermion parity even, and if we want to deduce the existence of a fermion parity even $V_{r}$ then we need to show that $\tilde{Q}_{r}$ has the same spectrum as $Q_{r}$ within each of the two fermion parity sectors, \emph{separately}. In general, the latter property may not hold even though $\tilde{Q}_{r}$ and $Q_r$ have the same total spectrum.

Conveniently this problem is solved by the neutral fermionic ancillas $\mathbb{C}^{1|1}$ with $Q = \mathrm{diag}(0,0)$, which we introduced earlier. The net effect of including these ancillas is to replace 
\begin{align}
P_{r} &\rightarrow P_{r} \otimes \begin{pmatrix} 1 & 0 \\ 0&  -1 \end{pmatrix} \nonumber \\
Q_{r} &\rightarrow Q_{r} \otimes \mathbbm{1} \nonumber \\
\tilde{Q}_{r} &\rightarrow \tilde{Q}_{r} \otimes \mathbbm{1} \nonumber \\
\end{align}
From these equations we can see that, after adding the ancillas, $Q_{r}$ has the same spectrum in the even parity sector as it does in the odd fermion parity sector. Of course, the same is true for $\tilde{Q}_{r}$. This result implies that $\tilde{Q}_{r}$ has the same spectrum as $\tilde{Q}_{r}$ within each of the two fermion parity sectors separately (since $\tilde{Q}_r, Q_r$ have the same total spectrum). We are then finished, since it follows immediately that there exists a fermion parity even $V_{r}$ obeying (\ref{Vexist2}). 

Before concluding, we should mention that the above argument needs to be modified in the special case where fermion parity is a subgroup of the $U(1)$ symmetry, i.e. $P = (-1)^Q$. The problem is that in this case, it is unphysical to add neutral fermionic ancillas because such ancillas are inconsistent with the underlying symmetry group. Fortunately, it is easy to get around this problem: in this case, we add \emph{charged} fermionic ancillas, i.e. $\mathbb{C}^{1|1}$ ancillas with $P = \mathrm{diag}(1,-1)$ and $Q = \mathrm{diag}(0,1)$. The key point is that $Q_{r}$ and $\tilde{Q}_{r}$ are guaranteed to have the same spectrum within in each of the two fermion parity sectors due to the relation $P = (-1)^Q$ together with the fact that they have the same total spectrum. Hence, in this case we can again deduce the existence of a fermion parity conserving $V_r$.

\section{Relation to cohomology classification}\label{scohomology}

We now discuss the relationship between our (bosonic) results and the group cohomology classification of Floquet symmetry protected topological (SPT) phases\cite{keyserlingk1DI, keyserlingk1DII, alldimensions, cohomology}. According to that work, (e.g. Ref.~\cite{cohomology}) the classification of bosonic $d$-dimensional Floquet SPT phases with an on-site, unitary, abelian symmetry $G$ is given by the cohomology group $H^{d+1}[G\times\mathbb{Z},U(1)]$. Using the Kunneth formula for group cohomology, this group can be split into two factors:
\begin{equation}\label{fspt}
H^{d+1}[G\times\mathbb{Z},U(1)]=H^{d+1}[G,U(1)]\times H^{d}[G,U(1)]
\end{equation}
These factors have a simple physical interpretation. The first factor, $H^{d+1}[G,U(1)]$, describes Floquet phases with $d$-dimensional ``SPT eigenstate order'': every eigenstate of the Floquet unitary $U_F$ looks like a $d$-dimensional (static) SPT ground state. In contrast, the second factor, $H^{d}[G,U(1)]$, describes Floquet phases in which $U_F$ is the \emph{identity} in the $d$-dimensional bulk, but has a nontrivial action on the $d-1$-dimensional edge: $U_F$ pumps a $d-1$ dimensional (static) SPT state to the edge, each period.

Specializing to the case $d=2$ and $G = U(1)$, the first factor evaluates to $\mathbb{Z}$ and the second factor is trivial; therefore, the cohomology classification predicts a $\mathbb{Z}$ classification coming purely from two dimensional SPT eigenstate order. This result is very different from the results described in this paper, so one may ask: what is the origin of this discrepancy?

We believe the discrepancy comes from two differences in how we define Floquet phases. The first difference is that, according to our definition (Definition 1 in Appendix~\ref{sphase}), Floquet phases with eigenstate order are trivial, while according to the cohomology definition (e.g. Definition 3 in Appendix~\ref{sphase}), these phases are non-trivial. This explains why the first factor in the cohomology classification (\ref{fspt}) does not appear in our classification, since the first factor in Eq.~\ref{fspt} corresponds to phases with SPT eigenstate order.\footnote{Actually, even we included eigenstate order, the $H^3[U(1),U(1)]=\mathbb{Z}$ factor would still not appear in our classification: this $\mathbb{Z}$ factor corresponds to the scenario where the eigenstates of $U_F$ are bosonic integer quantum Hall states, and this scenario is not possible if $U_F$ satisfies (\ref{mblcond}). \cite{nogo}}

To understand why the second factor in (\ref{fspt}) does not match our results either, we need to think about the role of \emph{ancillas} in our definitions of Floquet phases. In our definition of bosonic Floquet phases (Definition 1 in Appendix~\ref{sphase}), we did not mention ancillas at all. However, it is easy to show that our classification is stable to adding ancillas with an arbitrary symmetry representation (i.e. arbitrary $Q$ matrix), as long as these ancillas are finite dimensional.\footnote{Adding ancillas corresponds to stacking with a trivial phase with $\pi(z) = 1$. Since $\pi(z)$ is multiplicative under stacking, this operation has no effect on $\pi(z)$, and hence no effect on our classification.} On the other hand, if we were to allow \emph{infinite} dimensional ancillas, in particular ``quantum rotor'' ancillas with the $Q$ matrix $Q = \mathrm{diag}(...,-2,-1,0,1,2,...)$, then our classification would collapse completely: it is possible to show that if we included ancillas of this kind, there would be only one possible Floquet phase with $U(1)$ symmetry. The latter result is exactly what the second factor in (\ref{fspt}) predicts. Therefore, we believe the cohomology classification implicitly assumes that we are allowed to add ancillas with arbitrary symmetry representations -- finite or \emph{infinite} dimensional. This difference in the rules for ancillas explains why our classfiication is so much richer than the cohomology classification. 

\section{Discussion and next steps}\label{sdiscussion}
In this paper, we have derived a complete classification of $U(1)$ symmetric Floquet phases of interacting bosons and fermions in two spatial dimensions. According to our classification, each of these phases is uniquely labeled by a ratio of polynomials, $\pi(z) = a(z)/b(z)$, where $z$ is a formal parameter. In the bosonic case, the invariant $\pi(z)$ can be written as a product $\pi(z)=\frac{p}{q}\tilde{\pi}(z)$ where the GNVW index $\frac{p}{q}$ characterizes the flow of quantum information at the edge and $\tilde{\pi}(z)$ characterizes the flow of $U(1)$ charge at the edge. In the fermionic case, the invariant has a similar structure but with the bosonic GNVW index $\frac{p}{q}$ replaced by its fermionic counterpart, $\sqrt{2}^\zeta \frac{p}{q}$ where $\zeta=0,1$. In addition to our classification results, we have also shown that $\tilde{\pi}(z)$ is \emph{measureable}: it is directly related to the $U(1)$ current that flows at the boundary between two regions held at different chemical potentials.

Our work raises a number of interesting questions that deserve further study. We begin with a purely mathematical question. Recall that Eq.~\ref{requ} describes necessary and sufficient conditions for when a rational function $\pi(z)$ can be realized by a given charge matrix $Q$. In principle these conditions tell us the complete classification of Floquet phases for each $Q$, and we have worked out this classification in a number of examples. The problem is that we do not have \emph{systematic} way to compute this classification: that is, we do not have a general algorithm for finding the complete set of $\pi(z)$ obeying the conditions (\ref{requ}). It would be interesting to find such an algorithm.

Another question involves the recent paper, Ref.~\onlinecite{hierarchy}. In that paper, the authors constructed a set of ``higher order magnetization invariants" for 2D $U(1)$ symmetric Floquet systems that are \emph{partially} many-body localized, i.e. localized up to $n$-body terms. It would be interesting to understand the relationship between these higher order invariants and the invariants described here, in particular $\tilde{\pi}(z)$. 

Several papers have discussed 2D Floquet phases that have a fractional value of the GNVW index\cite{radical, dynamically}. To realize these phases it is necessary to replace the requirement that $U_F$ obeys the MBL condition (\ref{mblcond}) with the weaker requirement that $U_F^N$ obeys (\ref{mblcond}) for some finite integer $N$. It would interesting to investigate $U(1)$ symmetric analogs of these phases.

A natural direction for future work is to generalize our classification to 2D Floquet phases with \emph{discrete} symmetries. More specifically, consider Floquet phases with a discrete unitary on-site symmetry group $G$. By analogy with the $U(1)$ symmetric case discussed here, one might expect that classifying such phases is equivalent to classifying 1D $G$-symmetric locality preserving unitaries modulo 1D $G$-symmetric FDLUs. The latter classification problem was studied in Ref.~\onlinecite{kirby} and Ref.~\onlinecite{mpu}: in both cases the authors found that such locality preserving unitaries are classified by two indices: (i) the $\mathbb{Q}$-valued GNVW index, and (ii) a second index that takes values in the cohomology group $H^2[G,U(1)]$. Based on this result, one might guess that $G$-symmetric Floquet phases are also classified by these two indices. This guess is supported by the cohomology classification which predicts the $H^2[G,U(1)]$ factor (see Sec.~\ref{scohomology}). However, there is an important caveat here. As explained in Ref.~\onlinecite{mpu}, the above classification of 1D LPUs is only correct if we use a particular definition of equivalence in which we are allowed to add ancillas that transform under an arbitrary, finite dimensional representation of $G$. Ref.~\onlinecite{mpu} pointed out that there is another definition of equivalence  -- ``strong equivalence''  -- in which we are only allowed to add ancillas with the \emph{same} symmetry representation as the original sites. Ref.~\onlinecite{mpu} found that strong equivalence leads to a richer classification of locality preserving unitaries but did not work out this classification in generality. It would be interesting to explore the latter classification problem using the methods discussed in this paper.

Another natural direction is to consider higher dimensional systems. Recently there has been significant progress in the classification of LPUs in two and three dimensions in the absence of symmetry. In particular, it has been shown that the classification is trivial in 2D\cite{freedmanclassification,haahclifford} (apart from translations) but nontrivial in 3D\cite{haahnontrivial}. These results suggest that, in the absence of symmetry, there are no nontrivial Floquet phases in 3D, apart from layered phases consisting of stacks of 2D systems\cite{reiss3D}, but such phases do exist in 4D. An interesting question for further work would be to investigate how this classification changes if we introduce $U(1)$ symmetry.

\acknowledgments

We thank Kyle Kawagoe for useful discussions and for helping to prove the claim in Appendix~\ref{sex5app}. C.Z. and M.L. acknowledge the support of the Kadanoff Center for Theoretical Physics at the University of Chicago.
This work was supported by the Simons Collaboration on Ultra-Quantum Matter, which is a grant from the 
Simons Foundation (651440, M.L.), and the National Science Foundation Graduate Research Fellowship under Grant No. 1746045.

\appendix

\section{Relation to other definitions of Floquet phases}\label{sphase}

In this appendix we discuss the relation between our definition of Floquet phases (Sec.~\ref{smbl}) and other definitions of Floquet phases that have been discussed in the literature. Specifically, we compare the following three definitions:

\begin{enumerate}
\item{Two Floquet systems, $H_A(t)$ and $H_B(t)$, belong to the same phase if their boundary can be many-body localized in a symmetry-respecting way.}

\item{Two Floquet systems, $H_A(t)$ and $H_B(t)$, belong to the same phase if there exists a continuous, symmetric interpolation $\{H_s(t),0\leq s\leq 1\}$ with $H_0(t)=H_A(t)$ and $H_1(t)=H_B(t)$ such that the corresponding family of Floquet operators $U_F(s)$ obeys the MBL condition (\ref{mblcond}) for all $s$\cite{roy1D,alldimensions,chiralbosons,fermionic}.}

\item{Two Floquet systems, $H_A(t)$ and $H_B(t)$, belong to the same phase if there exists a continuous, symmetric interpolation for which $U_F(s)$ not only obeys the MBL condition (\ref{mblcond}) but also $U_F(s)$ commutes with a set of local, symmetric, commuting projectors $\{P_r(s)\}$ such that (i) $\{P_r(s)\}$ are \emph{complete} in the sense that their eigenvalues uniquely label every state in the Hilbert space, (ii)$\{P_r(s)\}$ are \emph{unique} in the sense that any two choices $\{P_r(s)\}$, $\{P_r'(s)\}$ commute with one another: $[P_r(s), P_r'(s)] = 0$ \cite{potter1D,keyserlingk1DI,cohomology,dynamically}.}
\end{enumerate}

Note that definition (3) imposes a stricter requirement on the interpolation than definition (2). For example, according to definition (3), the interpolation is not allowed to pass through the point $U_F(s)=\mathbbm{1}$ (since the projectors $P_r(s)$ are not unique in this case) while this is allowed in definition (2). Roughly speaking, the key difference between (2) and (3) is that definition (2) allows \emph{degeneracy} in Floquet spectrum during the interpolation, while definition (3) prohibits it. 

What is the relationship between these definitions? We conjecture that our definition (1) is equivalent to definition (2) -- that is, they give the same classification of Floquet phases\cite{dynamically}. On the other hand, definition (3) is different from the first two definitions and leads to a finer classification of Floquet phases. 

One class of Floquet phases that are trivial under definitions (1) and (2) and non-trivial under (3) are phases with ``eigenstate order'' \cite{huse_localization,mblspt}. For a concrete example, consider a Floquet system whose Hamiltonian $H$ is the \emph{static} toric code Hamiltonian (but with random coefficients). In this system, the Floquet unitary is simply $U_F = e^{-i H T}$, which means that the Floquet eigenstates are toric code eigenstates. In particular, every eigenstate carries toric code topological order. Under definition (3), this Floquet system belongs to a non-trivial phase: there is no way to interpolate to a trivial Floquet system in which the Floquet eigenstates carry trivial topological order, since according to definition (3), any interpolation would generate a corresponding interpolation for individual Floquet eigenstates, which is manifestly impossible. In contrast, under definitions (1) and (2), this Floquet system is trivial: in the case of definition (1), one can easily check that the boundary with the vacuum can be many-body localized, while for definition (2), one can construct an interpolation to the trivial Hamiltonian by simply tuning all the coefficients in the toric code Hamiltonian to zero. Another collection of examples of eigenstate order are Floquet systems in which the Floquet eigenstates carry symmetry protected topological (SPT) order. These Floquet phases are included in the cohomology classification of Floquet SPT phases and are classified by the $H^{d+1}[G,U(1)]$ factor in Eq.~\ref{fspt}. Again these phases are trivial under definitions (1) and (2) but are non-trivial under definition (3). 

All three of the above definitions can be modified by allowing us to add \emph{ancilla} sites to the Floquet system with an arbitrary Hilbert space structure and arbitrary symmetry action. For example, we can incorporate ancillas into definition (1) by allowing the addition of a one dimensional chain of ancillas along the boundary between $H_A$ and $H_B$. A priori, such ancillas could allow certain boundaries to be many-body localized which are otherwise not localizable. Likewise, we can modify definitions (2) and (3) by allowing the addition of a two dimensional lattice of ancilla sites. These ancilla sites are required to have trivial dynamics at the beginning and end of the interpolation, i.e. $s=0$ and $s=1$, but can have nontrivial dynamics during the middle of the interpolation, i.e. $0 < s < 1$. Such ancillas could facilitate interpolations which are otherwise impossible. For more discussion of ancillas, see Sec.~\ref{sfermionicnosymm} and Sec.~\ref{scohomology}.


\section{Proof that $U_{\mathrm{edge}}$ is well-defined}\label{sUr}
In this appendix, we prove that the edge unitary $U_{\mathrm{edge}}$ is \emph{well-defined} in the sense that different choices of the $\{U_r\}$ operators give rise to the same $U_{\mathrm{edge}}$ -- up to composition with a one dimensional FDLU. We first prove this result for systems without any symmetries. We then extend the result to the $U(1)$ symmetric case.

\subsection{No symmetry case}
We wish to show that different choices of the local unitaries $\{U_r\}$ lead to the same $U_{\mathrm{edge}}$ -- up to composition with a 1D FDLU. Our key tool is the following proposition. \\

\textbf{Proposition 1:} Let $\{U_r\}$ and $\{U'_r\}$ be two sets of unitaries that are (i) \emph{mutually commuting} in the sense that $[U_{r_1} , U_{r_2}] = [U'_{r_1}, U'_{r_2}] = 0$, and (ii) \emph{local} in the sense that each $U_r$, $U'_r$ is supported within a finite disk centered at $r$. Suppose also that
\begin{align}
V = \left(\prod_{r\in \mathbf{R}^-}U'_{r}\right)\left(\prod_{r\in \mathbf{R}^-}U_r^\dagger\right)
\end{align}
is supported within a finite distance of the $x$-axis. Then $V$ is a 1D FDLU. \\

Proposition 1 implies the claim because it is clear from the definition of $U_{\mathrm{edge}}$ (\ref{edgeu}) that two different choices of local unitaries, $\{U_r\}$ and $\{U_r'\}$, will give rise to edge unitaries $U_{\mathrm{edge}}$ and $U_{\mathrm{edge}}'$ that differ from one another by precisely the above operator $V$:
\begin{align}
U_{\mathrm{edge}}' = U_{\mathrm{edge}} \cdot V
\end{align}

We now prove Proposition 1. First, we note that we can assume without loss of generality that $\left(\prod_{r \in \mathbf{R}^-} U_r\right)$ commutes with $V$: indeed, we can guarantee this is the case by simply removing all $U_r$ operators from the product $\left(\prod_{r\in \mathbf{R}^-}U_r\right)$ whose region of support overlaps with the region of support of $V$. This operation only changes $V$ by a 1D FDLU so it will not affect our conclusions about $V$.

Next, since $\left(\prod_{r \in \mathbf{R}^-} U_r\right)$ commutes with $V$, it follows that $\left(\prod_{r \in \mathbf{R}^-} U_r\right)$ commutes with $\left(\prod_{r\in \mathbf{R}^-}U'_{r}\right)$. This, in turn, implies that the $n$th power of $V$ can be written in the form
\begin{align}
V^n = \left(\prod_{r\in \mathbf{R}^-}(U'_{r})^n \right)\left(\prod_{r\in \mathbf{R}^-}U_r^{n\dagger}\right)
\label{Vnid}
\end{align}
The above identity (\ref{Vnid}) is important because it implies that $V$ does not generate operator transport in the following sense: for any operator $O_r$ supported on site $r$, the conjugated operator, $V^{n \dagger} O_r V^{n}$, is supported within a finite distance of $r$, for arbitrarily large $n$. This follows from the fact that the $U_r$ operators (and also the $U_r'$ operators) are mutually commuting, local operators.

We now claim that since $V$ does not generate operator transport, it must be a 1D FDLU. We will prove this claim by showing $V$ has a trivial GNVW index:
\begin{align}
\mathrm{ind}(V) = 1
\label{indV}
\end{align}
We derive (\ref{indV}) as follows. First we relate $\mathrm{ind}(V)$ to $\mathrm{ind}(V^n)$ via
\begin{align}
\mathrm{ind}(V) =[\mathrm{ind}(V^n)]^{1/n}
\label{VVnid}
\end{align}
[This follows from the multiplicative property of the index: $\mathrm{ind}(U_1 U_2) = \mathrm{ind}(U_1) \mathrm{ind}(U_2)$].
Next, using the explicit formula for the GNVW index reviewed in Appendix~\ref{GNVWformapp}, together with the fact that $V^n$ only transports operators by a finite distance, one can easily show that 
\begin{align}
\frac{1}{c} \leq |\mathrm{ind}(V^n)|\leq c
\end{align}
for some constant $c > 0$ that does not depend on $n$ and is roughly of order $c \sim d^{(\mathrm{const.}) \xi \Delta y}$ where $\xi$ is the radius of the disks where the $U_r$'s are supported and $\Delta y$ is the width of the strip along the $x$-axis where $V$ is supported. Substituting this inequality into (\ref{VVnid}) and taking the limit $n \rightarrow \infty$, we deduce $\mathrm{ind}(V) = 1$, as we wished to show. 

\subsection{$U(1)$ symmetric case}
The proof in the $U(1)$ symmetric case follows the same logic as above. In particular, the key step is to prove the following analog of Proposition 1. \\

\textbf{Proposition 2:} Let $\{U_r\}$ and $\{U'_r\}$ be two sets of $U(1)$ symmetric unitaries that are mutually commuting and local. Suppose also that
\begin{align}
V = \left(\prod_{r\in \mathbf{R}^-}U'_{r}\right)\left(\prod_{r\in \mathbf{R}^-}U_r^\dagger\right)
\end{align}
is supported within a finite distance of the $x$-axis. Then $V$ is a 1D $U(1)$ symmetric FDLU. \\

The proof of Proposition 2 is identical to that of Proposition 1, except for the last step. In that step we need to prove \emph{two} identities  in order to conclude that $V$ is a $U(1)$ symmetric FDLU: $\mathrm{ind}(V) = 1$ and $\tilde{\pi}_V(z) = 1$. To show $\mathrm{ind}(V) = 1$, we can use the same argument as above, but an additional argument is needed to show $\tilde{\pi}_V(z) = 1$. We can establish this using a similar approach to above. First we relate $\tilde{\pi}_V(z)$ to $\tilde{\pi}_{V^n}(z)$ via
\begin{align}
\tilde{\pi}_V(z) =[\tilde{\pi}_{V^n}(z)]^{1/n}
\label{pipinid}
\end{align}
Next, using the definition of $\tilde{\pi}$ (Sec.~\ref{sdefs}) together with the fact that $V^n$ only transports operators by a finite distance, one can easily show that for any real $z \geq 0$, 
\begin{align}
\frac{1}{c(z)} \leq |[\tilde{\pi}_{V^n}(z)|\leq c(z)
\end{align}
for some constant $c(z)> 0$ that does not depend on $n$ and is roughly of order $c(z) \sim \left(f_Q(z)\right)^{(\mathrm{const.}) \xi \Delta y}$ where $\xi$ and $\Delta y$ are defined as above. Substituting this inequality into (\ref{pipinid}) and taking the limit $n \rightarrow \infty$, we deduce $\tilde{\pi}_V(z) = 1$ for all $z \geq 1$. Then, since $\tilde{\pi}_V(z)$ is a rational function, it follows that $\tilde{\pi}_V(z) \equiv 1$ for all $z$, as we wished to show.

\section{Equivalence between classification of 2D Floquet phases and 1D locality preserving unitaries}\label{sfdlu}

In this appendix, we prove that two Floquet systems $H_A(t)$ and $H_B(t)$ (without any symmetry) belong to the same phase if and only if their corresponding edge unitaries $U_{A,\mathrm{edge}}$ and $U_{B,\mathrm{edge}}$ differ by a 1D FDLU. (We do not include a separate discussion of the $U(1)$ symmetric case because the proof is identical except with Proposition 1 replaced by Proposition 2).

\subsection{Special case}
We start by proving a special case: we show that $H(t)$ belongs to the \emph{trivial} phase if and only if the edge unitary $U_{\mathrm{edge}}$ is a 1D FDLU. To make this statement precise, we need to explain what we mean by a ``trivial phase.'' Let $H_-(t)$ be the restriction of $H(t)$ to the lower half plane $\mathbf{R}^-= \{(x,y): y \leq 0\}$. We say that $H(t)$ belongs to the trivial phase if there exists a boundary Hamiltonian $H_{\mathrm{bd}}$ such that the Hamiltonian 
\begin{align}
H_{\mathrm{tot}}(t) &=H^-(t)+H_{\mathrm{bd}}(t)
\end{align}
is MBL -- that is, the corresponding Floquet unitary
\begin{align}
U_F^{\mathrm{tot}} &= \mathcal{T}e^{-i\int_0^T H_{\mathrm{tot}}(t) dt}
\label{uftot}
\end{align}
obeys the MBL condition (\ref{mblcond}).

We start by proving the ``if'' direction: we show that if $U_{\mathrm{edge}}$ is a 1D FDLU then $U_F^{\mathrm{tot}}$ obeys (\ref{mblcond}) for some $H_{\mathrm{bd}}$. The first step is to write $U_{\mathrm{edge}}$ as the time evolution of a 1D Hamiltonian acting near the $x$-axis:
\begin{equation}\label{edgeH}
U_{\mathrm{edge}}=\mathcal{T}e^{-i\int_0^T H_{\mathrm{edge}}dt}
\end{equation}
(This is possible since $U_{\mathrm{edge}}$ is a 1D FDLU). Then, using the definition of $U_{\mathrm{edge}}$ (\ref{edgeu}), we deduce that
\begin{equation}
\mathcal{T}e^{-i\int_0^T H^-(t)dt} = 
\left( \mathcal{T}e^{-i\int_0^TH_{\mathrm{edge}}dt} \right) \left( \prod_{r\in \mathbf{R}^-}U_r \right)
\end{equation}
where the $U_r$ are mutually commuting, local unitaries. 

Next, we rewrite the Floquet unitary (\ref{uftot}) as a product,
\begin{equation}\label{fullU}
U_F^{\mathrm{tot}} = U_{\mathrm{bd}} U^- 
\end{equation}
where
\begin{align}
U_{\mathrm{bd}} =  \mathcal{T}e^{-i\int_0^T \tilde{H}_{\mathrm{bd}}(t) dt}, \quad U^- =  \mathcal{T}e^{-i\int_0^T H^-(t) dt}
\end{align}
and
\begin{equation}
\tilde{H}_{\mathrm{bd}}(t)=U^-(T,t)H_{\mathrm{bd}}(t)U^{-,\dagger}(T,t)
\end{equation}
where $U^-(T,t)=\mathcal{T}e^{-i\int_t^T H^-(t)dt}$. Here Eq. (\ref{fullU}) is obtained by implementing the standard Dyson series method in the interaction picture, and accounts for commuting the terms in $H^-(t)$ and $H_{\mathrm{bd}}(t)$ through each other. 

Putting Eq. \ref{edgeH} and Eq. \ref{fullU} together, we derive
\begin{equation}
U_F^{\mathrm{tot}}= U_{\mathrm{bd}} \cdot \left(\mathcal{T} e^{-i\int_0^TH_{\mathrm{edge}}(t)dt}\right) \left(\prod_{r\in \mathbf{R}^-}U_r\right)
\label{uftot2}
\end{equation}
Now suppose we choose $H_{\mathrm{bd}}$ so that 
\begin{align}
\tilde{H}_{\mathrm{bd}}(t)=-H_{\mathrm{edge}}(T-t). 
\end{align}
(Note we can arrange this by choosing 
\begin{align}\label{Hbd}
H_{\mathrm{bd}}(t) = - U^{-.\dagger}(T,t) H_{\mathrm{edge}}(T-t) U^-(T,t)
\end{align}
This choice is allowed since $H_{\mathrm{bd}}(t)$ is manifestly local, periodic in time, and supported near the $x$-axis). Then (\ref{uftot2}) simplifies to
\begin{align*}
U_F^{\mathrm{tot}}=\prod_{r\in \mathbf{R}^-}U_{r}
\end{align*}
This proves $U_F^{\mathrm{tot}}$ obeys the MBL condition (\ref{mblcond}), as we wished to show.

Next we prove the "only if" direction: we assume that $U_F^{\mathrm{tot}}$ obeys the MBL condition (\ref{mblcond}), and then show that $U_{\mathrm{edge}}$ is a 1D FDLU. The first step is note that, by the definition of  $U_{\mathrm{edge}}$ (\ref{edgeu}), we have
\begin{align}
U_{\mathrm{edge}} = U^- \cdot \prod_{r\in \mathbf{R}^-}U_r^\dagger
\end{align}
where $U_r$ are mutually commuting, local unitaries.

Combining this equation with (\ref{fullU}), we deduce that
\begin{align}
U_{\mathrm{edge}} =U_{\mathrm{bd}}^\dagger \cdot U_F^{\mathrm{tot}} \cdot \prod_{r\in \mathbf{R}^-}U_r^\dagger
\end{align}
Next, using the fact that $U_F^{\mathrm{tot}}$ obeys the MBL condition (\ref{mblcond}), we know that
\begin{align}
U_F^{\mathrm{tot}} = \prod_{r\in \mathbf{R}^-}U'_{r}
\end{align}
for some mutually commuting, local unitaries $U'_{r}$.\footnote{Here, we do not assume any relation between the $U'_{r}$ unitaries and the $U_r$ unitaries.} Hence,
\begin{align}
U_{\mathrm{edge}}&=U_{\mathrm{bd}}^\dagger \cdot \left(\prod_{r\in \mathbf{R}^-}U'_{r}\right)\left(\prod_{r\in \mathbf{R}^-}U_r^\dagger\right)
\end{align}
To complete the proof we invoke Proposition 1 from Appendix~\ref{sUr}, which implies that
$\left(\prod_{r\in \mathbf{R}^-}U'_{r}\right)\left(\prod_{r\in \mathbf{R}^-}U_{r}^\dagger\right)$ is a 1D FDLU. Given that $U_{\mathrm{bd}}^\dagger(T)$ is manifestly a 1D FDLU, it follows immediately that $U_{\mathrm{edge}}$ is a FDLU, as we wished to show.

\subsection{General case}

We now move on to the general case: we show that the boundary between two Floquet systems $H_A(t)$ and $H_B(t)$ can be many-body localized if and only if
\begin{align}
U_{A,\mathrm{edge}} = U_{B,\mathrm{edge}} \cdot (\text{1D FDLU})
\label{uaubequiv}
\end{align}
We begin with the ``only if'' direction: we assume that the boundary between $H_A(t)$ and $H_B(t)$ can be many-body localized and we show that Eq. (\ref{uaubequiv}) holds. The first step is to use the``folding trick'' to map the boundary between $H_A^-, H_B^+$ onto the boundary between the tensor product of $H_A^-$ and $H_B^{\mathrm{op},-}$ and the vacuum. Here, ``$H_B^{\mathrm{op}}$'' denotes the Hamiltonian obtained from $H_B$ by reflecting about the $x$-axis. It is easy to see that the edge unitary for this tensor product Floquet system is $U_{A,\mathrm{edge}} \otimes U^B_{\mathrm{op}, \mathrm{edge}}$, so by the special case proved above, we know that
\begin{align}
U_{A,\mathrm{edge}} \otimes U^B_{\mathrm{op}, \mathrm{edge}}  = (\text{1D FDLU})
\label{eqab}
\end{align}
By the same reasoning, we know that
\begin{align}
U_{B,\mathrm{edge}} \otimes U^B_{\mathrm{op}, \mathrm{edge}} = (\text{1D FDLU})
\label{eqbb}
\end{align}
since the boundary between $H_B^-$ and $H_B^+$ can be many-body localized by assumption. If we now take the tensor product of both sides of Eq. (\ref{eqab}) with $U_{B,\mathrm{edge}}$ and use (\ref{eqbb}), it is not hard to show that
\begin{align}
(U_{A,\mathrm{edge}} \otimes \mathbbm{1} \otimes \mathbbm{1}) =   (U_{B,\mathrm{edge}} \otimes  \mathbbm{1} \otimes \mathbbm{1}) \cdot (\text{1D FDLU}) 
\end{align}
This implies the desired relation, Eq. (\ref{uaubequiv}). 

The proof of the ``if'' direction follows similar reasoning but in the reverse direction: if (\ref{uaubequiv}) holds, then using Eq. (\ref{eqbb}), we can deduce that Eq. (\ref{eqab}) holds. Then invoking the special case proved above, it follows that the boundary between the vacuum and the tensor product of  $H_A^-$ and $H_B^{\mathrm{op},-}$ can be many-body localized. Next, using the folding trick, we deduce that the boundary between $H_A^-, H_B^+$ can be many-body localized. This is what we wanted to show.

\section{Explicit formula for GNVW index}\label{GNVWformapp}

In this appendix, we review an explicit formula for the GNVW index, $\mathrm{ind}(U)$. This formula can also be viewed as a \emph{definition} of the GNVW index. Before presenting the formula, we first define a related quantity, $\eta(\mathcal{A}, \mathcal{B})$, which can be interpreted as an ``overlap'' between operator algebras  $\mathcal{A}, \mathcal{B}$. 

Let $\mathcal{A}, \mathcal{B}$ be two operator algebras consisting of operators acting on some finite dimensional Hilbert space. Let $\{O_a\}$ be a complete orthonormal basis of operators in $\mathcal{A}$ -- that is, a collection of operators such that  (i) $\{O_a\}$ is a complete basis for $\mathcal{A}$ and (ii) $\{O_a\}$ satisfies $\overline{\mathrm{Tr}}(O_a^\dagger O_{a'}) = \delta_{aa'}$ where $\overline{\mathrm{Tr}}$ is a normalized trace defined by $\overline{\mathrm{Tr}}(\mathbbm{1}) = 1$. Similarly, let $\{O_b\}$ be a complete orthonormal basis for $\mathcal{B}$. We then define the ``overlap'' between $\mathcal{A}, \mathcal{B}$, which is denoted by $\eta(\mathcal{A}, \mathcal{B})$, by
\begin{align}
\eta(\mathcal{A}, \mathcal{B}) = 
\sqrt{\sum_{O_a\in\mathcal{A},O_b\in\mathcal{B}}| \overline{\mathrm{Tr}}(O_{a}^\dagger O_{b})|^2}
\end{align}
One can check that $\eta(\mathcal{A}, \mathcal{B})$ only depends on the algebras $\mathcal{A}, \mathcal{B}$ and not on the choice of orthonormal bases $\{O_a\}, \{O_b\}$. Also, it is not hard to show that $\eta(\mathcal{A}, \mathcal{B}) \geq 1$ since the two algebras $\mathcal{A}, \mathcal{B}$ both contain the identity operator $\mathbbm{1}$. 

We are now ready to explain the formula for $\mathrm{ind}(U)$. Let $U$ be a locality preserving unitary, defined on a 1D spin chain, with a operating spreading length $\ell$. Choose any two adjacent intervals $A$, $B$ within the spin chain, with $A$ to the left of $B$ and such that $A$ and $B$ each have length larger than $\ell$. Let $\mathcal{A}, \mathcal{B}$ be the algebra of operators supported on $A, B$, respectively. Then,
$\mathrm{ind}(U)$ is given by\cite{GNVW,chiralbosons}:
\begin{align}
\mathrm{ind}(U)=\frac{\eta\left(U^\dagger\mathcal{A}U,\mathcal{B}\right)}{\eta\left(\mathcal{A},U^\dagger\mathcal{B}U\right)}
\label{GNVWform1}
\end{align}
One can check that $\mathrm{ind}(U)$ does not depend on the choice of $A, B$ and is therefore a well-defined function of the locality preserving unitary $U$. 

We note that Eq.~\ref{GNVWform1} differs from the formula for the index in Ref.~\onlinecite{chiralbosons} (Eq.~20) in that $U$ and $U^\dagger$ are switched.   As a result, our definition of $\mathrm{ind}(U)$ is the \emph{inverse} of the index defined in Ref.~\onlinecite{chiralbosons}. This discrepancy can be thought of as a difference in orientation conventions: in our convention, $\mathrm{ind}(U) = d$ for a unit translation $U$ that acts on single site operators as $U^\dagger O_r U = O_{r+1}$, while in Ref.~\onlinecite{chiralbosons}, $\mathrm{ind}(U) = d$ for a unit translation $U$ of the form $U^\dagger O_r U = O_{r-1}$.

\section{Proof that all $\pi(z)$'s in  Eq.~\ref{picycl} satisfy Eq.~\ref{requ}}\label{sex5app}

In this appendix, we derive a mathematical result that is useful for analyzing the example in Sec.~\ref{sex5}. To state this result, let $f_Q(z)=1+z+\cdots+z^{d-1}$, and let $\phi_k$ denote the $k$th cyclotomic polynomial. Also, let $\pi(z)$ be a rational function of the form given in Eq.~\ref{picycl}:
\begin{align}
\pi(z) = \prod_{k|d,k\neq 1}\phi_k(z)^{n_k}
\label{picyclapp}
\end{align}
We will show that every $\pi(z)$ of this form satisfies the two conditions in Eq. \ref{requ}.

To begin, we note that it is enough to prove the claim in the case where $\pi(z) = \phi_k(z)$ where $k \neq 1$ is a divisor of $d$, since the set of $\pi(z)$'s obeying condition (\ref{requ}) is closed under multiplication and inverses. To prove the claim in this special case, we use the following lemma: \\

\textbf{Lemma 1:} Let $\psi(z)$ be a polynomial with real coefficents with two properties: (i) $\psi(z) > 0$ for all $z \geq 0$ and (ii) $\psi(z)$ is a \emph{palindrome} in the sense that its coefficients obey $a_i=a_{m-i}$ where $\psi(z)=\sum_{i=0}^m a_i z^i$. Then, for sufficiently large integers $N$, the product $f_Q(z)^N \psi(z)$ is a polynomial with non-negative coefficients. \\

With the help of Lemma 1, the claim follows easily. Indeed, to see that $\phi_k(z)$ obeys the first condition in Eq. \ref{requ}, note that $\phi_k(z)$ satisfies properties (i) and (ii) in Lemma 1: that is, $\phi_k(z) > 0$ for $z \geq 0$ and also $\phi_k(z)$ is a palindrome. Applying Lemma 1 with $\psi(z) = \phi_k(z)$, we conclude that the product $f_Q(z)^N \phi_k(z)$ is a polynomial with non-negative integer coefficients for sufficiently large $N$. This shows that $\phi_k(z)$ obeys the first condition in Eq. \ref{requ}. Likewise, to see that $\phi_k(z)$ obeys the second condition in Eq. \ref{requ}, note that $f_Q(z)/\phi_k(z)$ also obeys properties (i)-(ii) in Lemma 1. Applying Lemma 1 with $\psi(z) =  f_Q(z)/\phi_k(z)$, we conclude that $f_Q(z)^{N+1}/ \phi_k(z)$ is a polynomial with non-negative integer coefficients for sufficiently large $N$. This proves the second condition in Eq. \ref{requ}.

All that remains is to prove Lemma 1. We do this using Polya's theorem about positive polynomials\cite{polya}\footnote{Polya's theorem is usually stated in terms of homogeneous polynomials in multiple variables. The theorem that we quote here is equivalent to Polya's theorem for the special case of homogeneous polynomials in two  variables.}: \\

\textbf{Polya's theorem:} Suppose $p(z)$ is a polynomial with real coefficients with the property that $p(z) > 0$ for all $z \geq 0$. Then, for all sufficiently large integers $N$, the product $(1+z)^N p(z) $ is a polynomial with positive coefficients. \\

To apply Polya's theorem, it is useful to separately consider the two cases where $d$ is even and $d$ is odd. First, suppose $d$ is even. In that case, $(1+z)=\phi_2(z)$ is a factor of $f_Q(z)$. In fact, $f_Q(z) = (1+z) \phi_{d,2}(z)$ where
\begin{equation}\label{secondnonneg}
\phi_{d,2} = 1+z^2+\cdots+z^{d-2}
\end{equation}
Lemma 1 now follows by writing 
\begin{align}
f_Q(z)^N \psi(z) = \phi_{d,2}(z)^N (1+z)^N \psi(z)
\end{align}
By Polya's theorem, the product $(1+z)^N \psi(z)$ has non-negative coeffiicents for sufficently large $N$, and therefore the whole expression $f_Q(z)^N \psi(z)$ must also have non-negative coefficients since $\phi_{d,2}(z)$ has non-negative coefficients. This proves Lemma 1 in the case where $d$ is even.

Now suppose $d$ is odd. In that case, we write
\begin{align}
f_Q(z)^{N} \psi(z) &=[1+z+\cdots z^{d-1}]^{N} \psi(z) \nonumber \\
&=[1+z(1+z+\cdots +z^{d-2})]^{N}\ \psi(z) \nonumber \\
&=[1+z(1+z)\phi_{d-1,2}(z)]^{N} \psi(z) \nonumber \\
&=\sum_{M=0}^{N}\begin{pmatrix}N\\M\end{pmatrix}[z(1+z)\phi_{d-1,2}(z)]^{M} \psi(z)
\end{align}
By Polya's theorem, there exists some $M_0$ such that the product $[z(1+z)\phi_{d-1,2}(z)]^{M} \psi(z)$ has non-negative coefficients for all $M \geq M_0$. Using this result, we will now show that if $N \geq 2[(d-1)M_0+m]$ then $f_Q(z)^{N} \psi(z)$ has non-negative coefficients. (Here $m$ is the degree of $\psi(z)$). To see this, first consider the coefficients of $z^i$ with $i \geq N/2$. It is easy to see that these coefficients come from terms of the form $[z(1+z)\phi_{d-1,2}(z)]^{M} \psi(z)$ where $M \geq M_0$, and are therefore guaranteed to be non-negative. Likewise, the coefficients of $z^i$ with $i < N/2$ are also guaranteed to be non-negative because $f_Q(z)^{N} \psi(z)$ is a palindrome with degree $\geq N$ (this follows from the fact that $\psi(z)$ and $f_Q(z)$ are both palindromes). Hence all coefficients are non-negative. This proves Lemma 1 in the case where $d$ is odd.

\section{Proof of Eq. \ref{UQU}}\label{sUQU}
In this appendix, we derive Eq.~\ref{UQU}: we show that for any $U(1)$ symmetric locality preserving unitary $U$ with an operating spreading length $\ell$, and any interval $A = [r_L, r_R]$  with length $r_R - r_L \geq 2 \ell - 1$, we can write $U^\dagger  Q_A U$ as a sum
\begin{align}
U^\dagger Q_A U = Q_A + O_L + O_R
\end{align}
where $O_L$ and $O_R$ are operators that are supported within the intervals $[r_L - \ell, r_L + \ell-1]$ and $[r_R - \ell+1, r_R + \ell]$.

To begin, we note that since $U$ is $U(1)$ symmetric, it commutes with the \emph{total} charge: that is,
\begin{align}
U^\dagger (Q_A + Q_{A^c})U = Q_A + Q_{A^c}
\end{align}
Rearranging terms we derive the identity
\begin{align}
U^\dagger Q_A U -Q_A= Q_{A^c} - U^\dagger Q_{A^c} U
\label{aacid}
\end{align}
Next, we observe that since $U$ is locality preserving with operating spreading length $\ell$, the operator on the left hand side is supported within the interval $[r_L - \ell, r_R + \ell]$, while the operator on the right hand side is supported within $[r_L+\ell,r_R -\ell]^c$. It follows that
both operators must be supported within the intersection of these two regions, namely 
\begin{align}
[r_L - \ell, r_L + \ell-1] \cup [r_R - \ell+1, r_R + \ell]
\label{intersectregion}
\end{align}
In particular, this means that 
\begin{align}
U^\dagger Q_A U = Q_A + O
\end{align}
where $O$ is supported in the region (\ref{intersectregion}). All that remains is to show that $O$ can be written as a sum $O = O_L + O_R$ where $O_L$ is supported in $[r_L - \ell, r_L + \ell-1]$ and $O_R$ is supported in $[r_R - \ell+1, r_R + \ell]$. To prove this, it suffices to show that 
\begin{align}
[[O, O_1], O_2] = 0
\label{oo1o2}
\end{align}
for any single site operator $O_1$ supported in $[r_L - \ell, r_L + \ell-1]$ and single site operator $O_2$ supported in $[r_R - \ell+1, r_R + \ell]$. The latter result (\ref{oo1o2}) follows from the fact that $O = Q_{A^c} - U^\dagger Q_{A^c} U$ can be written as a sum of operators, each supported on an interval that intersects at most one of the two intervals $[r_L - \ell, r_L + \ell-1]$, $[r_R - \ell+1, r_R + \ell]$  (which in turn follows from the fact that $U$ is locality preserving with operating spreading length $\ell$). 

\section{Stacking and composition}\label{scomp}
In this appendix, we prove that $\pi(z)$ is multiplicative under ``stacking'' (tensoring) and composition of 1D edge unitaries (or equivalently 2D MBL Floquet circuits). That is, we show that $\pi(z)$ obeys the following two identities:
\begin{enumerate}
\item{\textbf {Stacking}: $\pi_{U_1 \otimes U_2}(z) = \pi_{U_1}(z) \pi_{U_2}(z)$
}
\item{\textbf {Composition}: $\pi_{U_1 \cdot U_2}(z) = \pi_{U_1}(z) \pi_{U_2}(z)$ 
}
\end{enumerate}

We begin by showing that $\pi(z)$ is multiplicative under stacking. Consider two $U(1)$ symmetric locality preserving unitaries $U_1, U_2$ acting on two different 1D systems. Let $U_1 \otimes U_2$ denote the tensor product of these unitaries. It is easy to see that $\tilde{\pi}_{U_1 \otimes U_2}(z) = \tilde{\pi}_{U_1}(z) \tilde{\pi}_{U_2}(z)$ due to the trace being multiplicative under tensor product. Also, $\text{ind}(U_1 \otimes U_2) = \text{ind}(U_1) \text{ind}(U_2)$ as shown in Ref.~\onlinecite{GNVW}. Putting these two facts together, and using the definition of $\pi(z)$ (\ref{pidef}), it immediately follows that $\pi(z)$ is multiplicative under stacking.

We now show that $\pi(z)$ is multiplicative under composition. Let $U_1, U_2$ be two $U(1)$ symmetric LPUs acting on the same 1D system. To prove the composition property for $U_1$ and $U_2$, it is helpful to consider an enlarged Hilbert space which is a tensor product of two copies of this 1D system. Our strategy for proving the composition property is to first prove the identity
\begin{align}
\pi_{(U_1 U_2) \otimes\mathbbm{1}}(z) = \pi_{U_1 \otimes U_2}(z)
\label{compid}
\end{align}
where ``$\otimes$'' denotes the tensor product associated with the two copies, and where $\mathbbm{1}$ denotes the identity operator acting on the second copy. Once we prove this identity, the composition property will then follow immediately, using the fact that $\pi_{(U_1 U_2) \otimes \mathbbm{1}}(z) = \pi_{U_1 U_2}(z)$ together with the stacking property, $\pi_{U_1 \otimes U_2}(z) = \pi_{U_1}(z) \pi_{U_2}(z)$.

Before proving (\ref{compid}), we need to introduce two pieces of notation. First, we denote the unitary operator that exchanges the two copies of the 1D system by ``$ \mathrm{SWAP}$." The second piece of notation is that, in the equations that follow, we will denote $\pi_U(z)$ by $\pi(U)$, for brevity. 

First we note that
\begin{align}
\pi((U_1 U_2) \otimes\mathbbm{1}) &=\pi([U_1\otimes\mathbbm{1}] [U_2\otimes\mathbbm{1}]) \nonumber \\
&= \pi([U_1\otimes\mathbbm{1}] [\mathrm{SWAP}][\mathbbm{1} \otimes U_2][\mathrm{SWAP}]) \nonumber \\
&=\pi([U_1\otimes\mathbbm{1}][\mathbbm{1} \otimes U_2] [\overline{\mathrm{SWAP}}][\mathrm{SWAP}])
\label{compderiv0}
\end{align}
where 
\begin{align}
\overline{\mathrm{SWAP}}   \equiv  (\mathbbm{1} \otimes U_2^{-1} ) \mathrm{SWAP} (\mathbbm{1} \otimes U_2 )
\end{align}
Next we recall that $\pi(U) = \pi(U')$ if $U' U^{-1}$ is a $U(1)$ symmetric FDLU. This means we can drop
the factors of $\mathrm{SWAP}$ and $\overline{\mathrm{SWAP}}$ on the right hand side of (\ref{compderiv0}) since they are both $U(1)$ symmetric FDLUs. Hence
\begin{align}
\pi(U_1 U_2 \otimes\mathbbm{1}) =\pi([U_1\otimes\mathbbm{1}][\mathbbm{1} \otimes U_2])
\end{align}
implying the desired identity (\ref{compid}).

\section{Proof of Eq. \ref{ilrid}}\label{ilridproof}
In this appendix we prove Eq. \ref{ilrid}, by showing that
\begin{align}
[U(T)]^{n\dagger} Q^+ [U(T)]^{n} =  U_{\mathrm{edge}}^{n\dagger} Q^+ U_{\mathrm{edge}}^{n} + \mathcal{O}(1)
\label{ilridapp}
\end{align}
where $\mathcal{O}(1)$ denotes an \emph{operator} whose norm is bounded by a constant, independent of $n$.

The first step is to write $U(T)$ in terms of the edge and bulk unitaries:
\begin{equation}
U(T)=U_{\mathrm{edge}}\left(\prod_rU_r\right)
\end{equation}
where the $U_r$ operators are mutually commuting. In general the $U_r$ operators can be quasi-local (i.e. they can have exponential tails) but to simplify the proof we will assume that the $U_r$ operators are \emph{strictly} local. More specifically, we will assume that each $U_r$ operator is supported within a disk of radius at most $\xi$.

To proceed further, note that we can assume without loss of generality that $[U_{\mathrm{edge}},\prod_rU_r]=0$ since we can always remove all the $U_r$ operators  from the product $\prod_r U_r$ whose region of support overlaps with $U_{\mathrm{edge}}$, and then multiply $U_{\mathrm{edge}}$ by these $U_r$ operators to compensate.

Next, we claim that 
\begin{align}
[U(T)]^{n\dagger} Q^+ [U(T)]^{n} = V^{n \dagger} Q^+ V^{n}
\label{ulridapp2}
\end{align}
where 
\begin{align}
V=U_{\mathrm{edge}}\prod_{r_y = 0}U_{r},
\end{align}
and where the product over $U_r$ runs over $U_r$ operators whose region of support lies on both sides of the line $y =0$. To see this, consider a unitary $U_r$ whose region of support lies entirely on one side of $y=0$. Such an operator $U_r$ commutes with $Q^+$ so we can remove the operator $U_r$ from $U(T)$ without affecting $[U(T)]^{n\dagger} Q^+ [U(T)]^{n}$.  After removing all these $U_r$ operators, the result is that $U(T) \rightarrow V$. This justifies Eq.~ \ref{ulridapp2} above.

To proceed further, we claim that the following inequality holds:
\begin{align}
\bigg\| \left(\prod_{r_y=0}U_{r}^{n\dagger}\right)Q^+\left(\prod_{r_y=0}U_{r}^n\right)-Q^+ \bigg\| \leq 4 qw\xi
\label{urplusineq}
\end{align}
where $\|O \|$ denotes the operator norm, $w$ is the width of the annulus, $\xi$ is the radius of the region of the support of the unitaries $U_r$, and $q$ is the maximum eigenvalue of the single-site charge operator $Q_r$. To derive this inequality, let $Q_0^+$ denote the total charge contained in the region $A^+ \cap C_0$ where $C_0$ is region of support of $\prod_{r_y=0}U_{r}$. Observe that
\begin{align*}
\bigg\| \left(\prod_{r_y=0}U_{r}^{n\dagger}\right)Q^+\left(\prod_{r_y=0}U_{r}^n\right)-Q^+ \bigg\| &= \nonumber \\
\bigg\| \left(\prod_{r_y=0}U_{r}^{n\dagger}\right)Q_0^+\left(\prod_{r_y=0}U_{r}^n\right)-Q_0^+ \bigg\| &\leq
2 \|Q_0^+\| \nonumber \\
&\leq 4qw\xi
\end{align*}
where the first line follows from the fact that $Q^+ - Q_0^+$ commutes with $\prod_{r_y=0}U_{r}$, the second line follows from the triangle inequality, and the last line follows from the fact that $A^+ \cap C_0$ contains at most $2 w \xi$ sites, so $\|Q_0^+\| \leq 2qw\xi$.

With the inequality (\ref{urplusineq}), we now have everything we need to complete the proof. First, we note that the inequality (\ref{urplusineq}) implies that
\begin{align}
\| V^{n \dagger} Q^+ V^n- U_{\mathrm{edge}}^{n\dagger} Q^+ U_{\mathrm{edge}}^{n} \| \leq 4 qw\xi
\label{ulredgeineq}
\end{align}
Then, combining (\ref{ulridapp2}) and (\ref{ulredgeineq}), we deduce that
\begin{align}
\| U(T)^{n \dagger} Q^+ U(T)^n- U_{\mathrm{edge}}^{n\dagger} Q^+ U_{\mathrm{edge}}^{n} \| \leq 4 qw\xi
\end{align}
This implies Eq.~(\ref{ilridapp}).

\section{Three proofs that use results from Ref.~\onlinecite{GNVW}}\label{sgnvw}
In this appendix we prove three claims made in the main text -- from Sec.~\ref{ssatisfiesrequ}, Sec.~\ref{sedge}, and Sec.~\ref{ssatisfiesfermionic}. To prove these claims, we will need to use some results from Ref.~\onlinecite{GNVW}, which we will refer to as GNVW in the rest of this section. We begin by reviewing their setup and results.

\subsection{Review of some key results in GNVW}
Consider a one dimensional spin chain in which each spin can be in $D$ states\footnote{Here, we use the letter $D$ instead of $d$ because when we apply these results, $D$ will be of the form $D=d^n$: the sites on the spin chain will be superspins obtained by clustering together $n$ neighboring spins.}. Denote the sites where the spins are located by $r \in \mathbb{Z}$ and denote the algebra of operators supported on site $r$ by $\mathcal{A}_r$. Let $U$ be a locality preserving unitary that has an operator spreading length of at most $1$: that is, for any operator $O_r \in \mathcal{A}_r$, the conjugated operator $U^\dagger O_rU \in \mathcal{A}_{r-1} \otimes A_r\otimes A_{r+1}$.

GNVW derived several important constraints on how $U$ acts on the local algebras $\mathcal{A}_r$. To explain these constraints, it is useful to consider the action of $U$ on \emph{pairs} of neighboring sites $\{2r, 2r+1\}$. Observe that
\begin{equation*}
U^\dagger(\mathcal{A}_{2r}\otimes\mathcal{A}_{2r+1}) U\subset(\mathcal{A}_{2r-1}\otimes\mathcal{A}_{2r})\otimes(\mathcal{A}_{2r+1}\otimes\mathcal{A}_{2r+2})
\end{equation*}
This inclusion can be refined using the concept of \emph{support algebras}. The support algebra $\mathcal{R}=\mathbf{S}(\mathcal{A},\mathcal{B}_1)$ for an algebra $\mathcal{A}\in\mathcal{B}_1\otimes\mathcal{B}_2$ is the algebra of smallest dimension satisfying $\mathcal{A}\subset\mathcal{R}\otimes\mathcal{B}_2$. More concretely, if we choose a basis $\{e_\mu\}$ for $\mathcal{B}_2$ so that every operator $a\in\mathcal{A}$ has a unique expansion $a=\sum_{\mu}a_{\mu}\otimes e_\mu$ with $a_\mu\in\mathcal{B}_1$, then $\mathbf{S}(\mathcal{A},\mathcal{B}_1)$ is the algebra generated by all the elements $a_\mu$. 

Using the above definition, GNVW defined the following two sets of support algebras:
\begin{align}
\begin{split}
\mathcal{R}_{2r}&=\mathbf{S}( U^\dagger(\mathcal{A}_{2r}\otimes\mathcal{A}_{2r+1})U,\ \mathcal{A}_{2r-1}\otimes\mathcal{A}_{2r})\\
\mathcal{R}_{2r+1}&=\mathbf{S}( U^\dagger(\mathcal{A}_{2r}\otimes\mathcal{A}_{2r+1})U,\ \mathcal{A}_{2r+1}\otimes\mathcal{A}_{2r+2})
\label{supalgdef}
\end{split}
\end{align}
By construction, $U^\dagger(\mathcal{A}_{2r}\otimes\mathcal{A}_{2r+1})U\subset\mathcal{R}_{2r}\otimes\mathcal{R}_{2r+1}$. An important result in GNVW is that this inclusion is actually an equality: 
\begin{equation}
U^\dagger(\mathcal{A}_{2r}\otimes\mathcal{A}_{2r+1})U=\mathcal{R}_{2r}\otimes\mathcal{R}_{2r+1}
\label{AARR}
\end{equation}

Another important result of GNVW is that the support algebras $\mathcal{R}_r$ are isomorphic to finite-dimensional matrix algebras. We denote the rank of these algebras by $m(r)$.

To state the next result of GNVW, consider the two support algebras $\mathcal{R}_{2r+1}$ and $\mathcal{R}_{2r+2}$. By definition, both of these algebras are contained in $\mathcal{A}_{2r+1} \otimes \mathcal{A}_{2r+2}$. This means that $\mathcal{R}_{2r+1}$ and $\mathcal{R}_{2r+2}$ can be thought of as collections of operators acting on the space $\mathcal{H}_{2r+1} \otimes \mathcal{H}_{2r+2}$ where $\mathcal{H}_r$ is the $D$-dimensional Hilbert space associated with site $r$. What do these collections of operators look like? GNVW showed that there exists a basis for the two site Hilbert space $\mathcal{H}_{2r+1} \otimes \mathcal{H}_{2r+2}$ such that, in this basis, $\mathcal{R}_{2r+1}$ consists of matrices of the form 
\begin{align}
M_{m(2r+1)} \otimes \mathbbm{1}_{m(2r+2)}, 
\label{MtimesI}
\end{align}
and $\mathcal{R}_{2r+2}$ consists of matrices of the form 
\begin{align}
\mathbbm{1}_{m(2r+1)} \otimes M_{m(2r+2)}
\label{ItimesM}
\end{align}
where $M_{m(r)}$ denotes an arbitrary $m(r) \times m(r)$ matrix, and $\mathbbm{1}_{m(r)}$ denotes an $m(r) \times m(r)$ identity matrix. Equivalently, this result can be stated as the following operator algebra identity:
\begin{equation}\label{RRAA}
\mathcal{R}_{2r+1}\otimes\mathcal{R}_{2r+2} = \mathcal{A}_{2r+1}\otimes\mathcal{A}_{2r+2} 
\end{equation}

The last result that we will need from GNVW comes from considering the dimensions of the operator algebras in Eqs. (\ref{AARR}) and (\ref{RRAA}). In particular, if we equate the dimensions of the algebras on both sides of these identities, we see that
\begin{align*}
m(2r)m(2r+1) &= D^2 \nonumber \\
m(2r+1)m(2r+2) &= D^2
\end{align*}
Using these results, GNVW defined $\mathrm{ind}(U)$ as
\begin{equation}
\mathrm{ind}(U)= \frac{m(2r+1)}{D}=\frac{D}{m(2r+2)}
\label{indUdef}
\end{equation}

\subsection{Additional results in $U(1)$ symmetric case}
We now derive two additional results which we will need below. These results apply to the case where the locality preserving unitary is $U(1)$ symmetric -- that is, the case where $U$ commutes with $\sum_r Q_r$, where $Q_r$ is a Hermitian (charge) operator acting on site $r$ with smallest eigenvalue $0$. Some preliminary work was already made in this direction in Ref.~\onlinecite{kirby}.

To derive the first result, consider $U^\dagger(Q_{2r}+Q_{2r+1})U$. By the argument given in Appendix~\ref{sUQU}, we know that $U^\dagger(Q_{2r}+Q_{2r+1})U$ can be written as
\begin{align}
U^\dagger(Q_{2r}&+Q_{2r+1})U \nonumber \\
&= (Q_{2r} + O_{L,2r}) + (Q_{2r+1} + O_{R,2r+1})
\label{QQolr}
\end{align}
where $O_{L,2r} \in \mathcal{A}_{2r-1} \otimes \mathcal{A}_{2r}$ and $O_{R,2r+1} \in \mathcal{A}_{2r+1} \otimes \mathcal{A}_{2r+2}$.\footnote{Here, as in Sec.~\ref{sdefs}, we fix the ambiguity in $O_{L,2r}$ and $O_{R,2r+1}$ by choosing $O_{L,2r}$ and $O_{R,2r+1}$ so that the smallest eigenvalue of $Q_{2r} + O_{L,2r}$ and $Q_{2r+1} + O_{R,2r+1}$ is $0$.} Next, notice that since $Q_r \in \mathcal{A}_r$, we can further conclude that $Q_{2r}+O_{L,2r} \in \mathcal{A}_{2r-1} \otimes \mathcal{A}_{2r}$ and $Q_{2r+1}+ O_{R,2r+1} \in \mathcal{A}_{2r+1} \otimes \mathcal{A}_{2r+2}$. It then follows from the definitions of the support algebras (\ref{supalgdef}) that
\begin{align} 
Q_{2r} + O_{L,2r} \in \mathcal{R}_{2r}, \quad Q_{2r+1} + O_{R,2r+1} \in \mathcal{R}_{2r+1}
\label{QORsupport}
\end{align}
Eq.~\ref{QORsupport} is one of the two results that we will need about the $U(1)$ symmetric case. The other result is the following identity:
\begin{align}
O_{R,2r+1} = - O_{L, 2r+2}
\label{oloridapp} 
\end{align}
To derive (\ref{oloridapp}), consider $U$ acting on the operator $Q_{2r}+Q_{2r+1}+Q_{2r+2}+Q_{2r+3}$. By Eq.~(\ref{QQolr}), we have
\begin{align}
\begin{split}
U^\dagger&(Q_{2r}+Q_{2r+1}+Q_{2r+2}+Q_{2r+3})U\\
&=U^\dagger(Q_{2r}+Q_{2r+1})U+U^\dagger(Q_{2r+2}+Q_{2r+3})U\\
&=Q_{2r}+O_{L,2r}+Q_{2r+1}+O_{R,2r+1}+Q_{2r+2}\\
&+O_{L,2r+2}+Q_{2r+3}+O_{R,2r+3}
\label{QQQQ1}
\end{split}
\end{align}
At the same time, by the argument given in Appendix~\ref{sUQU}, we know that
\begin{align}
U^\dagger&(Q_{2r}+Q_{2r+1}+Q_{2r+2}+Q_{2r+3})U \nonumber \\
&= Q_{2r}+Q_{2r+1}+Q_{2r+2}+Q_{2r+3} + O_{L,2r} + O_{R,2r+3}
\label{QQQQ2}
\end{align}
where $O_{L,2r} \in \mathcal{A}_{2r-1} \otimes \mathcal{A}_{2r}$ and  $O_{R,2r+3} \in \mathcal{A}_{2r+3} \otimes \mathcal{A}_{2r+4}$.
Comparing the right hand sides of Eqs. (\ref{QQQQ1}) and (\ref{QQQQ2}), we deduce that $O_{R,2r+1} + O_{L,2r+2} = c \mathbbm{1}$ for some scalar $c$. In fact, it is easy to show that the constant $c=0$ given our convention that the smallest eigenvalues of $Q_{2r+1}$, $Q_{2r+2}$, and $Q_{2r+1} + O_{R,2r+1}$ and $Q_{2r+2} + O_{L,2r+2}$ are all $0$. This establishes Eq. (\ref{oloridapp}).

\subsection{Proofs of bosonic claims from the main text}\label{sGNVWproofs}
We are now ready to prove the two claims from the main text regarding bosonic systems. We start with the claim made in Eq.~\ref{qrorid} in Sec.~\ref{ssatisfiesrequ}, namely that the restriction of the operator $Q_R + O_R$ to the interval $R$ can be written, in an appropriate basis, as
\begin{align}
Q_{R}+O_R=O\otimes\mathbbm{1}
\end{align}
where $O$ is a matrix of dimension $[d^\ell \cdot \mathrm{ind}(U)]$ and $\mathbbm{1}$ is an identity matrix of dimension $d^\ell/ \mathrm{ind}(U)$. 

To prove this claim, we first cluster together groups of $\ell$ neighboring spins into superspins of dimension $D = d^\ell$. In this superspin representation, the interval $A$ corresponds to two sites $\{2r, 2r+1\}$, while the interval $R$ corresponds to the two sites $\{2r+1, 2r+2\}$. Also, the operator $Q_R+ O_R$ corresponds to $Q_{R,2r+1} +O_{R, 2r+1}$. Thus, in this new notation, the claim amounts to showing that the restriction of the operator $Q_{R,2r+1} +O_{R, 2r+1}$ to $\{2r+1, 2r+2\}$ can be written, in an appropriate basis, as $O \otimes \mathbbm{1}$ where $O$ has dimension $D \cdot \mathrm{ind}(U)$ and $\mathbbm{1}$ has dimension $D/ \mathrm{ind}(U)$. 

The latter claim follows from Eq.~(\ref{MtimesI}) and Eq.~(\ref{QORsupport}): combining these two equations, we immediately see that $Q_{R,2r+1} +O_{R, 2r+1}$ can be written in the form $O \otimes \mathbbm{1}$, where $O$ has dimension
$m(2r+1)$ and $\mathbbm{1}$ has dimension $m(2r+2)$. These dimensions are exactly what we want: according to (\ref{indUdef}), we have $m(2r+1) =  D \cdot \mathrm{ind}(U)$ and $m(2r+2) =  D/\mathrm{ind}(U)$.
 
We now move on to the second claim, made in Eq.~\ref{gnvwsplit} of Sec.~\ref{sedge}. According to this claim, the Hilbert space
$\mathcal{H}^R_{\mathrm{out}}$ can be decomposed as a tensor product $\mathcal{H}_1 \otimes \mathcal{H}_2$ in such a way that the restrictions of $(Q^+_{R, \mathrm{out}}+O_{R, \mathrm{out}}^{(n)})$ and $(Q^-_{R, \mathrm{out}}-O_{R, \mathrm{out}}^{(n)})$ to $\mathcal{H}^R_{\mathrm{out}}$ take the form $\mathcal{O}_1 \otimes \mathbbm{1}$ and $\mathbbm{1} \otimes \mathcal{O}_2$, respectively. 

To prove this claim, we first cluster all the spins in the four quadrants $A^+_{L,\mathrm{out}}$, $A^+_{R,\mathrm{out}}$, $A^-_{R,\mathrm{out}}$,  $A^-_{L,\mathrm{out}}$ into $4$ supersites $\{0, 1, 2, 3\}$, with one super site corresponding to each quadrant. In this clustering scheme, the charge operator $Q_{R,\mathrm{out}}^+$ corresponds to  $Q_{1}$, and $Q_{R,\mathrm{out}}^-$ corresponds to $Q_{2}$. Also, we can identify $\mathcal{H}^R_{\mathrm{out}}$ with the two site Hilbert space $\mathcal{H}_1 \otimes \mathcal{H}_2$ and we can identify $O_{R,\mathrm{out}}^{(n)}=O_{R,1} = -O_{L,2}$ (\ref{oloridapp}). The claim translates into showing that $Q_1 + O_{R,1}$ and $Q_2 - O_{L,2}$ can be written in the form $\mathcal{O}_1 \otimes \mathbbm{1}$ and $\mathbbm{1} \otimes \mathcal{O}_2$ for an appropriate tensor decomposition of $\mathcal{H}_1 \otimes \mathcal{H}_2$. This claim follows immediately from  Eq. (\ref{MtimesI}) and (\ref{ItimesM}) together with Eq. (\ref{QORsupport}).

\subsection{Fermionic systems}\label{sfermionicGNVW}

We now prove a claim about \emph{fermionic} systems that we made in Sec.~\ref{ssatisfiesfermionic}: let $U$ be a 1D $U(1)$ symmetric locality preserving unitary defined for some choice of $P, Q$. We will show that if $\mathrm{ind}^f(U) = \sqrt{2}^\zeta p/q$ where $\zeta = 1$, then $f_\chi(z) = 0$, where $f_\chi(z) = \mathrm{Tr}(z^Q P)$.

We start by reviewing two results from Ref.~\onlinecite{fermionic} regarding the extension of the GNVW analysis to fermionic systems. The first result is about the support algebras $\mathcal{R}_i$, which we define in the same way as in the bosonic case (\ref{supalgdef}). Specifically, the result that we need is that the $\mathcal{R}_r$'s have two different structures depending on $\zeta$: (1) if $\zeta = 0$, then the $\mathcal{R}_{r}$ are \emph{even} algebras, i.e. matrix algebras over a $\mathbb{Z}_2$-graded vector space, while (2) if $\zeta = 1$, then the $\mathcal{R}_r                                                                                     $ are \emph{odd} algebras, i.e. matrix algebras over an odd Clifford algebra ($\mathbb{C}\ell_n=\mathbb{C}\ell_1^{\otimes n}$ for $n$ odd).

The other result that we will need is that the $\mathcal{R}_r$ algebras commute with each other in the $\mathbb{Z}_2$ graded sense: that is, for any $O_r\in \mathcal{R}_r$ and $O_{r'}\in \mathcal{R}_{r'}$ with $r' \neq r$, the two operators $O_r, O_{r'}$ commute if at least one of them has even fermion parity and anti-commute if both have odd fermion parity.
  
Using these results, we can now prove the claim. Consider the support algebra $\mathcal{R}_{2r+1}$. Assuming $\zeta = 1$, we know that $\mathcal{R}_{2r+1}$ is odd. Therefore, like all odd algebras, $\mathcal{R}_{2r+1}$ contains an operator $\Gamma_{2r+1}$ that is (i) fermion parity odd, (ii) commutes with all the other elements of the algebra and (iii) satisfies $\Gamma_{2r+1}^2 = 1$, and $\Gamma_{2r+1}^\dagger =\Gamma_{2r+1}$.

Next note that $Q_{2r+1} + O_{R,2r+1} \in \mathcal{R}_{2r+1}$ by the same reasoning as in the bosonic case (\ref{QORsupport}). Hence, $\Gamma_{2r+1}$ commutes with  $Q_{2r+1} + O_{R,2r+1}$. Also, $\Gamma_{2r+1}$ commutes with $Q_{2r+2} + O_{L,2r+2}$ since $Q_{2r+2} + O_{L,2r+2} \in \mathcal{R}_{2r+2}$ and the different $\mathcal{R}_r$'s commute with each other in the $\mathbb{Z}_2$ graded sense. Putting this all together, and using $O_{R,2r+1} = -O_{L,2r+2}$ (\ref{oloridapp}), we conclude that $\Gamma_{2r+1}$ commutes with $Q_{2r+1} + Q_{2r+2}$. 

To complete the argument, note that since $Q_{2r+1} + Q_{2r+2}$ commutes with $\Gamma_{2r+1}$, it follows that $Q_{2r+1} + Q_{2r+2}$ has the same spectrum in the even and odd fermion parity sectors. This implies that $\mathrm{Tr}(z^{Q_{2r+1} + Q_{2r+2}} P_{2r+1} P_{2r+2}) = 0$. Hence, $f_\chi(z)^2 = 0$, which implies that $f_\chi(z)= 0$. This proves the claim.

\bibliography{u1floquetbib}

\begin{thebibliography}{45}%
\makeatletter
\providecommand \@ifxundefined [1]{%
 \@ifx{#1\undefined}
}%
\providecommand \@ifnum [1]{%
 \ifnum #1\expandafter \@firstoftwo
 \else \expandafter \@secondoftwo
 \fi
}%
\providecommand \@ifx [1]{%
 \ifx #1\expandafter \@firstoftwo
 \else \expandafter \@secondoftwo
 \fi
}%
\providecommand \natexlab [1]{#1}%
\providecommand \enquote  [1]{``#1''}%
\providecommand \bibnamefont  [1]{#1}%
\providecommand \bibfnamefont [1]{#1}%
\providecommand \citenamefont [1]{#1}%
\providecommand \href@noop [0]{\@secondoftwo}%
\providecommand \href [0]{\begingroup \@sanitize@url \@href}%
\providecommand \@href[1]{\@@startlink{#1}\@@href}%
\providecommand \@@href[1]{\endgroup#1\@@endlink}%
\providecommand \@sanitize@url [0]{\catcode `\\12\catcode `\$12\catcode
  `\&12\catcode `\#12\catcode `\^12\catcode `\_12\catcode `\%12\relax}%
\providecommand \@@startlink[1]{}%
\providecommand \@@endlink[0]{}%
\providecommand \url  [0]{\begingroup\@sanitize@url \@url }%
\providecommand \@url [1]{\endgroup\@href {#1}{\urlprefix }}%
\providecommand \urlprefix  [0]{URL }%
\providecommand \Eprint [0]{\href }%
\providecommand \doibase [0]{http://dx.doi.org/}%
\providecommand \selectlanguage [0]{\@gobble}%
\providecommand \bibinfo  [0]{\@secondoftwo}%
\providecommand \bibfield  [0]{\@secondoftwo}%
\providecommand \translation [1]{[#1]}%
\providecommand \BibitemOpen [0]{}%
\providecommand \bibitemStop [0]{}%
\providecommand \bibitemNoStop [0]{.\EOS\space}%
\providecommand \EOS [0]{\spacefactor3000\relax}%
\providecommand \BibitemShut  [1]{\csname bibitem#1\endcsname}%
\let\auto@bib@innerbib\@empty
\bibitem [{\citenamefont {Harper}\ \emph {et~al.}(2020)\citenamefont {Harper},
  \citenamefont {Roy}, \citenamefont {Rudner},\ and\ \citenamefont
  {Sondhi}}]{floquetreview}%
  \BibitemOpen
  \bibfield  {author} {\bibinfo {author} {\bibfnamefont {F.}~\bibnamefont
  {Harper}}, \bibinfo {author} {\bibfnamefont {R.}~\bibnamefont {Roy}},
  \bibinfo {author} {\bibfnamefont {M.~S.}\ \bibnamefont {Rudner}}, \ and\
  \bibinfo {author} {\bibfnamefont {S.}~\bibnamefont {Sondhi}},\ }\href
  {\doibase 10.1146/annurev-conmatphys-031218-013721} {\bibfield  {journal}
  {\bibinfo  {journal} {Annual Review of Condensed Matter Physics}\ }\textbf
  {\bibinfo {volume} {11}},\ \bibinfo {pages} {345} (\bibinfo {year}
  {2020})}\BibitemShut {NoStop}%
\bibitem [{\citenamefont {Rudner}\ and\ \citenamefont
  {Lindner}(2020)}]{rudnerband}%
  \BibitemOpen
  \bibfield  {author} {\bibinfo {author} {\bibfnamefont {M.~S.}\ \bibnamefont
  {Rudner}}\ and\ \bibinfo {author} {\bibfnamefont {N.~H.}\ \bibnamefont
  {Lindner}},\ }\href {\doibase 10.1038/s42254-020-0170-z} {\bibfield
  {journal} {\bibinfo  {journal} {Nature Reviews Physics}\ ,\ \bibinfo {pages}
  {1}} (\bibinfo {year} {2020})}\BibitemShut {NoStop}%
\bibitem [{\citenamefont {Abanin}\ \emph {et~al.}(2019)\citenamefont {Abanin},
  \citenamefont {Altman}, \citenamefont {Bloch},\ and\ \citenamefont
  {Serbyn}}]{abanincolloquium}%
  \BibitemOpen
  \bibfield  {author} {\bibinfo {author} {\bibfnamefont {D.~A.}\ \bibnamefont
  {Abanin}}, \bibinfo {author} {\bibfnamefont {E.}~\bibnamefont {Altman}},
  \bibinfo {author} {\bibfnamefont {I.}~\bibnamefont {Bloch}}, \ and\ \bibinfo
  {author} {\bibfnamefont {M.}~\bibnamefont {Serbyn}},\ }\href {\doibase
  10.1103/RevModPhys.91.021001} {\bibfield  {journal} {\bibinfo  {journal}
  {Reviews of Modern Physics}\ }\textbf {\bibinfo {volume} {91}},\ \bibinfo
  {pages} {021001} (\bibinfo {year} {2019})}\BibitemShut {NoStop}%
\bibitem [{\citenamefont {Lazarides}\ \emph {et~al.}(2015)\citenamefont
  {Lazarides}, \citenamefont {Das},\ and\ \citenamefont {Moessner}}]{fate}%
  \BibitemOpen
  \bibfield  {author} {\bibinfo {author} {\bibfnamefont {A.}~\bibnamefont
  {Lazarides}}, \bibinfo {author} {\bibfnamefont {A.}~\bibnamefont {Das}}, \
  and\ \bibinfo {author} {\bibfnamefont {R.}~\bibnamefont {Moessner}},\ }\href
  {\doibase 10.1103/PhysRevLett.115.030402} {\bibfield  {journal} {\bibinfo
  {journal} {Phys. Rev. Lett.}\ }\textbf {\bibinfo {volume} {115}},\ \bibinfo
  {pages} {030402} (\bibinfo {year} {2015})}\BibitemShut {NoStop}%
\bibitem [{\citenamefont {Ponte}\ \emph
  {et~al.}(2015{\natexlab{a}})\citenamefont {Ponte}, \citenamefont
  {Papi\ifmmode~\acute{c}\else \'{c}\fi{}}, \citenamefont {Huveneers},\ and\
  \citenamefont {Abanin}}]{ponte}%
  \BibitemOpen
  \bibfield  {author} {\bibinfo {author} {\bibfnamefont {P.}~\bibnamefont
  {Ponte}}, \bibinfo {author} {\bibfnamefont {Z.}~\bibnamefont
  {Papi\ifmmode~\acute{c}\else \'{c}\fi{}}}, \bibinfo {author} {\bibfnamefont
  {F.~m.~c.}\ \bibnamefont {Huveneers}}, \ and\ \bibinfo {author}
  {\bibfnamefont {D.~A.}\ \bibnamefont {Abanin}},\ }\href {\doibase
  10.1103/PhysRevLett.114.140401} {\bibfield  {journal} {\bibinfo  {journal}
  {Phys. Rev. Lett.}\ }\textbf {\bibinfo {volume} {114}},\ \bibinfo {pages}
  {140401} (\bibinfo {year} {2015}{\natexlab{a}})}\BibitemShut {NoStop}%
\bibitem [{\citenamefont {Abanin}\ \emph {et~al.}(2016)\citenamefont {Abanin},
  \citenamefont {Roeck},\ and\ \citenamefont {Huveneers}}]{abanintheory}%
  \BibitemOpen
  \bibfield  {author} {\bibinfo {author} {\bibfnamefont {D.~A.}\ \bibnamefont
  {Abanin}}, \bibinfo {author} {\bibfnamefont {W.~D.}\ \bibnamefont {Roeck}}, \
  and\ \bibinfo {author} {\bibfnamefont {F.}~\bibnamefont {Huveneers}},\ }\href
  {\doibase https://doi.org/10.1016/j.aop.2016.03.010} {\bibfield  {journal}
  {\bibinfo  {journal} {Annals of Physics}\ }\textbf {\bibinfo {volume}
  {372}},\ \bibinfo {pages} {1 } (\bibinfo {year} {2016})}\BibitemShut
  {NoStop}%
\bibitem [{\citenamefont {Bordia}\ \emph {et~al.}(2017)\citenamefont {Bordia},
  \citenamefont {L{\"u}schen}, \citenamefont {Schneider}, \citenamefont
  {Knap},\ and\ \citenamefont {Bloch}}]{bordia}%
  \BibitemOpen
  \bibfield  {author} {\bibinfo {author} {\bibfnamefont {P.}~\bibnamefont
  {Bordia}}, \bibinfo {author} {\bibfnamefont {H.}~\bibnamefont {L{\"u}schen}},
  \bibinfo {author} {\bibfnamefont {U.}~\bibnamefont {Schneider}}, \bibinfo
  {author} {\bibfnamefont {M.}~\bibnamefont {Knap}}, \ and\ \bibinfo {author}
  {\bibfnamefont {I.}~\bibnamefont {Bloch}},\ }\href {\doibase
  10.1038/nphys4020} {\bibfield  {journal} {\bibinfo  {journal} {Nature
  Physics}\ }\textbf {\bibinfo {volume} {13}},\ \bibinfo {pages} {460}
  (\bibinfo {year} {2017})}\BibitemShut {NoStop}%
\bibitem [{\citenamefont {D'Alessio}\ and\ \citenamefont
  {Rigol}(2014)}]{rigollongtime}%
  \BibitemOpen
  \bibfield  {author} {\bibinfo {author} {\bibfnamefont {L.}~\bibnamefont
  {D'Alessio}}\ and\ \bibinfo {author} {\bibfnamefont {M.}~\bibnamefont
  {Rigol}},\ }\href {\doibase 10.1103/PhysRevX.4.041048} {\bibfield  {journal}
  {\bibinfo  {journal} {Phys. Rev. X}\ }\textbf {\bibinfo {volume} {4}},\
  \bibinfo {pages} {041048} (\bibinfo {year} {2014})}\BibitemShut {NoStop}%
\bibitem [{\citenamefont {Lazarides}\ \emph {et~al.}(2014)\citenamefont
  {Lazarides}, \citenamefont {Das},\ and\ \citenamefont
  {Moessner}}]{lazaridesgeneric}%
  \BibitemOpen
  \bibfield  {author} {\bibinfo {author} {\bibfnamefont {A.}~\bibnamefont
  {Lazarides}}, \bibinfo {author} {\bibfnamefont {A.}~\bibnamefont {Das}}, \
  and\ \bibinfo {author} {\bibfnamefont {R.}~\bibnamefont {Moessner}},\ }\href
  {\doibase 10.1103/PhysRevE.90.012110} {\bibfield  {journal} {\bibinfo
  {journal} {Phys. Rev. E}\ }\textbf {\bibinfo {volume} {90}},\ \bibinfo
  {pages} {012110} (\bibinfo {year} {2014})}\BibitemShut {NoStop}%
\bibitem [{\citenamefont {Ponte}\ \emph
  {et~al.}(2015{\natexlab{b}})\citenamefont {Ponte}, \citenamefont {Chandran},
  \citenamefont {Papić},\ and\ \citenamefont {Abanin}}]{ponteergodic}%
  \BibitemOpen
  \bibfield  {author} {\bibinfo {author} {\bibfnamefont {P.}~\bibnamefont
  {Ponte}}, \bibinfo {author} {\bibfnamefont {A.}~\bibnamefont {Chandran}},
  \bibinfo {author} {\bibfnamefont {Z.}~\bibnamefont {Papić}}, \ and\ \bibinfo
  {author} {\bibfnamefont {D.~A.}\ \bibnamefont {Abanin}},\ }\href {\doibase
  https://doi.org/10.1016/j.aop.2014.11.008} {\bibfield  {journal} {\bibinfo
  {journal} {Annals of Physics}\ }\textbf {\bibinfo {volume} {353}},\ \bibinfo
  {pages} {196 } (\bibinfo {year} {2015}{\natexlab{b}})}\BibitemShut {NoStop}%
\bibitem [{\citenamefont {Khemani}\ \emph {et~al.}(2016)\citenamefont
  {Khemani}, \citenamefont {Lazarides}, \citenamefont {Moessner},\ and\
  \citenamefont {Sondhi}}]{khemaniphase}%
  \BibitemOpen
  \bibfield  {author} {\bibinfo {author} {\bibfnamefont {V.}~\bibnamefont
  {Khemani}}, \bibinfo {author} {\bibfnamefont {A.}~\bibnamefont {Lazarides}},
  \bibinfo {author} {\bibfnamefont {R.}~\bibnamefont {Moessner}}, \ and\
  \bibinfo {author} {\bibfnamefont {S.~L.}\ \bibnamefont {Sondhi}},\ }\href
  {\doibase 10.1103/PhysRevLett.116.250401} {\bibfield  {journal} {\bibinfo
  {journal} {Phys. Rev. Lett.}\ }\textbf {\bibinfo {volume} {116}},\ \bibinfo
  {pages} {250401} (\bibinfo {year} {2016})}\BibitemShut {NoStop}%
\bibitem [{\citenamefont {von Keyserlingk}\ and\ \citenamefont
  {Sondhi}(2016{\natexlab{a}})}]{keyserlingk1DI}%
  \BibitemOpen
  \bibfield  {author} {\bibinfo {author} {\bibfnamefont {C.~W.}\ \bibnamefont
  {von Keyserlingk}}\ and\ \bibinfo {author} {\bibfnamefont {S.~L.}\
  \bibnamefont {Sondhi}},\ }\href {\doibase 10.1103/PhysRevB.93.245145}
  {\bibfield  {journal} {\bibinfo  {journal} {Phys. Rev. B}\ }\textbf {\bibinfo
  {volume} {93}},\ \bibinfo {pages} {245145} (\bibinfo {year}
  {2016}{\natexlab{a}})}\BibitemShut {NoStop}%
\bibitem [{\citenamefont {von Keyserlingk}\ and\ \citenamefont
  {Sondhi}(2016{\natexlab{b}})}]{keyserlingk1DII}%
  \BibitemOpen
  \bibfield  {author} {\bibinfo {author} {\bibfnamefont {C.~W.}\ \bibnamefont
  {von Keyserlingk}}\ and\ \bibinfo {author} {\bibfnamefont {S.~L.}\
  \bibnamefont {Sondhi}},\ }\href {\doibase 10.1103/PhysRevB.93.245146}
  {\bibfield  {journal} {\bibinfo  {journal} {Phys. Rev. B}\ }\textbf {\bibinfo
  {volume} {93}},\ \bibinfo {pages} {245146} (\bibinfo {year}
  {2016}{\natexlab{b}})}\BibitemShut {NoStop}%
\bibitem [{\citenamefont {Else}\ and\ \citenamefont
  {Nayak}(2016)}]{cohomology}%
  \BibitemOpen
  \bibfield  {author} {\bibinfo {author} {\bibfnamefont {D.~V.}\ \bibnamefont
  {Else}}\ and\ \bibinfo {author} {\bibfnamefont {C.}~\bibnamefont {Nayak}},\
  }\href {\doibase 10.1103/PhysRevB.93.201103} {\bibfield  {journal} {\bibinfo
  {journal} {Phys. Rev. B}\ }\textbf {\bibinfo {volume} {93}},\ \bibinfo
  {pages} {201103} (\bibinfo {year} {2016})}\BibitemShut {NoStop}%
\bibitem [{\citenamefont {Potter}\ \emph {et~al.}(2016)\citenamefont {Potter},
  \citenamefont {Morimoto},\ and\ \citenamefont {Vishwanath}}]{potter1D}%
  \BibitemOpen
  \bibfield  {author} {\bibinfo {author} {\bibfnamefont {A.~C.}\ \bibnamefont
  {Potter}}, \bibinfo {author} {\bibfnamefont {T.}~\bibnamefont {Morimoto}}, \
  and\ \bibinfo {author} {\bibfnamefont {A.}~\bibnamefont {Vishwanath}},\
  }\href {\doibase 10.1103/PhysRevX.6.041001} {\bibfield  {journal} {\bibinfo
  {journal} {Phys. Rev. X}\ }\textbf {\bibinfo {volume} {6}},\ \bibinfo {pages}
  {041001} (\bibinfo {year} {2016})}\BibitemShut {NoStop}%
\bibitem [{\citenamefont {Roy}\ and\ \citenamefont {Harper}(2016)}]{roy1D}%
  \BibitemOpen
  \bibfield  {author} {\bibinfo {author} {\bibfnamefont {R.}~\bibnamefont
  {Roy}}\ and\ \bibinfo {author} {\bibfnamefont {F.}~\bibnamefont {Harper}},\
  }\href {\doibase 10.1103/PhysRevB.94.125105} {\bibfield  {journal} {\bibinfo
  {journal} {Phys. Rev. B}\ }\textbf {\bibinfo {volume} {94}},\ \bibinfo
  {pages} {125105} (\bibinfo {year} {2016})}\BibitemShut {NoStop}%
\bibitem [{\citenamefont {Po}\ \emph {et~al.}(2016)\citenamefont {Po},
  \citenamefont {Fidkowski}, \citenamefont {Morimoto}, \citenamefont {Potter},\
  and\ \citenamefont {Vishwanath}}]{chiralbosons}%
  \BibitemOpen
  \bibfield  {author} {\bibinfo {author} {\bibfnamefont {H.~C.}\ \bibnamefont
  {Po}}, \bibinfo {author} {\bibfnamefont {L.}~\bibnamefont {Fidkowski}},
  \bibinfo {author} {\bibfnamefont {T.}~\bibnamefont {Morimoto}}, \bibinfo
  {author} {\bibfnamefont {A.~C.}\ \bibnamefont {Potter}}, \ and\ \bibinfo
  {author} {\bibfnamefont {A.}~\bibnamefont {Vishwanath}},\ }\href {\doibase
  10.1103/PhysRevX.6.041070} {\bibfield  {journal} {\bibinfo  {journal} {Phys.
  Rev. X}\ }\textbf {\bibinfo {volume} {6}},\ \bibinfo {pages} {041070}
  (\bibinfo {year} {2016})}\BibitemShut {NoStop}%
\bibitem [{\citenamefont {Harper}\ and\ \citenamefont
  {Roy}(2017)}]{harperorder}%
  \BibitemOpen
  \bibfield  {author} {\bibinfo {author} {\bibfnamefont {F.}~\bibnamefont
  {Harper}}\ and\ \bibinfo {author} {\bibfnamefont {R.}~\bibnamefont {Roy}},\
  }\href {\doibase 10.1103/PhysRevLett.118.115301} {\bibfield  {journal}
  {\bibinfo  {journal} {Phys. Rev. Lett.}\ }\textbf {\bibinfo {volume} {118}},\
  \bibinfo {pages} {115301} (\bibinfo {year} {2017})}\BibitemShut {NoStop}%
\bibitem [{\citenamefont {Fidkowski}\ \emph {et~al.}(2019)\citenamefont
  {Fidkowski}, \citenamefont {Po}, \citenamefont {Potter},\ and\ \citenamefont
  {Vishwanath}}]{fermionic}%
  \BibitemOpen
  \bibfield  {author} {\bibinfo {author} {\bibfnamefont {L.}~\bibnamefont
  {Fidkowski}}, \bibinfo {author} {\bibfnamefont {H.~C.}\ \bibnamefont {Po}},
  \bibinfo {author} {\bibfnamefont {A.~C.}\ \bibnamefont {Potter}}, \ and\
  \bibinfo {author} {\bibfnamefont {A.}~\bibnamefont {Vishwanath}},\ }\href
  {\doibase 10.1103/PhysRevB.99.085115} {\bibfield  {journal} {\bibinfo
  {journal} {Phys. Rev. B}\ }\textbf {\bibinfo {volume} {99}},\ \bibinfo
  {pages} {085115} (\bibinfo {year} {2019})}\BibitemShut {NoStop}%
\bibitem [{\citenamefont {Duschatko}\ \emph {et~al.}(2018)\citenamefont
  {Duschatko}, \citenamefont {Dumitrescu},\ and\ \citenamefont
  {Potter}}]{tracking}%
  \BibitemOpen
  \bibfield  {author} {\bibinfo {author} {\bibfnamefont {B.~R.}\ \bibnamefont
  {Duschatko}}, \bibinfo {author} {\bibfnamefont {P.~T.}\ \bibnamefont
  {Dumitrescu}}, \ and\ \bibinfo {author} {\bibfnamefont {A.~C.}\ \bibnamefont
  {Potter}},\ }\href {\doibase 10.1103/PhysRevB.98.054309} {\bibfield
  {journal} {\bibinfo  {journal} {Phys. Rev. B}\ }\textbf {\bibinfo {volume}
  {98}},\ \bibinfo {pages} {054309} (\bibinfo {year} {2018})}\BibitemShut
  {NoStop}%
\bibitem [{\citenamefont {Gross}\ \emph {et~al.}(2012)\citenamefont {Gross},
  \citenamefont {Nesme}, \citenamefont {Vogts},\ and\ \citenamefont
  {Werner}}]{GNVW}%
  \BibitemOpen
  \bibfield  {author} {\bibinfo {author} {\bibfnamefont {D.}~\bibnamefont
  {Gross}}, \bibinfo {author} {\bibfnamefont {V.}~\bibnamefont {Nesme}},
  \bibinfo {author} {\bibfnamefont {H.}~\bibnamefont {Vogts}}, \ and\ \bibinfo
  {author} {\bibfnamefont {R.~F.}\ \bibnamefont {Werner}},\ }\href {\doibase
  10.1007/s00220-012-1423-1} {\bibfield  {journal} {\bibinfo  {journal}
  {Communications in Mathematical Physics}\ }\textbf {\bibinfo {volume}
  {310}},\ \bibinfo {pages} {419} (\bibinfo {year} {2012})}\BibitemShut
  {NoStop}%
\bibitem [{\citenamefont {Kitagawa}\ \emph {et~al.}(2010)\citenamefont
  {Kitagawa}, \citenamefont {Berg}, \citenamefont {Rudner},\ and\ \citenamefont
  {Demler}}]{kitagawa2010}%
  \BibitemOpen
  \bibfield  {author} {\bibinfo {author} {\bibfnamefont {T.}~\bibnamefont
  {Kitagawa}}, \bibinfo {author} {\bibfnamefont {E.}~\bibnamefont {Berg}},
  \bibinfo {author} {\bibfnamefont {M.}~\bibnamefont {Rudner}}, \ and\ \bibinfo
  {author} {\bibfnamefont {E.}~\bibnamefont {Demler}},\ }\href {\doibase
  10.1103/PhysRevB.82.235114} {\bibfield  {journal} {\bibinfo  {journal} {Phys.
  Rev. B}\ }\textbf {\bibinfo {volume} {82}},\ \bibinfo {pages} {235114}
  (\bibinfo {year} {2010})}\BibitemShut {NoStop}%
\bibitem [{\citenamefont {Rudner}\ \emph {et~al.}(2013)\citenamefont {Rudner},
  \citenamefont {Lindner}, \citenamefont {Berg},\ and\ \citenamefont
  {Levin}}]{anomalousedge}%
  \BibitemOpen
  \bibfield  {author} {\bibinfo {author} {\bibfnamefont {M.~S.}\ \bibnamefont
  {Rudner}}, \bibinfo {author} {\bibfnamefont {N.~H.}\ \bibnamefont {Lindner}},
  \bibinfo {author} {\bibfnamefont {E.}~\bibnamefont {Berg}}, \ and\ \bibinfo
  {author} {\bibfnamefont {M.}~\bibnamefont {Levin}},\ }\href {\doibase
  10.1103/PhysRevX.3.031005} {\bibfield  {journal} {\bibinfo  {journal} {Phys.
  Rev. X}\ }\textbf {\bibinfo {volume} {3}},\ \bibinfo {pages} {031005}
  (\bibinfo {year} {2013})}\BibitemShut {NoStop}%
\bibitem [{\citenamefont {Titum}\ \emph {et~al.}(2016)\citenamefont {Titum},
  \citenamefont {Berg}, \citenamefont {Rudner}, \citenamefont {Refael},\ and\
  \citenamefont {Lindner}}]{afai}%
  \BibitemOpen
  \bibfield  {author} {\bibinfo {author} {\bibfnamefont {P.}~\bibnamefont
  {Titum}}, \bibinfo {author} {\bibfnamefont {E.}~\bibnamefont {Berg}},
  \bibinfo {author} {\bibfnamefont {M.~S.}\ \bibnamefont {Rudner}}, \bibinfo
  {author} {\bibfnamefont {G.}~\bibnamefont {Refael}}, \ and\ \bibinfo {author}
  {\bibfnamefont {N.~H.}\ \bibnamefont {Lindner}},\ }\href {\doibase
  10.1103/PhysRevX.6.021013} {\bibfield  {journal} {\bibinfo  {journal} {Phys.
  Rev. X}\ }\textbf {\bibinfo {volume} {6}},\ \bibinfo {pages} {021013}
  (\bibinfo {year} {2016})}\BibitemShut {NoStop}%
\bibitem [{\citenamefont {Nathan}\ \emph {et~al.}(2017)\citenamefont {Nathan},
  \citenamefont {Rudner}, \citenamefont {Lindner}, \citenamefont {Berg},\ and\
  \citenamefont {Refael}}]{magnetization}%
  \BibitemOpen
  \bibfield  {author} {\bibinfo {author} {\bibfnamefont {F.}~\bibnamefont
  {Nathan}}, \bibinfo {author} {\bibfnamefont {M.~S.}\ \bibnamefont {Rudner}},
  \bibinfo {author} {\bibfnamefont {N.~H.}\ \bibnamefont {Lindner}}, \bibinfo
  {author} {\bibfnamefont {E.}~\bibnamefont {Berg}}, \ and\ \bibinfo {author}
  {\bibfnamefont {G.}~\bibnamefont {Refael}},\ }\href {\doibase
  10.1103/PhysRevLett.119.186801} {\bibfield  {journal} {\bibinfo  {journal}
  {Phys. Rev. Lett.}\ }\textbf {\bibinfo {volume} {119}},\ \bibinfo {pages}
  {186801} (\bibinfo {year} {2017})}\BibitemShut {NoStop}%
\bibitem [{\citenamefont {Kundu}\ \emph {et~al.}(2020)\citenamefont {Kundu},
  \citenamefont {Rudner}, \citenamefont {Berg},\ and\ \citenamefont
  {Lindner}}]{kundubias}%
  \BibitemOpen
  \bibfield  {author} {\bibinfo {author} {\bibfnamefont {A.}~\bibnamefont
  {Kundu}}, \bibinfo {author} {\bibfnamefont {M.}~\bibnamefont {Rudner}},
  \bibinfo {author} {\bibfnamefont {E.}~\bibnamefont {Berg}}, \ and\ \bibinfo
  {author} {\bibfnamefont {N.~H.}\ \bibnamefont {Lindner}},\ }\href {\doibase
  10.1103/PhysRevB.101.041403} {\bibfield  {journal} {\bibinfo  {journal}
  {Phys. Rev. B}\ }\textbf {\bibinfo {volume} {101}},\ \bibinfo {pages}
  {041403} (\bibinfo {year} {2020})}\BibitemShut {NoStop}%
\bibitem [{\citenamefont {Nathan}\ \emph
  {et~al.}(2019{\natexlab{a}})\citenamefont {Nathan}, \citenamefont {Abanin},
  \citenamefont {Berg}, \citenamefont {Lindner},\ and\ \citenamefont
  {Rudner}}]{nathanafi}%
  \BibitemOpen
  \bibfield  {author} {\bibinfo {author} {\bibfnamefont {F.}~\bibnamefont
  {Nathan}}, \bibinfo {author} {\bibfnamefont {D.}~\bibnamefont {Abanin}},
  \bibinfo {author} {\bibfnamefont {E.}~\bibnamefont {Berg}}, \bibinfo {author}
  {\bibfnamefont {N.~H.}\ \bibnamefont {Lindner}}, \ and\ \bibinfo {author}
  {\bibfnamefont {M.~S.}\ \bibnamefont {Rudner}},\ }\href {\doibase
  10.1103/PhysRevB.99.195133} {\bibfield  {journal} {\bibinfo  {journal} {Phys.
  Rev. B}\ }\textbf {\bibinfo {volume} {99}},\ \bibinfo {pages} {195133}
  (\bibinfo {year} {2019}{\natexlab{a}})}\BibitemShut {NoStop}%
\bibitem [{\citenamefont {Glorioso}\ \emph {et~al.}(2019)\citenamefont
  {Glorioso}, \citenamefont {Gromov},\ and\ \citenamefont {Ryu}}]{eft}%
  \BibitemOpen
  \bibfield  {author} {\bibinfo {author} {\bibfnamefont {P.}~\bibnamefont
  {Glorioso}}, \bibinfo {author} {\bibfnamefont {A.}~\bibnamefont {Gromov}}, \
  and\ \bibinfo {author} {\bibfnamefont {S.}~\bibnamefont {Ryu}},\ }\href
  {https://arxiv.org/abs/1908.03217} {\bibfield  {journal} {\bibinfo  {journal}
  {arXiv preprint arXiv:1908.03217}\ } (\bibinfo {year} {2019})}\BibitemShut
  {NoStop}%
\bibitem [{\citenamefont {Hastings}(2013)}]{kirby}%
  \BibitemOpen
  \bibfield  {author} {\bibinfo {author} {\bibfnamefont {M.~B.}\ \bibnamefont
  {Hastings}},\ }\href {\doibase 10.1103/PhysRevB.88.165114} {\bibfield
  {journal} {\bibinfo  {journal} {Phys. Rev. B}\ }\textbf {\bibinfo {volume}
  {88}},\ \bibinfo {pages} {165114} (\bibinfo {year} {2013})}\BibitemShut
  {NoStop}%
\bibitem [{\citenamefont {Gong}\ \emph {et~al.}(2020)\citenamefont {Gong},
  \citenamefont {S\"underhauf}, \citenamefont {Schuch},\ and\ \citenamefont
  {Cirac}}]{mpu}%
  \BibitemOpen
  \bibfield  {author} {\bibinfo {author} {\bibfnamefont {Z.}~\bibnamefont
  {Gong}}, \bibinfo {author} {\bibfnamefont {C.}~\bibnamefont {S\"underhauf}},
  \bibinfo {author} {\bibfnamefont {N.}~\bibnamefont {Schuch}}, \ and\ \bibinfo
  {author} {\bibfnamefont {J.~I.}\ \bibnamefont {Cirac}},\ }\href {\doibase
  10.1103/PhysRevLett.124.100402} {\bibfield  {journal} {\bibinfo  {journal}
  {Phys. Rev. Lett.}\ }\textbf {\bibinfo {volume} {124}},\ \bibinfo {pages}
  {100402} (\bibinfo {year} {2020})}\BibitemShut {NoStop}%
\bibitem [{\citenamefont {Roy}\ and\ \citenamefont
  {Harper}(2017)}]{alldimensions}%
  \BibitemOpen
  \bibfield  {author} {\bibinfo {author} {\bibfnamefont {R.}~\bibnamefont
  {Roy}}\ and\ \bibinfo {author} {\bibfnamefont {F.}~\bibnamefont {Harper}},\
  }\href {\doibase 10.1103/PhysRevB.95.195128} {\bibfield  {journal} {\bibinfo
  {journal} {Phys. Rev. B}\ }\textbf {\bibinfo {volume} {95}},\ \bibinfo
  {pages} {195128} (\bibinfo {year} {2017})}\BibitemShut {NoStop}%
\bibitem [{\citenamefont {Bachmann}\ \emph {et~al.}(2020)\citenamefont
  {Bachmann}, \citenamefont {Bols}, \citenamefont {De~Roeck},\ and\
  \citenamefont {Fraas}}]{bachmanncharge}%
  \BibitemOpen
  \bibfield  {author} {\bibinfo {author} {\bibfnamefont {S.}~\bibnamefont
  {Bachmann}}, \bibinfo {author} {\bibfnamefont {A.}~\bibnamefont {Bols}},
  \bibinfo {author} {\bibfnamefont {W.}~\bibnamefont {De~Roeck}}, \ and\
  \bibinfo {author} {\bibfnamefont {M.}~\bibnamefont {Fraas}},\ }\href
  {\doibase 10.1007/s00220-019-03537-x} {\bibfield  {journal} {\bibinfo
  {journal} {Communications in Mathematical Physics}\ }\textbf {\bibinfo
  {volume} {375}},\ \bibinfo {pages} {1249} (\bibinfo {year}
  {2020})}\BibitemShut {NoStop}%
\bibitem [{\citenamefont {Thouless}(1983)}]{Thouless}%
  \BibitemOpen
  \bibfield  {author} {\bibinfo {author} {\bibfnamefont {D.~J.}\ \bibnamefont
  {Thouless}},\ }\href {\doibase 10.1103/PhysRevB.27.6083} {\bibfield
  {journal} {\bibinfo  {journal} {Phys. Rev. B}\ }\textbf {\bibinfo {volume}
  {27}},\ \bibinfo {pages} {6083} (\bibinfo {year} {1983})}\BibitemShut
  {NoStop}%
\bibitem [{\citenamefont {Hastings}\ and\ \citenamefont
  {Michalakis}(2015)}]{HastingsMichalakis}%
  \BibitemOpen
  \bibfield  {author} {\bibinfo {author} {\bibfnamefont {M.~B.}\ \bibnamefont
  {Hastings}}\ and\ \bibinfo {author} {\bibfnamefont {S.}~\bibnamefont
  {Michalakis}},\ }\href {\doibase 10.1007/s00220-014-2167-x} {\bibfield
  {journal} {\bibinfo  {journal} {Communications in Mathematical Physics}\
  }\textbf {\bibinfo {volume} {334}},\ \bibinfo {pages} {433} (\bibinfo {year}
  {2015})}\BibitemShut {NoStop}%
\bibitem [{\citenamefont {Kapustin}\ and\ \citenamefont
  {Fidkowski}(2019)}]{nogo}%
  \BibitemOpen
  \bibfield  {author} {\bibinfo {author} {\bibfnamefont {A.}~\bibnamefont
  {Kapustin}}\ and\ \bibinfo {author} {\bibfnamefont {L.}~\bibnamefont
  {Fidkowski}},\ }\href {\doibase 10.1007/s00220-019-03444-1} {\bibfield
  {journal} {\bibinfo  {journal} {Communications in Mathematical Physics}\ ,\
  \bibinfo {pages} {1}} (\bibinfo {year} {2019})}\BibitemShut {NoStop}%
\bibitem [{\citenamefont {Nathan}\ \emph
  {et~al.}(2019{\natexlab{b}})\citenamefont {Nathan}, \citenamefont {Abanin},
  \citenamefont {Lindner}, \citenamefont {Berg},\ and\ \citenamefont
  {Rudner}}]{hierarchy}%
  \BibitemOpen
  \bibfield  {author} {\bibinfo {author} {\bibfnamefont {F.}~\bibnamefont
  {Nathan}}, \bibinfo {author} {\bibfnamefont {D.~A.}\ \bibnamefont {Abanin}},
  \bibinfo {author} {\bibfnamefont {N.~H.}\ \bibnamefont {Lindner}}, \bibinfo
  {author} {\bibfnamefont {E.}~\bibnamefont {Berg}}, \ and\ \bibinfo {author}
  {\bibfnamefont {M.~S.}\ \bibnamefont {Rudner}},\ }\href
  {https://arxiv.org/abs/1907.12228} {\bibfield  {journal} {\bibinfo  {journal}
  {arXiv:1907.12228}\ } (\bibinfo {year} {2019}{\natexlab{b}})}\BibitemShut
  {NoStop}%
\bibitem [{\citenamefont {Po}\ \emph {et~al.}(2017)\citenamefont {Po},
  \citenamefont {Fidkowski}, \citenamefont {Vishwanath},\ and\ \citenamefont
  {Potter}}]{radical}%
  \BibitemOpen
  \bibfield  {author} {\bibinfo {author} {\bibfnamefont {H.~C.}\ \bibnamefont
  {Po}}, \bibinfo {author} {\bibfnamefont {L.}~\bibnamefont {Fidkowski}},
  \bibinfo {author} {\bibfnamefont {A.}~\bibnamefont {Vishwanath}}, \ and\
  \bibinfo {author} {\bibfnamefont {A.~C.}\ \bibnamefont {Potter}},\ }\href
  {\doibase 10.1103/PhysRevB.96.245116} {\bibfield  {journal} {\bibinfo
  {journal} {Phys. Rev. B}\ }\textbf {\bibinfo {volume} {96}},\ \bibinfo
  {pages} {245116} (\bibinfo {year} {2017})}\BibitemShut {NoStop}%
\bibitem [{\citenamefont {Potter}\ and\ \citenamefont
  {Morimoto}(2017)}]{dynamically}%
  \BibitemOpen
  \bibfield  {author} {\bibinfo {author} {\bibfnamefont {A.~C.}\ \bibnamefont
  {Potter}}\ and\ \bibinfo {author} {\bibfnamefont {T.}~\bibnamefont
  {Morimoto}},\ }\href {\doibase 10.1103/PhysRevB.95.155126} {\bibfield
  {journal} {\bibinfo  {journal} {Phys. Rev. B}\ }\textbf {\bibinfo {volume}
  {95}},\ \bibinfo {pages} {155126} (\bibinfo {year} {2017})}\BibitemShut
  {NoStop}%
\bibitem [{\citenamefont {Freedman}\ and\ \citenamefont
  {Hastings}(2020)}]{freedmanclassification}%
  \BibitemOpen
  \bibfield  {author} {\bibinfo {author} {\bibfnamefont {M.}~\bibnamefont
  {Freedman}}\ and\ \bibinfo {author} {\bibfnamefont {M.~B.}\ \bibnamefont
  {Hastings}},\ }\href {\doibase 10.1007/s00220-020-03735-y} {\bibfield
  {journal} {\bibinfo  {journal} {Communications in Mathematical Physics}\
  }\textbf {\bibinfo {volume} {376}},\ \bibinfo {pages} {1171} (\bibinfo {year}
  {2020})}\BibitemShut {NoStop}%
\bibitem [{\citenamefont {Haah}(2019)}]{haahclifford}%
  \BibitemOpen
  \bibfield  {author} {\bibinfo {author} {\bibfnamefont {J.}~\bibnamefont
  {Haah}},\ }\href {https://arxiv.org/abs/1907.02075} {\bibfield  {journal}
  {\bibinfo  {journal} {arXiv:1907.02075}\ } (\bibinfo {year}
  {2019})}\BibitemShut {NoStop}%
\bibitem [{\citenamefont {Haah}\ \emph {et~al.}(2018)\citenamefont {Haah},
  \citenamefont {Fidkowski},\ and\ \citenamefont {Hastings}}]{haahnontrivial}%
  \BibitemOpen
  \bibfield  {author} {\bibinfo {author} {\bibfnamefont {J.}~\bibnamefont
  {Haah}}, \bibinfo {author} {\bibfnamefont {L.}~\bibnamefont {Fidkowski}}, \
  and\ \bibinfo {author} {\bibfnamefont {M.~B.}\ \bibnamefont {Hastings}},\
  }\href {https://arxiv.org/abs/1812.01625} {\bibfield  {journal} {\bibinfo
  {journal} {arXiv preprint arXiv:1812.01625}\ } (\bibinfo {year}
  {2018})}\BibitemShut {NoStop}%
\bibitem [{\citenamefont {Reiss}\ \emph {et~al.}(2018)\citenamefont {Reiss},
  \citenamefont {Harper},\ and\ \citenamefont {Roy}}]{reiss3D}%
  \BibitemOpen
  \bibfield  {author} {\bibinfo {author} {\bibfnamefont {D.}~\bibnamefont
  {Reiss}}, \bibinfo {author} {\bibfnamefont {F.}~\bibnamefont {Harper}}, \
  and\ \bibinfo {author} {\bibfnamefont {R.}~\bibnamefont {Roy}},\ }\href
  {\doibase 10.1103/PhysRevB.98.045127} {\bibfield  {journal} {\bibinfo
  {journal} {Phys. Rev. B}\ }\textbf {\bibinfo {volume} {98}},\ \bibinfo
  {pages} {045127} (\bibinfo {year} {2018})}\BibitemShut {NoStop}%
\bibitem [{\citenamefont {Huse}\ \emph {et~al.}(2013)\citenamefont {Huse},
  \citenamefont {Nandkishore}, \citenamefont {Oganesyan}, \citenamefont {Pal},\
  and\ \citenamefont {Sondhi}}]{huse_localization}%
  \BibitemOpen
  \bibfield  {author} {\bibinfo {author} {\bibfnamefont {D.~A.}\ \bibnamefont
  {Huse}}, \bibinfo {author} {\bibfnamefont {R.}~\bibnamefont {Nandkishore}},
  \bibinfo {author} {\bibfnamefont {V.}~\bibnamefont {Oganesyan}}, \bibinfo
  {author} {\bibfnamefont {A.}~\bibnamefont {Pal}}, \ and\ \bibinfo {author}
  {\bibfnamefont {S.~L.}\ \bibnamefont {Sondhi}},\ }\href {\doibase
  10.1103/PhysRevB.88.014206} {\bibfield  {journal} {\bibinfo  {journal} {Phys.
  Rev. B}\ }\textbf {\bibinfo {volume} {88}},\ \bibinfo {pages} {014206}
  (\bibinfo {year} {2013})}\BibitemShut {NoStop}%
\bibitem [{\citenamefont {Chandran}\ \emph {et~al.}(2014)\citenamefont
  {Chandran}, \citenamefont {Khemani}, \citenamefont {Laumann},\ and\
  \citenamefont {Sondhi}}]{mblspt}%
  \BibitemOpen
  \bibfield  {author} {\bibinfo {author} {\bibfnamefont {A.}~\bibnamefont
  {Chandran}}, \bibinfo {author} {\bibfnamefont {V.}~\bibnamefont {Khemani}},
  \bibinfo {author} {\bibfnamefont {C.~R.}\ \bibnamefont {Laumann}}, \ and\
  \bibinfo {author} {\bibfnamefont {S.~L.}\ \bibnamefont {Sondhi}},\ }\href
  {\doibase 10.1103/PhysRevB.89.144201} {\bibfield  {journal} {\bibinfo
  {journal} {Phys. Rev. B}\ }\textbf {\bibinfo {volume} {89}},\ \bibinfo
  {pages} {144201} (\bibinfo {year} {2014})}\BibitemShut {NoStop}%
\bibitem [{\citenamefont {Powers}\ and\ \citenamefont {Reznick}(2001)}]{polya}%
  \BibitemOpen
  \bibfield  {author} {\bibinfo {author} {\bibfnamefont {V.}~\bibnamefont
  {Powers}}\ and\ \bibinfo {author} {\bibfnamefont {B.}~\bibnamefont
  {Reznick}},\ }\href {\doibase https://doi.org/10.1016/S0022-4049(00)00155-9}
  {\bibfield  {journal} {\bibinfo  {journal} {Journal of pure and applied
  algebra}\ }\textbf {\bibinfo {volume} {164}},\ \bibinfo {pages} {221}
  (\bibinfo {year} {2001})}\BibitemShut {NoStop}%
\end{thebibliography}%

\end{document}